\documentclass[showkeys,aps,pre,preprint,showpacs,preprintnumbers,amsmath,amssymb]{revtex4-1}
\usepackage{amsmath}
\usepackage{amsthm}
\usepackage{amssymb}
\usepackage[cp1251]{inputenc}
\usepackage[german,english]{babel}
\usepackage[showlabels,sections,floats,textmath,displaymath]{preview}

\newtheorem{prop}{Proposition}
\newtheorem*{rem}{Remark}

\begin{document}

\title{Asymptotic Solutions of the Kinetic Boltzmann Equation, 
Multicomponent Non-equilibrium Gas Dynamics and Turbulence}

\author{S.~A.~Serov}
 \email{serov@vniief.ru}
 \affiliation{
Institute of Theoretical and Mathematical Physics, Sarov, Russia}
\author{S.~S.~Serova}
 \affiliation{St. Petersburg State University, 
   St. Petersburg, Russia}

\date{\today}

\begin{abstract}
In the article correct method for the kinetic Boltzmann equation
asymptotic solution is formulated, the Hilbert's and Enskog's methods
are discussed. 
The equations system of multicomponent non-equilibrium gas dynamics is
derived, that corresponds to the first order in the approximate
(asymptotic) method for solution of the system of kinetic Boltzmann
equations. 
It is shown, that the velocity distribution functions of particles, obtained by the
proposed method and by Enskog's method, within Enskog's approach, are
equivalent up to the infinitesimal first order terms of the asymptotic
expansion, but, generally speaking, differ in the next order.
Interpretation of turbulent gas flow is proposed, as stratified 
on components gas flow, which is described by the derived equations
system of multicomponent non-equilibrium gas dynamics.
\end{abstract}


\maketitle

\section{Introduction}
\label{sec:introduction}

In 1912 Hilbert considered the kinetic Boltzmann equation for
one-component gas as an example of integral equation and proposed a
"recipe" for its approximate (asymptotic) solution (see
\cite{hilbert12}, Chapter~XXII). 
Hilbert's "recipe" was inconvenient for practical use, because the
five arbitrary functional parameters of the first and the following
approximations of the velocity distribution function had to be found
by solving the differential equations in partial derivatives
(equations of gas dynamics of the first and higher orders).  
Five years later Enskog proposed to use zero conditions, conditions
(\ref{r_condition1_in})-(\ref{r_condition3n_in}) below with zero
right-hand sides, to determine the five arbitrary functional parameters
of the first and following approximations of the velocity distribution
function. 
As we will see in the section \ref {sec:remarks} (in case of
one-component gas), the imposition of the zero conditions leads, in
fact, to using different comparison scales in the asymptotic expansion
of the velocity distribution function and in the asymptotic expansion
of the particle number density, the mean (mass) velocity and the
temperature, that are derived from the asymptotic expansion of the
velocity distribution function by integration over velocities with
different weighting functions. As a result of paralogism of the method
of successive approximations (one has to set variable coefficients of
the same terms of the unified comparison scale equal to each other)
partial time derivatives vanish in the necessary conditions of
solutions existence of integral equations of higher orders (see below)
and with them terms of gas-dynamic equations, corresponding to viscosity,
heat conduction, \ldots \ vanish.  
Enskog "improved" the situation by the introducing
(see, for example, \cite{chap52}, Chapter~7, \S ~1, Section~5) of the
unsubstantiated expansion of partial time derivative.

The approach of Struminskii, who had proposed in 1974 in
\cite{strum74} his approximate(asymptotic) method of solution of the
system of kinetic Boltzmann equations for multicomponent gas, differs
from the approach of Enskog to asymptotic solution of the Boltzmann
equations system for gas mixture in that, how the infinitesimal
parameter is introduced in the Boltzmann equations system for gas
mixture, i.e. the solution is constructing in another asymptotic
limit. 
It could be noted, that approaches, that are similar to Struminskii's
approach to asymptotic solution of the kinetic Boltzmann equations
system, were regarded earlier in the kinetic theory of plasma (for
instance, see \cite{gurov66}, \S ~7.5).  
In substance, Struminskii's method of solution of kinetic equations
system is the same as Enskog's method (Struminskii used the partial
time derivative expansion, as Enskog did). 

After some preliminary remarks in section \ref{sec:remarks}, in section 
\ref{sec:method} there will be proposed \textit{the correct method of
  asymptotic solution of the kinetic Boltzmann equations system for
  multicomponent gas mixture} for the approach, that combines Enskog's
and Struminskii's approaches; in particular, it will be shown, how one
has to modify Enskog's method -- in addition to asymptotic expansion
of the velocity distribution function of $i$-component particles of gas mixture
it is necessary to determine and to use the expansion of the particle
number density $n_{i}$ of $i$-component, mean mass velocity
${\boldsymbol{u}}$ and temperature $T$ of the gas mixture. 
 
Further, in the section \ref{sec:order1gdes} 
\textit{the system of infinitesimal first order equations of multicomponent
non-equilibrium gas dynamics}, appearing during the process of the
solution of the system of Boltzmann equations by successive
approximations method in the section \ref{sec:method} as necessary
condition of the existence of approximate (asymptotic) solution of the
integral equations system, is considered in more detail.

In section \ref{sec:order1df} 
\textit{the expressions for the velocity distribution function of particles of gas 
mixture in the infinitesimal first order} are obtained for the
approach, that Enskog's approach and 
Struminskii's approach combines, in particular it is shown, that the
velocity distribution functions, that are derived within Enskog's
approach by Enskog's method and the proposed method, are equivalent up
to the infinitesimal first order terms of the asymptotic expansion (inclusive),
but, generally speaking, differ in the next order of the asymptotic
expansion; the difference is a possible reason, why the transition to
a higher order approximation in Enskog's method does not lead to any
essential improvements of the result. 

In section \ref{sec:turbulence} \textit{an interpretation of
turbulent flows is proposed within the multicomponent gas dynamics}. 

Struminskii's article \cite{strum74}, and, as a result, the followings
articles, that referred to the paper \cite{strum74} (for example,
\cite{strum82}), contained errors in calculation of collision
integrals%
. 
In appendix \ref{sec:integration} general analytic expressions of
these collision integrals (that depend on the interaction
cross-section) are derived in general case, when separate components
(with Maxwell velocity distribution function of particles) have different
mean velocities and temperatures. 

For the elementary model of an interaction of gas mixture particles --
the interaction of particles as rigid spheres -- collision integrals
can be simply calculated completely, the expressions for the
interaction potential of rigid spheres are written in appendix
\ref{sec:integrals}.  

Notations, used below, are close to notations in \cite{chap52}; it is
assumed, that all regarded functions are continuous and continuously
differentiable so many times as it is necessary, if their derivatives
are considered, and all regarded integrals converge.

\section{Some remarks}
\label{sec:remarks}

In kinetic theory of gases and gas dynamics, closely connected with it,
there are no small dimensionless constants, similar, for example, to the
fine structure constant in quantum electrodynamics, in terms of which it would 
be possible to construct perturbation theory expansions, paying special
attention to convergence of derivable expansions. 
In particular applications of kinetic theory of gases and gas dynamics
it is possible often to make up from parameters of the problem a
variable dimensionless value (in gas dynamics such variable
dimensionless values are named numbers: the Mach number, the Reynolds
number \ldots ), 
we will denote this value by $\theta$, and to consider \textit{asymptotic expansions
with variable coefficients} (\cite{bourbaki2004}, Chapter~V, \S ~2,
Section~5) of physical values in terms of integral powers of $\theta$ -- 
$\left\{ {\theta ^r } \right\}_{r \ge 0} $, when $\theta $ tends to
$0$ (most important asymptotic expansions, further just those
asymptotic expansions are considered; 
generally speaking, it is possible to use also integer negative powers
$r$, but we shall not do it), or more rarely 
$\left\{ {\theta ^{-r} } \right\}_{r \ge 0} $, when $\theta $ tends to 
$+ \infty $ or to $- \infty $; changing $\theta $ to ${1\!\left/ \theta \right.}$, 
it is possible to switch easily from one of these cases to another. 
In the theory of asymptotic expansions variable coefficients are
considered to be comparable to the unit (and may depend on $\theta $
in general case, but not in this article) and functions of a 
\textit{comparison scale} ($\left\{ {\theta ^r } \right\}_{r \ge 0} $)
-- to be infinitesimal. 
If an asymptotic solution to specified precision (${\theta ^r }$)
relative to a  comparison scale 
($\left\{ {\theta ^r } \right\}_{r \ge 0}$) exists, it is  
\textit{unique} 
(see, for example, \cite{bourbaki2004}, Chapter~V, \S ~2, Section~2)
and, if asymptotic expansions are considered at $\theta \to 0$, is
\textit{exact}, at least, 
at $\theta = 0$, therefore it is convenient to consider the asymptotic expansions at 
$\theta \to 0$ (and integer ${r \ge 0}$).

In applications of the theory of asymptotic expansions, solving
physical equations by the successive approximations method, it is
convenient not to extract the dimensionless parameter $\theta ^r$
from a term of asymptotic expansion 
$\theta ^r K^{\left( r \right)} \left( x \right)$ as a really small 
multiplier, but to use $\theta ^r$ as an 
\textit{indicator of infinitesimal order} of 
corresponding (infinitely) small variable coefficient 
$K^{\left( r \right)}\left( x \right)$, setting in the resultant expressions 
$\theta =1$;
with such agreement the formal "small" parameter $\theta $ can be
introduced in an equation arbitrarily, but the meaning of obtained
asymptotic solutions of such equation with small parameter is
determined by physical validity of the introduction of the small
parameter in the original equation. 

By method of successive approximations we obtain, generally speaking,
an \textit{asymptotic} solution of the problem.
If some additional conditions exist (for example, initial conditions
or boundary conditions for a differential equation), the conditions
must be expanded in asymptotic series, and the obtained equations
system have to be solved, setting variable coefficient at the same
terms of \textit{unified} comparison scale equal to each other.  
If the system is solved successfully, we obtain an asymptotic
solution of the initial equation, that meets (asymptotically) the
additional conditions. 
In some cases the found asymptotic solution turns out to be 
\textit{regular} solution (\cite{lomov81}, Chapter~1), i.e. it depends
analytically on $\theta $, or even exact solution of the problem. 
For example, the equation 
(see \cite{cercignani75}, Chapter~V, \S ~2 or 
\cite{resibois77}, Chapter~IV, \S ~7.1)
\begin{eqnarray}
\theta \,\frac{df}{dt}+f=0 ,\qquad
f\left( 0\right) =0
\label{example0}
\end{eqnarray}
with zero initial condition has asymptotic solution 
$f^{\left( r\right)}\left( t\right) \equiv 0$, $\left( r=0,1,2\ldots
\right)$ relative to the comparison scale, formed of functions 
$\theta ^{r}$, 
that is also the exact solution of the problem.
But the problem 
\begin{eqnarray}
\theta \,\frac{df}{dt}+f=0 ,\qquad
f\left( 0\right) =1
\label{example1}
\end{eqnarray}
has no analogous asymptotic solution, 
because the zero order asymptotic solution of the differential equation 
$f^{\left( 0\right)}\left( t\right) \equiv 0$ 
contradicts to the initial condition $f^{\left( 0\right)}\left( 0\right) =1$.
It is not a serious drawback
of the method of successive approximations.
In the problem (\ref{example1}) 
one can introduce new function $g\left( t\right)$:
\begin{eqnarray}
f\left( t\right) = \exp {\left( {-t/\theta }\right)}\, g\left( t\right)
\label{new_function}
\end{eqnarray}
(cf. with \cite{lomov81}; 
though not all considered by Lomov expansions are the asymptotic expansions 
with variable coefficients, and, accordingly, Lomov's approach in toto
seems not quite correct, 
in the monograph \cite{lomov81} it is actually shown, that similar
replacements of functions allow to obtain asymptotic solutions for a
wide class of problems) 
and obtain asymptotic solution of the problem by the successive
approximations method 
$g^{\left( 0\right)}\left( t\right) \equiv \mathrm{const} =1$, 
$g^{\left( r\right)}\left( t\right) \equiv \mathrm{const} =0$ 
$\left( r=1,2\ldots \right)$, that is again the exact solution.

Criticism of the successive approximations method in
\cite{cercignani75}, Chapter~V, \S ~2 and \cite{resibois77},
Chapter~IV, \S ~7.1 reflects, possibly, dissatisfaction of the authors 
with the introduction of the unsubstantiated expansion of partial time
derivative in Enskog's method for the kinetic Boltzmann equation
asymptotic solution, see below. 

Hilbert noted in \cite{hilbert12}, Chapter~XXII, that the
expansion of the velocity distribution function
\begin{eqnarray}
F  = \frac{\Phi }
{\lambda } + {\Psi }  + X \lambda +\dotsb 
\label{f_seriesHilbert}
\end{eqnarray}
[similar to (\ref{f_seriesEnscog}) below; 
Hilbert considered one-component gas only;
we use Hilbert's notations here, however, the meaning of the notations
is sufficiently understandable from context] 
is the power series in (a small parameter) $\lambda $, 
satisfying to the Boltzmann equation and such, that expressions 
[cf. with 
(\ref{condition1})-(\ref{condition3n_in}) and
(\ref{n_series})-(\ref{T_series_in}) below]
\begin{eqnarray}
\int \psi ^{\left( i\right)} F\,d {\omega }
&=&\frac{1}{\lambda }
\int \psi ^{\left( i\right)} {\Phi }\,d {\omega }
+
\int \psi ^{\left( i\right)} {\Psi }\,d {\omega }
\nonumber\\
&&+ \,\lambda 
\int \psi ^{\left( i\right)} X\,d {\omega }
+\dotsb 
\qquad {\left( i=1,2,3,4,5\right)} 
\label{fint_seriesHilbert}
\end{eqnarray}
for $t=t_0$ change to power series
\begin{eqnarray}
\Lambda ^{\left( i \right)}  = \frac{{f^{\left( i \right)} }}
{\lambda } + g^{\left( i \right)}  + \lambda h^{\left( i \right)} 
+\dotsb 
\qquad {\left( i=1,2,3,4,5\right)} ,
\label{f0int_seriesHilbert}
\end{eqnarray}
and in the theorem, completing the work, he has formulated a "recipe" to
get an asymptotic solution for the Boltzmann equation, in which he
proposed to determine five arbitrary functional parameters of
functions ${\Phi }, {\Psi }, X\ldots $ 
"\textit{from five partial differential equations}"
[analogous (\ref{r_eq_in_condiition}), (\ref{r_eq_out_condiition}) below], 
"\textit{setting at $t=t_{0} $}"  
\begin{eqnarray}
\int \psi ^{\left( i\right)} {\Phi }\,d {\omega }
&=&{\lambda }\Lambda ^{\left( i \right)} 
\qquad {\left( i=1,2,3,4,5\right)} ,
\label{Phi_conditionHilbert}
\\
\int \psi ^{\left( i\right)} \Psi \,d {\omega }
&=&0 
\!\!\qquad \qquad {\left( i=1,2,3,4,5\right)} ,
\label{Psi_conditionHilbert}
\\
\int \psi ^{\left( i\right)} X\,d {\omega }
&=&0 
\!\!\qquad \qquad {\left( i=1,2,3,4,5\right)} .
\label{X_conditionHilbert}
\end{eqnarray}
In notations from (\ref{psi}),
(\ref{condition1})-(\ref{r_condition3n_in}) below, Hilbert
proposed simply to set special \textit{initial values} 
\begin{eqnarray}
n\left( {\boldsymbol{r}},t_0\right) 
&=&
n^{\left( 0\right)}\left( {\boldsymbol{r}},t_0\right),
\label{n0_Hilbert} \\
{\boldsymbol{u}} \left( {\boldsymbol{r}},t_0\right) 
&=& 
{\boldsymbol{u}}^{\left( 0\right)}\left( {\boldsymbol{r}},t_0\right) ,
\label{u0_Hilbert} \\
T\left( {\boldsymbol{r}},t_0\right)  
&=& 
T^{\left( 0\right)}\left( {\boldsymbol{r}},t_0\right) 
\label{T0_Hilbert}
\end{eqnarray}
or
\begin{eqnarray}
\left. {\int \psi ^{\left( l\right)}
f^{\left( r\right)}\,d {\boldsymbol{c}}} \right|_{t = t_0 } 
\ {\overset{\boldsymbol{r}}{\equiv}}\ 0 
\qquad {\left( l=1,2,3\right)} 
\label{psi0_r}
\end{eqnarray}
for $r=1,2\ldots $ .

"For the further substantiation of the gas theory" it might be as well 
to supplement Hilbert's theorem with explicit expressions
of five arbitrary functional parameters of the functions 
$f _ {i} ^ {\left( r\right)} $, found on the $r$-step
($r=0,1,2\ldots $) 
of the successive approximations method, 
through through the physical parameters of the gas  
[see (\ref{beta_1_r_in})-(\ref{beta_3_r_in}),
(\ref{beta_1_r_out})-(\ref{beta_3_r_out}) below];
Hilbert had not done it.

Enskog supplemented Hilbert's
"recipe". 
However, at that Enskog made a logical mistake. 
He used zero
conditions (\ref{psi0_r}) \textit{identically} at any $t$, 
not at $t=t_{0}$ only 
(see \cite{chap52}, Chapter~7, \S ~1, Section~1):  
\begin{eqnarray}
{\int \psi ^{\left( l\right)}
f^{\left( r\right)}\,d {\boldsymbol{c}}}
\ {\overset{\boldsymbol{r}{,}\,t}{\equiv}}\ 0 
\qquad {\left( l=1,2,3\right)} 
\label{psi_r_Enskog}
\end{eqnarray}
for $r=1,2\ldots $ .
From the viewpoint of the theory of asymptotic expansions, Enskog
[instead of (\ref{n_series})-(\ref{T_series_in}) below] supposed
\begin{eqnarray}
n\left( {\boldsymbol{r}},t,\theta \right) 
&=&
{\theta ^{\,0}}\,n\left( {\boldsymbol{r}},t,\theta \right) 
+{\theta ^1}\,0+{\theta ^2}\,0 +\dotsb ,
\label{n_Enskog} \\
{\boldsymbol{u}} \left( {\boldsymbol{r}},t,\theta \right) 
&=& 
{\theta ^{\,0}}\,{\boldsymbol{u}} \left( {\boldsymbol{r}},t,\theta \right) 
+{\theta ^1}\,0+{\theta ^2}\,0 +\dotsb ,
\label{u_Enskog} \\
T\left( {\boldsymbol{r}},t,\theta \right) 
&=&
{\theta ^{\,0}}\,T\left( {\boldsymbol{r}},t,\theta \right) 
+{\theta ^1}\,0+{\theta ^2}\,0 +\dotsb .
\label{T_Enskog}
\end{eqnarray}
If $n$, ${\boldsymbol{u}}$ and $T$ did not depend on $\boldsymbol{r}$ 
and $t$, that would mean,
that Enskog used different scales of comparison in the method of
successive approximations at the same time 
$\left \{ n\left( \theta \right) ,\theta ^1,\theta ^2 \ldots \right \}$,
$\left \{ {\boldsymbol{u}}\left( \theta \right) ,\theta ^1,\theta ^2
\ldots \right \}$,
$\left \{ T\left( \theta \right) ,\theta ^1,\theta ^2 
\ldots \right \}$,
that is in itself wrong.
In general case,
when $n$, ${\boldsymbol {u}}$ and $T$ depend on 
$\boldsymbol{r}$ and $t$, sums
(\ref{n_Enskog})-(\ref{T_Enskog}) can not even
be considered as asymptotic expansions with variable
coefficients. 

Violation of logic of the successive approximations method shows itself
immediately in vanishing of partial time derivatives in gas-dynamics
equations systems of the ${\left( r+1\right)}$-order 
[analogous (\ref{r_eq_in_condiition}) 
${\left( r=1,2,\ldots \right)}$ below, 
the necessary conditions of solutions existence of the integral higher orders
equations in Enskog's approach], according to
(\ref{psi_r_Enskog}),
\begin{eqnarray}
\int \psi ^{\left( l\right)} 
\frac{\partial f^{\left( r\right)}}{\partial t}d {\boldsymbol{c}}
=
\frac{\partial }{\partial t}
\int \psi ^{\left( l\right)} f^{\left( r\right)}d {\boldsymbol{c}}
=0
\qquad {\left( l=1,2,3\right)} ,
\label{timeD}
\end{eqnarray}
and with them terms of gas-dynamic equations, corresponding to viscosity,
heat conduction, \ldots , vanish.

Hence Enskog should have concluded, that the use of
(\ref{psi_r_Enskog}) is incorrect, as in proof of 
theorem by contradiction, but instead of that Enskog proposed to use
the unsubstantiated expansion of partial time
derivative
\begin{eqnarray}
\frac{\partial }
{{\partial t}} = 
\sum\limits_{r=0}^\infty  {\theta ^r } \frac{{\partial _r }}
{{\partial t}} . 
\label{t_series_Enskog}
\end{eqnarray}

\section{Correct method of solution of the kinetic Boltzmann equations system}
\label{sec:method}

The Boltzmann equations system, that describes change of
dependent on $t$ and spatial coordinates, prescribed by radius-vector
${\boldsymbol{r}}$, the velocity distribution functions
$f_{i}\left( t, {\boldsymbol{r}}, {\boldsymbol{c}}_{i} \right)$ due to 
collision with particles of other 
components of mixture of rarefied monatomic gases, where
${\boldsymbol{c}}_{i}$ are the velocities of 
particles of $i$-component of the mixture
\{see \cite{chap52}, Chapter~8, (1.1);
discussion of the derivation of the Boltzmann equations system and
its applicability range see, for example, in
\cite{chap52}, Chapters~3 and 18, \cite{hirsch54}, Chapter~7, \S ~1,
\cite{ferziger72}, Chapter~3  
and the Bogolyubov paper \cite{bog46};
below the \textit{central interaction} of molecules are considered only, when
the force, with which each molecule acts on the other, is directed
along the line, connecting the centers of the molecules\}, 
could be written as:  
\begin{eqnarray}
\frac{\partial f_{i}}{\partial t}+
{\boldsymbol{c}}_{i}\cdot \frac{\partial f_{i}}{\partial \boldsymbol{r}}
+ \frac{{\boldsymbol{X}}_{i}}{m_{i}}\cdot \frac{\partial f_{i}}
{\partial {\boldsymbol{c}}_{i}}
&=&
\sum\limits_{{j} \in {N} } \iiint
\left( f_{i}^{\prime }f_{j}^{\prime }-f_{i}f_{j} \right)
g_{ij}\, b\,db \,d\epsilon \,d{\boldsymbol{c}}_{j} \nonumber\\*
&=&
\sum\limits_{{j} \in {N} } \iint
\left( f_{i}^{\prime }f_{j}^{\prime }-f_{i}f_{j} \right)
k_{ij} \,d{\mathbf{k}} 
\,d{\boldsymbol{c}}_{j} 
\qquad {\left( {i} \in {N} \right)} ;
\label{boltzmannend}
\end{eqnarray}
in (\ref{boltzmannend}) $N$ is a set of indexes, that are
numbering components of the mixture; 
${\boldsymbol{X}}_{i}$ is an external force,
which acts on the molecule of the $i$-component; 
$m_{i}$ is the  mass of the molecule of the $i$-component; 
$g_{ij}$ is the modulus of the relative velocity of colliding
particles
${\boldsymbol{g}}_{ij}={\boldsymbol{c}}_{i}-{\boldsymbol{c}}_{j}$; 
$b$ is the impact distance, 
$\epsilon $ is the azimuth angle,
${\mathbf{k}}$ is the unit vector, directed to the center of mass of the
colliding particles from the point of closest approach --
see \cite{chap52}, Chapter~3, Fig. 3; 
the scalar function
$k_{ij} \left( {\boldsymbol{g}}_{ij} ,{\mathbf{k}} \right)$ is
determined by equality
\begin{eqnarray}
g_{ij}\, b\,db \,d\epsilon 
\ {\overset{\mathrm{def}}{=}}\ 
k_{ij} \,d{\mathbf{k}} ;
\label{def_k_ij}
\end{eqnarray}
by prime in (\ref{boltzmannend}) and below the velocities
and the functions of velocities after the collision are denoted. 

Let us introduce following notations: 
\begin{eqnarray}
J_{i}\left( f_{i},f\right) &=&
\iint
\left( f_{i}f-f_{i}^{\prime }f^{\prime } \right)
k_{i} \,d {\mathbf{k}}
\,d {\boldsymbol{c}} ,
\label{J_i}
\\
J_{ij}\left( f_{i},f_{j}\right) &=& 
\iint
\left( f_{i}f_{j}-f_{i}^{\prime }f_{j}^{\prime } \right)
k_{ij} \,d {\mathbf{k}} 
\,d {\boldsymbol{c}}_{j} ;
\label{J_ij}
\end{eqnarray}
to differ velocities of colliding molecules of the
same kind in (\ref{J_i}) the one velocity is
denoted by ${\boldsymbol{c}}_{j}$ and the other is denoted by
${\boldsymbol{c}}$ (without any index) and the index of the
corresponding velocity distribution function $f$ is omitted.  

In Enskog's approach the differential parts of the Boltzmann equations
(\ref{boltzmannend}), that are denoted by ${{\mathcal D}_i}{f_i}$
below, are considered to be small as compared with 
the right-hand sides of equations (\ref{boltzmannend}) -- see
\cite{chap52}, Chapter~7, \S ~1, Section~5 -- therefore the 
indicator of infinite smallness ${\theta }$ is formally introduced in the
Boltzmann equations system in the following way:  
\begin{eqnarray}
{\theta }{{\mathcal D}_i}{f_i} =
- \sum\limits_{j}J_{ij}\left( f_{i},f_{j}\right) 
\qquad {\left( {i} \in {N} \right)}.
\label{epsenskog}
\end{eqnarray}
In Struminskii's approach to the asymptotic solution of the Boltzmann
equations system the differential parts of the Boltzmann equations
(\ref{boltzmannend}) and the
collision integrals of the particles of $i$-component with the
particles of the other components are considered to be small as
compared with the collision integral of the particles of $i$-component
between each other, 
therefore the indicator of infinitesimality ${\theta }$ is introduced in
the Boltzmann equations system in another way:  
\begin{eqnarray}
{\theta }{{\mathcal D}_i}{f_i} =
- J_{i}\left( f_{i},f\right)
- {\theta }\sum\limits_{j\neq i}J_{ij}\left( f_{i},f_{j}\right) 
\qquad {\left( {i} \in {N} \right)}.
\label{epsstrum}
\end{eqnarray}

It is possible to combine Enskog's approach with Struminskii's
approach. 
For this purpose we divide the set of mixture components $N$
into two subsets: the subset of components, that we call formally
\textit{inner components} (we could consider the case, when there are
some subsets of inner components, but this case does not fundamentally
differ from the one, considered below, the only difference is that the
notation become more complicated) and the subset of components, that
we call \textit{external components}. 
To differ the two groups of
mixture components we denote the subset of indexes of inner components
$\skew4\hat{N}$ as well as the indexes of inner components
${\skew5\hat{i}} \in \skew4\hat{N}$ by symbol "$\,\hat{}\,$" and the
subset of indexes of external components $\skew4\check{N}$ as well as
the indexes of external components 
${\skew5\check{i}} \in \skew4\check{N}$ by symbol "$\,\check{}\,$"; 
the intersection of the sets $\skew4\hat{N}$ and
$\skew4\check{N}$ is the empty set -- 
$\skew4\hat{N} \cap \skew4\check{N} = \emptyset$
and the union of these sets is the set of indexes of all mixture
components $\skew4\hat{N} \cup \skew4\check{N} = N$; 
if an assertion concerns both kinds of components the symbols
"$\,\hat{}\,$" and "$\,\check{}\,$" will be omitted. 
In new notations the
Boltzmann equations system can be rewritten as:
\begin{subequations}
\label{epseq}
 \begin{eqnarray}
{\theta }{{\mathcal D}_{\skew5\hat{i}}}{f_{\skew5\hat{i}}} &=&
- \sum\limits_{{\skew5\hat{j}} \in \skew4\hat{N}}
J_{{\skew5\hat{i}}{\skew5\hat{j}}}\left( f_{\skew5\hat{i}},f_{\skew5\hat{j}}\right)  
- {\theta }\sum\limits_{{\skew5\check{j}} \in  \skew4\check{N}} 
J_{{\skew5\hat{i}}{\skew5\check{j}}}\left( f_{\skew5\hat{i}},f_{\skew5\check{j}}\right) 
\qquad {\left( {{\skew5\hat{i}} \in \skew4\hat{N}}\right)},
\label{epsinner}
 \\
{\theta }{{\mathcal D}_{\skew5\check{i}}}{f_{\skew5\check{i}}} &=&
- J_{\skew5\check{i}}\left( f_{\skew5\check{i}},f\right)
- {\theta }\sum\limits_{j\neq {\skew5\check{i}}} 
J_{{\skew5\check{i}}j}\left( f_{\skew5\check{i}},f_{j}\right) 
\qquad {\left( {{\skew5\check{i}} \in \skew4\check{N}}\right)}.
\label{epsouter}
 \end{eqnarray}
\end{subequations}

Let us write the asymptotic expansion of the velocity distribution
function $f_{i}$ of particles of $i$-component as formal series of
successive approximations in powers of $\theta $:  
\begin{eqnarray}
f_{i}=f_{i}^{\left( 0\right)}+\theta f_{i}^{\left( 1\right)}+
\theta ^{2}f_{i}^{\left( 2\right)}+\dotsb .
\label{f_series}
\end{eqnarray}
The differential parts of the equations (\ref{epseq}) are written as: 
\begin{eqnarray}
{\mathcal D}_i f_i  &=& 
\left( { \frac{\partial }
{\partial t} + {\boldsymbol{c}}_{i}\cdot \frac{\partial }{\partial \boldsymbol{r}} + 
\frac{{\boldsymbol{X}}_{i}}{m_{i}}\cdot \frac{\partial }
{\partial {\boldsymbol{c}}_{i}}} \right)\left( {f_i^{\left( 0 \right)}  + 
\theta f_i^{\left( 1 \right)}  + \dotsb } \right) \nonumber\\*
&=&{\mathcal D}_i^{\left( 0 \right)}
+  \theta {\mathcal D}_i^{\left( 1 \right)}
+  \theta ^2 {\mathcal D}_i^{\left( 2 \right)} + \dotsb , 
\label{Dseries}
\end{eqnarray}
where
\begin{eqnarray}
{\mathcal D}_i^{\left( r \right)}
=\frac{\partial f_{i}^{\left( r\right)}}{\partial t}+
{\boldsymbol{c}}_{i}\cdot \frac{\partial 
f_{i}^{\left( r\right)}}{\partial \boldsymbol{r}} +
\frac{{\boldsymbol{X}}_{i}}{m_{i}}\cdot \frac{\partial 
f_{i}^{\left( r\right)}}
{\partial {\boldsymbol{c}}_{i}} 
\qquad {\left( r=0,1,2\ldots \right)} ,
\label{D_r}
\end{eqnarray}
-- cf. with \cite{chap52}, Chapter~7, \S ~1, Sections~4,~5 and \cite{strum74}. 
In (\ref{Dseries})-(\ref{D_r}) the partial
time derivative expansion (\ref{t_series_Enskog}) is not used in
contrast to that, how it was made by Enskog and further by
Struminskii. 
As result, described below method for solution of the system of
kinetic Boltzmann equations differ fundamentally from Enskog's method
and Struminskii's method.

Substituting (\ref{f_series}) and
(\ref{Dseries}) in (\ref{epsinner}) and equating
coefficients at the same powers of $\theta $ to each other, 
we obtain the equations system of the method of successive approximations
for finding the velocity distribution functions of inner components particles of
gas mixture $f_{\skew5\hat{i}}^{\left( r\right)}$; 
taking introduced notations (\ref{J_i}),
(\ref{J_ij}) and (\ref{D_r}) into account, the system can be rewritten as:  
\begin{eqnarray}
\sum\limits_{{\skew5\hat{j}} \in \skew4\hat{N}}
J_{{\skew5\hat{i}}{\skew5\hat{j}}}
\left( f_{\skew5\hat{i}}^{\left( 0\right)},f_{\skew5\hat{j}}^{\left( 0 \right)}\right)=0
\qquad {\left( {{\skew5\hat{i}} \in \skew4\hat{N}}\right)},
\label{sys_eq_series_0_in}
\end{eqnarray}
\begin{eqnarray}
{\mathcal D}_{\skew5\hat{i}}^{\left( r-1 \right)} 
&+& \sum\limits_{{\skew5\hat{j}} \in \skew4\hat{N}}
J_{{\skew5\hat{i}}{\skew5\hat{j}}}
\left( f_{\skew5\hat{i}}^{\left( r\right)},f_{\skew5\hat{j}}^{\left( 0 \right)}\right)
+ \sum\limits_{{\skew5\hat{j}} \in \skew4\hat{N}}
\sum\limits_{s = 1}^{r-1} 
J_{{\skew5\hat{i}}{\skew5\hat{j}}}
\left( f_{\skew5\hat{i}}^{\left( r-s\right)},f_{\skew5\hat{j}}^{\left( s\right)}\right)
+ \sum\limits_{{\skew5\hat{j}} \in \skew4\hat{N}}
J_{{\skew5\hat{i}}{\skew5\hat{j}}}
\left( f_{\skew5\hat{i}}^{\left( 0\right)},f_{\skew5\hat{j}}^{\left( r\right)}\right)
\nonumber\\
&&+ \sum\limits_{{\skew5\check{j}} \in  \skew4\check{N}} 
\sum\limits_{s = 0}^{r-1} 
J_{{\skew5\hat{i}}{\skew5\check{j}}}
\left( f_{\skew5\hat{i}}^{\left( r-1-s\right)},f_{\skew5\check{j}}^{\left( s\right)}\right)  
=0
\qquad {\left( {{\skew5\hat{i}} \in \skew4\hat{N}},\ r=1,2\ldots \right)}.
\label{sys_eq_series_r_in}
\end{eqnarray}

Similarly substituting (\ref{f_series}) and
(\ref{Dseries}) in (\ref{epsouter}) and equating
coefficients at the same powers of $\theta $ to each other, we
obtain the equations system of the method of successive approximations
for finding the velocity distribution functions of particles of external components
of gas mixture $f_{\skew5\check{i}}^{\left( r\right)}$:
\begin{eqnarray}
J_{\skew5\check{i}}\left(
  f_{\skew5\check{i}}^{\left( {0} \right)},f^{\left( {0} \right)}\right)=0
\qquad {\left( {{\skew5\check{i}} \in \skew4\check{N}}\right)},
\label{sys_eq_series_0_out}
\end{eqnarray}
\begin{eqnarray}
{\mathcal D}_{\skew5\check{i}}^{\left( r-1 \right)} 
&+&
J_{\skew5\check{i}}\left(
  f_{\skew5\check{i}}^{\left( {r} \right)},f^{\left( {0} \right)}\right) 
+ 
\sum\limits_{s = 1}^{r-1} 
J_{\skew5\check{i}}\left(
  f_{\skew5\check{i}}^{\left( {r-s} \right)},f^{\left( s \right)}\right) 
+ J_{\skew5\check{i}}\left(
  f_{\skew5\check{i}}^{\left( {0} \right)},f^{\left( {r} \right)}\right) 
\nonumber\\
&&+ 
\sum\limits_{j\neq {\skew5\check{i}}} 
\sum\limits_{s = 0}^{r-1} 
J_{{\skew5\check{i}}j}\left( f_{\skew5\check{i}}^{\left( r-1-s\right)},
f_{j}^{\left( s\right)}\right)
=0
\qquad {\left( {{\skew5\check{i}} \in \skew4\check{N}},\ r=1,2\ldots \right)}.
\label{sys_eq_series_r_out}
\end{eqnarray}

It could be noted also, that in original Hilbert's paper
\cite{hilbert12} and Enskog's
paper (see. \cite{chap52}, Chapter~7, \S ~1, Sections~4) the
formal parameter $\theta $ was not directly introduced in the Boltzmann
equation, but in the series of successive approximations for the
velocity distribution function the parameter $\theta $ was introduced
in another way (not as in (\ref{f_series})):
\begin{eqnarray}
f_{i}^{\text{H,\,E}}=\frac{1}{\theta }\,f_{i}=
\frac{1}{\theta }\,f_{i}^{\left( 0\right)}+
f_{i}^{\left( 1\right)}+
\theta f_{i}^{\left( 2\right)}+ \dotsb ;
\label{f_seriesEnscog}
\end{eqnarray}
Obviously, with the same result it is possible to use the expansion
(\ref{f_series}) for the velocity distribution function, but to enter
the multiplier $\theta $ in the Boltzmann equation, as it was made in
(\ref{epsenskog}). 
Successive  approximations 
$f_{i}^{\left( 0\right)}, f_{i}^{\left( 1\right)}, f_{i}^{\left( 2\right)}\ldots $, 
that are calculated within Enskog's 
approach, turn out to be (inversely) ordered by \textit{the number
density of molecules} of the mixture $n$: $f_{i}^{\left( 0\right)}$ is
proportional to $n$, $f_{i}^{\left( 1\right)}$ does not depend
directly on $n$ etc. 
Thus, there is a physical justification to use the method
of successive approximations to find an asymptotic solution of the
Boltzmann equation. 
For Struminskii's approach it is difficult to
determine explicitly a small physical variable to construct
asymptotic expansions in powers of the variable.  

Speaking about an order of approximation below, we assume the order to
be equal to the value of index $r$ in (\ref{sys_eq_series_r_in}),
(\ref{sys_eq_series_r_out}). 
According to (\ref{J_ij}), (\ref{sys_eq_series_0_in}), 
in zero order approximation we have
the following system of integral equations to find the velocity
distribution functions of particles of inner components of gas mixture
$f_{\skew5\hat{i}}^{\left( 0\right)}$: 
\begin{eqnarray}
\sum\limits_{{\skew5\hat{j}} \in \skew4\hat{N}}
J_{{\skew5\hat{i}}{\skew5\hat{j}}}
\left( f_{\skew5\hat{i}}^{\left( 0\right)},f_{\skew5\hat{j}}^{\left( 0\right)}\right)
=\sum\limits_{{\skew5\hat{j}} \in \skew4\hat{N}}
\iint
\left( f_{\skew5\hat{i}}^{\left( 0\right)}
       f_{\skew5\hat{j}}^{\left( 0\right)}
-f_{\skew5\hat{i}}^{\left( 0\right) \prime }
 f_{\skew5\hat{j}}^{\left( 0\right) \prime } \right)
k_{{\skew5\hat{i}}{\skew5\hat{j}}} \,d{\mathbf{k}}
\,d{\boldsymbol{c}}_{\skew5\hat{j}} = 0 
\qquad {\left( {{\skew5\hat{i}} \in \skew4\hat{N}}\right)}.
\label{system0_eq_in}
\end{eqnarray}

In the kinetic theory of gases the following equality is often used: 
\begin{eqnarray}
\iiiint \phi _{i}f_{i}^{\prime }f_{j}^{\prime }
\,g_{ij}\, b\,db \,d\epsilon 
\,d\boldsymbol{c}_{i}\,d{\boldsymbol{c}}_{j} 
=\iiiint \phi _{i}^{\prime }f_{i}f_{j}
\,g_{ij}\, b\,db \,d\epsilon 
\,d\boldsymbol{c}_{i}\,d{\boldsymbol{c}}_{j} 
\label{equality}
\end{eqnarray}
-- in (\ref{equality}), as well as in
(\ref{boltzmannend}) velocities and velocity functions of
particles after collision are denoted by prime. 
To each collision of two gas molecules, that transforms the velocities of the
molecules before the collision
$\boldsymbol{c}_{i}$,~$\boldsymbol{c}_{j}$ into 
$\boldsymbol{c}_{i}^{\,\prime}$,~$\boldsymbol{c}_{j}^{\,\prime}$, 
the collision of two gas
molecules corresponds bijectively, that transforms the velocities of
the molecules 
$\boldsymbol{c}_{i}^{\,\prime}, \boldsymbol{c}_{j}^{\,\prime} \to 
-\boldsymbol{c}_{i}^{\,\prime}, -\boldsymbol{c}_{j}^{\,\prime} \to 
-\boldsymbol{c}_{i}, -\boldsymbol{c}_{j} \to 
\boldsymbol{c}_{i}, \boldsymbol{c}_{j}$, 
and for considered central interactions it follows
from the energy conservation law, that magnitudes of relative velocities
of molecules before and after the collision are equal 
\begin{eqnarray}
g_{ij}=g_{ij}^{\,\prime},
\label{g_ij_equality}
\end{eqnarray}
from the angular momentum conservation law it follows, that the impact
parameters are also equal
\begin{eqnarray}
b=b^{\,\prime}
\label{b_equality}
\end{eqnarray}
for the collisions;
thus, the equality (\ref{equality})
follows directly from that the Jacobian determinant of the transformation
of not primed velocities into primed velocities
equals to unity; 
the equality does not depend on form of the functions
$\phi , f$, the only requirement is that integrals are determined and
converge -- cf. with \cite{chap52}, Chapter~3, \S ~5, Section~3.

Following Hecke's idea (see \cite{hilbert12}, Chapter~XXII), we
multiply both sides of the equation (\ref{system0_eq_in}) on
$\ln f_{\skew5\hat{i}}$, integrate over ${\boldsymbol{c}}_{\skew5\hat{i}}$,
sum over ${\skew5\hat{i}}$ and transform integrals, taking (\ref{def_k_ij}), (\ref{equality}) into
account; as a
result we obtain:  
\begin{eqnarray}
\frac{1}{4} \sum\limits_{{\skew5\hat{i}}, {\skew5\hat{j}} \in \skew4\hat{N}}
\iiint \ln \left( 
\frac
{f_{\skew5\hat{i}}^{\left( 0\right)} f_{\skew5\hat{j}}^{\left( 0\right)}}
{f_{\skew5\hat{i}}^{\left( 0\right) \prime }f_{\skew5\hat{j}}^{\left( 0\right) \prime }} 
\right)
\left( 
f_{\skew5\hat{i}}^{\left( 0\right)}
f_{\skew5\hat{j}}^{\left( 0\right)}
- f_{\skew5\hat{i}}^{\left( 0\right) \prime }
  f_{\skew5\hat{j}}^{\left( 0\right) \prime }
\right)
k_{{\skew5\hat{i}}{\skew5\hat{j}}} \,d{\mathbf{k}}
\,d\boldsymbol{c}_{\skew5\hat{i}}\,d{\boldsymbol{c}}_{\skew5\hat{j}} =0 .
\label{H-equality_in}
\end{eqnarray}
Integrands in (\ref{H-equality_in}) can not be (strictly) less
than zero, therefore the sum in (\ref{H-equality_in}) can be
equal to zero under the condition only, that all integrands for all
values of integration variables are vanished (all considered functions
are assumed to be continuous in each point of their definition
domain), i.e. 
\begin{eqnarray}
f_{\skew5\hat{i}}^{\left( 0\right) \prime } f_{\skew5\hat{j}}^{\left( 0\right) \prime }
\equiv 
f_{\skew5\hat{i}}^{\left( 0\right)} f_{\skew5\hat{j}}^{\left( 0\right)} 
\label{res_H-equality_in}
\end{eqnarray}
or 
\begin{eqnarray}
  \ln f_{\skew5\hat{i}}^{\left( 0\right) \prime } 
+ \ln f_{\skew5\hat{j}}^{\left( 0\right) \prime } 
- \ln f_{\skew5\hat{i}}^{\left( 0\right)}
- \ln f_{\skew5\hat{j}}^{\left( 0\right)} \equiv 0 .
\label{res_ln_H-equality_in}
\end{eqnarray}
Hence, $\ln f_{\skew5\hat{i}}^{\left( 0\right)}$ should be expressed
linearly in terms of \textit{summational invariants} of collision
\begin{subequations}
\label{psi}
 \begin{eqnarray}
 \psi _{i}^{\left( 1\right)}
 &=& m_{i},
 \label{psi_1}
 \\
 {\boldsymbol{\psi}}_{i}^{\left( 2\right)}
 &=& m_{i}{\boldsymbol{c}}_{i},
 \label{psi_2}
 \\
 \psi _{i}^{\left( 3\right)}
 &=& 
 \frac{1}{2}\, m_{i}{c}_{i}^{2};
 \label{psi_3}
 \end{eqnarray}
\end{subequations}
for a collision of the ${\skew5\hat{i}}$-molecule with the 
${\skew5\hat{j}}$-molecule
conservation of the collision invariant 
$\psi _{\skew5\hat{i}}^{\left( l\right)}$ is expressed by equality:
\begin{eqnarray}
\psi _{\skew5\hat{i}}^{\left( l\right) \prime }+ \psi _{\skew5\hat{j}}^{\left( l\right) \prime }-
\psi _{\skew5\hat{i}}^{\left( l\right)}-\psi _{\skew5\hat{j}}^{\left( l\right)} = 0 
\qquad {\left( l=1,2,3\right)} .
\label{add_inv_in}
\end{eqnarray}

Linear combination of the summational invariants 
$\psi _{\skew5\hat{i}}^{\left( l\right)}$ is an additive invariant.
There are no other summational invariants, that depend on velocities of molecules and
are linear independent of $\psi _{\skew5\hat{i}}^{\left( l\right)}$: 
six scalar unknown quantities (six components of velocities of the
molecules after the collision 
$\boldsymbol{c}_{\skew5\hat{i}}^{\,\prime}$,~$\boldsymbol{c}_{\skew5\hat{j}}^{\,\prime }$)
are completely determined by six
known components of velocities of the molecules before the collision
$\boldsymbol{c}_{\skew5\hat{i}}$,~$\boldsymbol{c}_{\skew5\hat{j}}$ 
by two free geometric parameters,
that define a collision, for example, such as impact parameter $b$
and azimuth angle $\epsilon $ (see above), that gives us two
constraint equations, and from four scalar equations
(\ref{add_inv_in}), corresponding to the conservation of energy and three
components of momentum;
Existence of summational invariant of collision, depending on
velocities of molecules and linearly independent of 
$\psi _{\skew5\hat{i}}^{\left( l\right)}$, 
would give similar to (\ref{add_inv_in}) 
redundant constraint equation for velocities of molecules after
collision and velocities of molecules before collision. 

Therefore, 
\begin{eqnarray}
\ln f_{\skew5\hat{i}}^{\left( 0\right)}=\alpha _{\skew5\hat{i}}^{\left( 1,0\right)}+
\boldsymbol{\alpha }_{\skew5\hat{\phantom{i}}}^
{\left( 2,0\right)}\cdot m_{\skew5\hat{i}}\,\boldsymbol{c}_{\skew5\hat{i}}+ 
{\alpha }_{\skew5\hat{\phantom{i}}}^{\left( 3,0\right)}
\frac{1}{2}\,m_{\skew5\hat{i}}\,c_{\skew5\hat{i}}^{2} ,
\label{ln_f_1_in}
\end{eqnarray}
where $\alpha _{\skew5\hat{i}}^{\left( 1,0\right)}$ and 
${\alpha }_{\skew5\hat{\phantom{i}}}^{\left( 3,0\right)}$  
are some, independent of $\boldsymbol{c}_{\skew5\hat{i}}$, scalar functions of spatial
coordinates (that are defined by the radius vector $\boldsymbol{r}$) and time $t$,
and $\boldsymbol{\alpha }_{\skew5\hat{\phantom{i}}}^{\left( 2,0\right)}$ is a vector function of
$\boldsymbol{r}$ and $t$ 
[as it follows from the equations
(\ref{add_inv_in}) for $l=2,3$, the functions 
$\boldsymbol{\alpha }_{\skew5\hat{\phantom{i}}}^{\left( 2,0\right)}$
and ${\alpha }_{\skew5\hat{\phantom{i}}}^{\left( 3,0\right)}$ are the same
for all inner components of mixture]. 
Or 
\begin{eqnarray}
\ln f_{\skew5\hat{i}}^{\left( 0\right)}=\ln \alpha
_{\skew5\hat{i}}^{\left( 0,0\right)}
+{\alpha }_{\skew5\hat{\phantom{i}}}^{\left( 3,0\right)} \frac{1}{2}\,m_{\skew5\hat{i}}
\left( \boldsymbol{c}_{\skew5\hat{i}}+\boldsymbol{\alpha }_{\skew5\hat{\phantom{i}}}^{\left( 2,0\right)}
{\left/{\alpha }_{\skew5\hat{\phantom{i}}}^{\left( 3,0\right)}\right.} \right) ^{2} ,
\label{ln_f_0}
\end{eqnarray}
where $\alpha _{\skew5\hat{i}}^{\left( 0,0\right)}$ is a new scalar
function of $\boldsymbol{r}$ and $t$.
I.e. the general solution of the equations system (\ref{system0_eq_in})
can be written as a set of the Maxwell functions:  
\begin{eqnarray}
f_{\skew5\hat{i}}^{\left( 0\right)}=
\beta _{\skew5\hat{i}}^{\left( 1,0\right)}
\left( \frac{m_{\skew5\hat{i}}}{2\pi k
\beta _{\skew5\hat{\phantom{i}}}^{\left( 3,0\right)}}\right) 
^{3 \left/ 2 \right.}
e^{-\frac{m_{\skew5\hat{i}}\left( {\boldsymbol{c}}_{\skew5\hat{i}}-
\boldsymbol{\beta }_{\skew5\hat{\phantom{i}}}^{\left( 2,0\right)}\right) ^{2}} 
{2k\beta _{\skew5\hat{\phantom{i}}}^{\left( 3,0\right)}}} 
\qquad {\left( {{\skew5\hat{i}} \in \skew4\hat{N}}\right)},
\label{maxwell_f0_in}
\end{eqnarray}
where $k$ is the Boltzmann constant,
\begin{eqnarray}
\beta _{\skew5\hat{i}}^{\left( 1,0\right)}&=&
\alpha _{\skew5\hat{i}}^{\left( {0,0} \right)} \left( -{\frac{{2\pi }}
{{m_{\skew5\hat{i}}\, \alpha _{\skew5\hat{\phantom{i}}}^
{\left( {3,0} \right)} }}} \right)^{3 \left/ 2 \right.} 
, 
\label{beta1_in}
\\
\boldsymbol{\beta }_{\skew5\hat{\phantom{i}}}^{\left( 2,0\right)} &=&
- \frac{{\boldsymbol{\alpha }_{\skew5\hat{\phantom{i}}}^{\left( 2,0\right)}}}
{{\alpha _{\skew5\hat{\phantom{i}}}^{\left( {3,0} \right)} }},
\label{beta2_in}
\\
\beta _{\skew5\hat{\phantom{i}}}^{\left( 3,0 \right)}  &=& 
- \frac{1}{{k\alpha _{\skew5\hat{\phantom{i}}}^{\left( {3,0} \right)} }}.
\label{beta3_in}
\end{eqnarray}

Particle number density $n_{i}$ of the $i$-component, mean mass
velocity ${\boldsymbol{u}}_{\skew5\hat{\phantom{i}}}$ and temperature
$T_{\skew5\hat{\phantom{i}}}$ of inner
components of mixture are introduced by definitions:  
\begin{eqnarray}
n_{i} \,&{\overset{\mathrm{def}}{=}}&\, 
\int f_{i}\,d {\boldsymbol{c}}_{i} ,
\label{condition1}
\\
{\boldsymbol{u}}_{\skew5\hat{\phantom{i}}}
\sum\limits_{{\skew5\hat{i}} \in \skew4\hat{N}} 
n_{\skew5\hat{i}}\,m_{\skew5\hat{i}} \,&{\overset{\mathrm{def}}{=}}&\, 
\sum\limits_{{\skew5\hat{i}} \in \skew4\hat{N}} 
\int m_{\skew5\hat{i}} \,{\boldsymbol{c}}_{\skew5\hat{i}}
f_{\skew5\hat{i}}\,d {\boldsymbol{c}}_{\skew5\hat{i}} , 
\label{condition2_in}
\\
\frac{3}{2}\, kT_{\skew5\hat{\phantom{i}}}
\sum\limits_{{\skew5\hat{i}} \in \skew4\hat{N}} 
n_{\skew5\hat{i}} 
\,&{\overset{\mathrm{def}}{=}}&\, 
\sum\limits_{{\skew5\hat{i}} \in \skew4\hat{N}} 
\int \frac{1}{2}\, m_{\skew5\hat{i}} 
\left( {\boldsymbol{c}}_{\skew5\hat{i}}-{\boldsymbol{u}}_{\skew5\hat{\phantom{i}}}
\right) ^{2} f_{\skew5\hat{i}}
\,d {\boldsymbol{c}}_{\skew5\hat{i}} , 
\label{condition3_in}
\end{eqnarray}
in \ref{condition3_in} $k$ is the Boltzmann constant.
From (\ref{condition1})-(\ref{condition3_in}) the equality is obtained: 
\begin{eqnarray}
\frac{3}{2}\, kT_{\skew5\hat{\phantom{i}}}
\sum\limits_{{\skew5\hat{i}} \in \skew4\hat{N}} 
n_{\skew5\hat{i}} 
+ 
\frac{1}{2}\,{u}_{\skew5\hat{\phantom{i}}}^{2} 
\sum\limits_{{\skew5\hat{i}} \in \skew4\hat{N}} 
n_{\skew5\hat{i}}\,m_{\skew5\hat{i}} =
\sum\limits_{{\skew5\hat{i}} \in \skew4\hat{N}} 
\int \frac{1}{2}\, m_{\skew5\hat{i}} \,{c}_{\skew5\hat{i}}^{2}
f_{\skew5\hat{i}} 
\,d {\boldsymbol{c}}_{\skew5\hat{i}},
\label{condition3n_in}
\end{eqnarray}
that is convenient to use below instead of definition
(\ref{condition3_in}).

According to definitions (\ref{condition1}), (\ref{condition2_in}),
(\ref{condition3n_in}), in addition
to the asymptotic expansion (\ref{f_series}) it is
\textit{necessary} to determine asymptotic expansions for particle
number density $n_{i}$ of the $i$-component 
\begin{eqnarray}
n_{i}=n_{i}^{\left( 0\right)}+\theta n_{i}^{\left( 1\right)}+
\theta ^{2}n_{i}^{\left( 2\right)}+\dotsb ,
\label{n_series}
\end{eqnarray}
mean mass velocity ${\boldsymbol{u}}_{\skew5\hat{\phantom{i}}}$
\begin{eqnarray}
{\boldsymbol{u}}_{\skew5\hat{\phantom{i}}}
={\boldsymbol{u}}_{\skew5\hat{\phantom{i}}}^{\left( 0\right)}+
\theta {\boldsymbol{u}}_{\skew5\hat{\phantom{i}}}^{\left( 1\right)}+
\theta ^{2}{\boldsymbol{u}}_{\skew5\hat{\phantom{i}}}^{\left( 2\right)}+\dotsb 
\label{u_series_in}
\end{eqnarray}
and temperature $T_{\skew5\hat{\phantom{i}}}$ of inner components of mixture 
\begin{eqnarray}
T_{\skew5\hat{\phantom{i}}}
=T_{\skew5\hat{\phantom{i}}}^{\left( 0\right)}
+\theta T_{\skew5\hat{\phantom{i}}}^{\left( 1\right)}+
\theta ^{2}T_{\skew5\hat{\phantom{i}}}^{\left( 2\right)}+\dotsb .
\label{T_series_in}
\end{eqnarray}

Substituting (\ref{f_series}) and
(\ref{n_series})-(\ref{T_series_in}) in
(\ref{condition1}), (\ref{condition2_in}), (\ref{condition3n_in}) and 
equating terms of the same
infinitesimal order we obtain 
${\mathrm{Card}}\left( \skew4\hat{N}\right) +4$ scalar relations, 
that connect asymptotic expansions (\ref{f_series}) and
(\ref{n_series})-(\ref{T_series_in}):  
\begin{eqnarray}
\int f_{\skew5\hat{i}}^{\left( r\right)}
d{\boldsymbol{c}}_{\skew5\hat{i}}
&=&n_{\skew5\hat{i}}^{\left( r\right)} 
\qquad {\left( {{\skew5\hat{i}} \in \skew4\hat{N}} \right)} ,
\label{r_condition1_in}
\end{eqnarray}
\begin{eqnarray}
\sum\limits_{{\skew5\hat{i}} \in \skew4\hat{N}} 
\int
m_{\skew5\hat{i}}
{\boldsymbol{c}}_{\skew5\hat{i}}f_{\skew5\hat{i}}^{\left( r\right)}
d{\boldsymbol{c}}_{\skew5\hat{i}}
&=&
\sum\limits_{{\skew5\hat{i}} \in \skew4\hat{N}} 
m_{\skew5\hat{i}}\, {\left( n_{\skew5\hat{i}}
{\boldsymbol{u}}_{\skew5\hat{\phantom{i}}}
\right)}^{\left( {r} \right)}
=
\sum\limits_{{\skew5\hat{i}} \in \skew4\hat{N}} 
m_{\skew5\hat{i}} \sum\limits_{s = 0}^r 
{n_{\skew5\hat{i}}^{\left( r - s \right)} } 
{\boldsymbol{u}}_{\skew5\hat{\phantom{i}}}^{\left( {s} \right)} 
=
\sum\limits_{s = 0}^r 
{\hat{\rho }^{\left( r - s \right)} } 
{\boldsymbol{u}}_{\skew5\hat{\phantom{i}}}^{\left( {s} \right)} ,
\label{r_condition2_in}
\end{eqnarray}
\begin{eqnarray}
\sum\limits_{{\skew5\hat{i}} \in \skew4\hat{N}} 
\int
\frac{1}{2}\, m_{\skew5\hat{i}} {c}_{\skew5\hat{i}}^{2}
f_{\skew5\hat{i}}^{\left( r\right)}
d {\boldsymbol{c}}_{\skew5\hat{i}}
&=&
\frac{3}{2}\, k 
\sum\limits_{{\skew5\hat{i}} \in \skew4\hat{N}} 
{\left( n_{\skew5\hat{i}} 
T_{\skew5\hat{\phantom{i}}} \right)}^{\left( {r} \right)}
+ 
\frac{1}{2} 
\sum\limits_{{\skew5\hat{i}} \in \skew4\hat{N}} 
m_{\skew5\hat{i}} 
{\left( n_{\skew5\hat{i}} 
u_{\skew5\hat{\phantom{i}}}^{2} \right)}^{\left( {r} \right)}
\nonumber\\
&=&
\frac{3}{2}\, k 
\sum\limits_{{\skew5\hat{i}} \in \skew4\hat{N}} 
\sum\limits_{s = 0}^r 
{n_{\skew5\hat{i}}^{\left( r - s \right)} } 
{T_{\skew5\hat{\phantom{i}}}^{\left( {s} \right)}}
+ 
\frac{1}{2} 
\sum\limits_{{\skew5\hat{i}} \in \skew4\hat{N}} 
m_{\skew5\hat{i}} 
\sum\limits_{s = 0}^{r} \sum\limits_{q = 0}^{s} 
n_{\skew5\hat{i}}^{\left( {r-s}\right)}
{\boldsymbol{u}}_{\skew5\hat{\phantom{i}}}^{\left( {s-q} \right)}\cdot 
{\boldsymbol{u}}_{\skew5\hat{\phantom{i}}}^{\left( {q} \right)}
\nonumber\\
&=&
\frac{3}{2}\, k 
\sum\limits_{s = 0}^r 
{\hat{n}^{\left( r - s \right)} } 
{T_{\skew5\hat{\phantom{i}}}^{\left( {s} \right)}}
+ 
\frac{1}{2} 
\sum\limits_{s = 0}^{r} \sum\limits_{q = 0}^{s} 
{\hat{\rho }^{\left( r - s \right)} } 
{\boldsymbol{u}}_{\skew5\hat{\phantom{i}}}^{\left( {s-q} \right)}\cdot 
{\boldsymbol{u}}_{\skew5\hat{\phantom{i}}}^{\left( {q} \right)} .
\label{r_condition3n_in}
\end{eqnarray}
In (\ref{r_condition2_in}), (\ref{r_condition3n_in}) the notations are introduced  
\begin{eqnarray}
{\hat{\rho }^{\left( r-s \right)} } &=&
\sum\limits_{{\skew5\hat{i}} \in \skew4\hat{N}} 
m_{\skew5\hat{i}} \,
n_{\skew5\hat{i}}^{\left( {r-s}\right)} ,
\label{rho_def_in}
\\
{\hat{n}^{\left( r-s \right)} } &=&
\sum\limits_{{\skew5\hat{i}} \in \skew4\hat{N}} 
n_{\skew5\hat{i}}^{\left( {r-s}\right)} .
\label{n_def_in}
\end{eqnarray}

In particular, for $r=0$ from
(\ref{r_condition1_in})-(\ref{r_condition3n_in})
we obtain expressions for arbitrary functions
$\beta _{\skew5\hat{i}}^{\left( 1,0\right)}\left( {\boldsymbol{r}},t\right)$,
$\boldsymbol{\beta }_{\skew5\hat{\phantom{i}}}^{\left( 2,0\right)}
\left( {\boldsymbol{r}},t\right)$ and 
$\beta _{\skew5\hat{\phantom{i}}}^{\left( 3,0\right)}\left( {\boldsymbol{r}},t\right)$
in (\ref{maxwell_f0_in}) through the zero order approximations to
local values of the ${\skew5\hat{i}}$-component number
density, the mean mass velocity and the temperature of inner
components of the mixture: 
\begin{subequations}
 \label{param_f_in}
 \begin{eqnarray}
  \beta _{\skew5\hat{i}}^{\left( 1,0\right)}\left( {\boldsymbol{r}},t\right) 
&=&
  n_{\skew5\hat{i}}^{\left( 0\right)}\left( {\boldsymbol{r}},t\right),
\label{param_fa_in} \\
  \boldsymbol{\beta }_{\skew5\hat{\phantom{i}}}^{\left( 2,0\right)}\left(
    {\boldsymbol{r}},t\right) 
&=& {\boldsymbol{u}}_{\skew5\hat{\phantom{i}}}^{\left( 0\right)}
  \left( {\boldsymbol{r}},t\right) ,
\label{param_fb_in} \\
  \beta _{\skew5\hat{\phantom{i}}}^{\left( 3,0\right)}\left( {\boldsymbol{r}},t\right) 
&=& T_{\skew5\hat{\phantom{i}}}^{\left( 0\right)}\left( {\boldsymbol{r}},t\right).
\label{param_fc_in}
 \end{eqnarray}
\end{subequations}

According to (\ref{J_i}), (\ref{sys_eq_series_0_out}), 
zero order integral equations, from which the velocity
distribution functions $f_{\skew5\check{i}}^{\left( 0\right)}$ of particles of
outer components of the mixture are found: 
\begin{eqnarray}
J_{\skew5\check{i}}\left( f_{\skew5\check{i}}^{\left( 0\right)},f^{\left( 0\right)}\right) =
\iint
\left( f_{\skew5\check{i}}^{\left( 0\right)}f^{\left( 0\right)}
-f_{\skew5\check{i}}^{\left( 0\right) \prime }f^{\left( 0\right) \prime } \right)
k_{i} \,d {\mathbf{k}}
\,d {\boldsymbol{c}}  = 0 
\qquad {\left( {{\skew5\check{i}} \in \skew4\check{N}}\right)},
\label{system0_eq_out}
\end{eqnarray}
-- are simpler than equations (\ref{system0_eq_in}) and differ
actually from (\ref{system0_eq_in}) only by lack of summation over
components. 
Therefore, similarly (\ref{maxwell_f0_in}), the general solution of the
equations system \eqref{system0_eq_out} 
can be written as a set of the Maxwell functions:
\begin{eqnarray}
f_{\skew5\check{i}}^{\left( 0\right)}=
\beta _{\skew5\check{i}}^{\left( 1,0\right)}
\left( \frac{m_{\skew5\check{i}}}{2\pi k
\beta _{\skew5\check{i}}^{\left( 3,0\right)}}\right) 
^{3 \left/ 2 \right.}
e^{-\frac{m_{\skew5\check{i}}\left( {\boldsymbol{c}}_{\skew5\check{i}}-
\boldsymbol{\beta }_{\skew5\check{i}}^{\left( 2,0\right)}\right) ^{2}} 
{2k\beta _{\skew5\check{i}}^{\left( 3,0\right)}}}
\qquad {\left( {{\skew5\check{i}} \in \skew4\check{N}}\right)},
\label{maxwell_f0_out}
\end{eqnarray}
where $\beta _{\skew5\check{i}}^{\left( 1,0\right)}$ and 
${\beta }_{\skew5\check{i}}^{\left( 3,0\right)}$  
are some, independent of $\boldsymbol{c}_{\skew5\check{i}}$, scalar
functions of spatial coordinates, defined by the radius vector
$\boldsymbol{r}$, and time $t$, and 
$\boldsymbol{\beta }_{\skew5\check{i}}^{\left( 2,0\right)}$ is a
vector function of $\boldsymbol{r}$ and $t$. 

Let's add to the definition of the number density of particles of
$i$-component definitions of mean velocity ${\boldsymbol{u}}_{\skew5\check{i}}$ and temperature
$T_{\skew5\check{i}}$ of outer component of mixture:
\begin{eqnarray}
{\boldsymbol{u}}_{\skew5\check{i}}
\,n_{\skew5\check{i}}\,m_{\skew5\check{i}} \,&{\overset{\mathrm{def}}{=}}&\, 
\int m_{\skew5\check{i}} \,{\boldsymbol{c}}_{\skew5\check{i}}
f_{\skew5\check{i}}\,d {\boldsymbol{c}}_{\skew5\check{i}} , 
\label{condition2_out}
\\
\frac{3}{2}\, kT_{\skew5\check{i}}
\,n_{\skew5\check{i}} 
\,&{\overset{\mathrm{def}}{=}}&\, 
\int \frac{1}{2}\, m_{\skew5\check{i}} 
\left( {\boldsymbol{c}}_{\skew5\check{i}}-{\boldsymbol{u}}_{\skew5\check{i}}
\right) ^{2} f_{\skew5\check{i}}\,d {\boldsymbol{c}}_{\skew5\check{i}} ;
\label{condition3_out}
\end{eqnarray}
from (\ref{condition1}), (\ref{condition2_out}),
(\ref{condition3_out}) the equality is obtained: 
\begin{eqnarray}
\frac{3}{2}\, kT_{\skew5\check{i}}
\,n_{\skew5\check{i}} 
+ 
\frac{1}{2}\,{u}_{\skew5\check{i}}^{2} 
\,n_{\skew5\check{i}}\,m_{\skew5\check{i}} =
\int \frac{1}{2}\, m_{\skew5\check{i}} \,{c}_{\skew5\check{i}}^{2}
f_{\skew5\check{i}} 
\,d {\boldsymbol{c}}_{\skew5\check{i}},
\label{condition3n_out}
\end{eqnarray}
that is convenient to use below instead of definition
(\ref{condition3_out}).

Let's enter similar (\ref{u_series_in})-(\ref{T_series_in}) asymptotic 
expansions of outer ${\skew5\check{i}}$-component mean velocity 
${\boldsymbol{u}}_{\skew5\check{i}}$
\begin{eqnarray}
{\boldsymbol{u}}_{\skew5\check{i}}
={\boldsymbol{u}}_{\skew5\check{i}}^{\left( 0\right)}+
\theta {\boldsymbol{u}}_{\skew5\check{i}}^{\left( 1\right)}+
\theta ^{2}{\boldsymbol{u}}_{\skew5\check{i}}^{\left( 2\right)}+\dotsb 
\label{u_series_out}
\end{eqnarray}
and outer ${\skew5\check{i}}$-component temperature $T_{\skew5\check{i}}$
\begin{eqnarray}
T_{\skew5\check{i}}
=T_{\skew5\check{i}}^{\left( 0\right)}
+\theta T_{\skew5\check{i}}^{\left( 1\right)}+
\theta ^{2}T_{\skew5\check{i}}^{\left( 2\right)}+\dotsb .
\label{T_series_out}
\end{eqnarray}

Substituting (\ref{f_series}), (\ref{n_series}), 
(\ref{u_series_out}), (\ref{T_series_out}) in
(\ref{condition1}), (\ref{condition2_out}), (\ref{condition3n_out}) and 
equating terms of the same
infinitesimal order we obtain for each ${\skew5\check{i}}$
5 (scalar) relations, 
that connect asymptotic expansions (\ref{f_series}),
(\ref{n_series}), (\ref{u_series_out}), (\ref{T_series_out}):  
\begin{eqnarray}
\int f_{\skew5\check{i}}^{\left( r\right)}
d{\boldsymbol{c}}_{\skew5\check{i}}
&=&n_{\skew5\check{i}}^{\left( r\right)} ,
\label{r_condition1_out}
\end{eqnarray}
\begin{eqnarray}
\int
m_{\skew5\check{i}}
{\boldsymbol{c}}_{\skew5\check{i}}f_{\skew5\check{i}}^{\left( r\right)}
d{\boldsymbol{c}}_{\skew5\check{i}}
&=&
m_{\skew5\check{i}}\, {\left( n_{\skew5\check{i}}
{\boldsymbol{u}}_{\skew5\check{i}}
\right)}^{\left( {r} \right)}
=
m_{\skew5\check{i}} \sum\limits_{s = 0}^r 
{n_{\skew5\check{i}}^{\left( r - s \right)} } 
{\boldsymbol{u}}_{\skew5\check{i}}^{\left( {s} \right)} 
=
\sum\limits_{s = 0}^r 
{{\rho }_{\skew5\check{i}}^{\left( r-s \right)} }
{\boldsymbol{u}}_{\skew5\check{i}}^{\left( {s} \right)} ,
\label{r_condition2_out}
\end{eqnarray}
\begin{eqnarray}
\int
\frac{1}{2}\, m_{\skew5\check{i}} {c}_{\skew5\check{i}}^{2}
f_{\skew5\check{i}}^{\left( r\right)}
d {\boldsymbol{c}}_{\skew5\check{i}}
&=&
\frac{3}{2}\, k 
{\left( n_{\skew5\check{i}} 
T_{\skew5\check{i}} \right)}^{\left( {r} \right)}
+ 
\frac{1}{2} 
m_{\skew5\check{i}} 
{\left( n_{\skew5\check{i}} 
u_{\skew5\check{i}}^{2} \right)}^{\left( {r} \right)}
\nonumber\\
&=&
\frac{3}{2}\, k 
\sum\limits_{s = 0}^r 
{n_{\skew5\check{i}}^{\left( r - s \right)} } 
{T_{\skew5\check{i}}^{\left( {s} \right)}}
+ 
\frac{1}{2} 
m_{\skew5\check{i}} 
\sum\limits_{s = 0}^{r} \sum\limits_{q = 0}^{s} 
n_{\skew5\check{i}}^{\left( {r-s}\right)}
{\boldsymbol{u}}_{\skew5\check{i}}^{\left( {s-q} \right)}\cdot 
{\boldsymbol{u}}_{\skew5\check{i}}^{\left( {q} \right)}
\nonumber\\
&=&
\frac{3}{2}\, k 
\sum\limits_{s = 0}^r 
{n_{\skew5\check{i}}^{\left( r - s \right)} } 
{T_{\skew5\check{i}}^{\left( {s} \right)}}
+ 
\frac{1}{2} 
\sum\limits_{s = 0}^{r} \sum\limits_{q = 0}^{s} 
{{\rho }_{\skew5\check{i}}^{\left( r-s \right)} }
{\boldsymbol{u}}_{\skew5\check{i}}^{\left( {s-q} \right)}\cdot 
{\boldsymbol{u}}_{\skew5\check{i}}^{\left( {q} \right)},
\label{r_condition3n_out}
\end{eqnarray}
cf. with (\ref{r_condition1_in})-(\ref{r_condition3n_in}).
In (\ref{r_condition2_out}), (\ref{r_condition3n_out}) the notation is
used 
\begin{eqnarray}
{{\rho }_{\skew5\check{i}}^{\left( r-s \right)} }
=
m_{\skew5\check{i}} \,
{n_{\skew5\check{i}}^{\left( r - s \right)} } .
\label{rho_def_out}
\end{eqnarray}
For $r=0$ from
(\ref{r_condition1_out})-(\ref{r_condition3n_out})
we obtain expressions for arbitrary functions
$\beta _{\skew5\check{i}}^{\left( 1,0\right)}\left( {\boldsymbol{r}},t\right)$,
$\boldsymbol{\beta }_{\skew5\check{i}}^{\left( 2,0\right)}
\left( {\boldsymbol{r}},t\right)$ and 
$\beta _{\skew5\check{i}}^{\left( 3,0\right)}\left( {\boldsymbol{r}},t\right)$
in (\ref{maxwell_f0_out}) through the zero order approximations to
local values of the number
density, the mean velocity and the temperature of outer
${\skew5\check{i}}$-component of the mixture:
\begin{subequations}
 \label{param_f_out}
 \begin{eqnarray}
  \beta _{\skew5\check{i}}^{\left( 1,0\right)}\left( {\boldsymbol{r}},t\right) 
&=&
  n_{\skew5\check{i}}^{\left( 0\right)}\left( {\boldsymbol{r}},t\right),
\label{param_fa_out} \\
  \boldsymbol{\beta }_{\skew5\check{i}}^{\left( 2,0\right)}\left(
    {\boldsymbol{r}},t\right) 
&=& {\boldsymbol{u}}_{\skew5\check{i}}^{\left( 0\right)}
  \left( {\boldsymbol{r}},t\right) ,
\label{param_fb_out} \\
  \beta _{\skew5\check{i}}^{\left( 3,0\right)}\left( {\boldsymbol{r}},t\right) 
&=& T_{\skew5\check{i}}^{\left( 0\right)}\left( {\boldsymbol{r}},t\right).
\label{param_fc_out}
 \end{eqnarray}
\end{subequations}

For $r \geq 1$ the velocity distribution functions of inner components of
gas mixture $f_{\skew5\hat{i}}^{\left( r\right)}$
are found from the integral equations system (\ref{sys_eq_series_r_in}),
which, taking (\ref{J_ij}) and (\ref{res_H-equality_in}) into account,
can be rewritten in the form
\begin{eqnarray}
{\mathcal D}_{\skew5\hat{i}}^{\left( r-1 \right)} 
&+& \sum\limits_{{\skew5\hat{j}} \in \skew4\hat{N}}
\sum\limits_{s = 1}^{r-1} 
J_{{\skew5\hat{i}}{\skew5\hat{j}}}
\left( f_{\skew5\hat{i}}^{\left( r-s\right)},f_{\skew5\hat{j}}^{\left( s\right)}\right)
+ \sum\limits_{{\skew5\check{j}} \in  \skew4\check{N}} 
\sum\limits_{s = 0}^{r-1} 
J_{{\skew5\hat{i}}{\skew5\check{j}}}
\left( f_{\skew5\hat{i}}^{\left( r-1-s\right)},f_{\skew5\check{j}}^{\left( s\right)}\right)  
\nonumber\\
&=& 
-\sum\limits_{{\skew5\hat{j}} \in \skew4\hat{N}}
J_{{\skew5\hat{i}}{\skew5\hat{j}}}
\left( f_{\skew5\hat{i}}^{\left( 0\right)}
\chi _{\skew5\hat{i}}^{\left( r\right)},f_{\skew5\hat{j}}^{\left( 0\right)}\right)
-\sum\limits_{{\skew5\hat{j}} \in \skew4\hat{N}}
J_{{\skew5\hat{i}}{\skew5\hat{j}}}
\left( f_{\skew5\hat{i}}^{\left( 0\right)},f_{\skew5\hat{j}}^{\left( 0\right)}
\chi _{\skew5\hat{j}}^{\left( r\right)}\right)
\nonumber\\
&=&
-\sum\limits_{{\skew5\hat{j}} \in \skew4\hat{N}}
\iint
f_{\skew5\hat{i}}^{\left( 0\right)}f_{\skew5\hat{j}}^{\left( 0\right)}
\left( 
\chi _{\skew5\hat{i}}^{\left( r\right)} + \chi _{\skew5\hat{j}}^{\left( r\right)} -
\chi _{\skew5\hat{i}}^{\left( r\right) \prime } - \chi _{\skew5\hat{j}}^{\left( r\right)\prime } 
\right)
k_{{\skew5\hat{i}}{\skew5\hat{j}}} \,d{\mathbf{k}}
\,d{\boldsymbol{c}}_{\skew5\hat{j}}
\qquad {\left( {{\skew5\hat{i}} \in \skew4\hat{N}} \right)},
\label{r_eq_in}
\end{eqnarray}
in (\ref{r_eq_in}) functions 
$f_{\skew5\hat{i}}^{\left( r\right)}$ are written as
$f_{\skew5\hat{i}}^{\left( r\right)}=f_{\skew5\hat{i}}^{\left( 0\right)}
\chi _{\skew5\hat{i}}^{\left( r\right)}$, where 
$\chi _{\skew5\hat{i}}^{\left( r\right)}$ are new unknown functions.

The left-hand sides of equations (\ref{r_eq_in}) involves 
functions, that are known from the previous step of the successive
approximations method.
Unknown functions $\chi _{\skew5\hat{i}}^{\left( r\right)}$ enter,
linearly, only into the right-hand sides of equations
(\ref{r_eq_in}). 
Therefore the general solution of the system of equations 
(\ref{sys_eq_series_r_in}) is a family of functions of a form 
$\{{f_{\skew5\hat{i}}^{\left( r\right)}=
\Xi _{\skew5\hat{i}}^{\left( r\right)} + \xi _{\skew5\hat{i}}^{\left( r\right)}}\}
_{{\skew5\hat{i}} \in \skew4\hat{N}}$, where 
$\{\Xi _{\skew5\hat{i}}^{\left( r\right)}=f_{\skew5\hat{i}}^{\left( 0\right)} 
\mathit{\Phi } _{\skew5\hat{i}}^{\left( r\right)}\}
_{{\skew5\hat{i}} \in \skew4\hat{N}}$,
$\{\xi _{\skew5\hat{i}}^{\left( r\right)}=f_{\skew5\hat{i}}^{\left( 0\right)} 
\phi _{\skew5\hat{i}}^{\left( r\right)}\}
_{{\skew5\hat{i}} \in \skew4\hat{N}}$, 
a family of functions
$\{{\mathit{\Phi } _{\skew5\hat{i}}^{\left( r\right)}}\}
_{{\skew5\hat{i}} \in \skew4\hat{N}}$ is
a particular solution of the system of inhomogeneous equations 
(\ref{r_eq_in}) and
a family of functions
$\{\phi _{\skew5\hat{i}}^{\left( r\right)}\}
_{{\skew5\hat{i}} \in \skew4\hat{N}}$ is 
the general solution of the system of homogeneous equations 
\begin{eqnarray}
0=\sum\limits_{{\skew5\hat{j}} \in \skew4\hat{N}}
\iint
f_{\skew5\hat{i}}^{\left( 0\right)}f_{\skew5\hat{j}}^{\left( 0\right)}
\left( 
\phi _{\skew5\hat{i}}^{\left( r\right)} + \phi _{\skew5\hat{j}}^{\left( r\right)} -
\phi _{\skew5\hat{i}}^{\left( r\right) \prime } - \phi _{\skew5\hat{j}}^{\left( r\right)\prime } 
\right)
k_{{\skew5\hat{i}}{\skew5\hat{j}}} \,d{\mathbf{k}}
\,d{\boldsymbol{c}}_{\skew5\hat{j}} 
\qquad {\left( {{\skew5\hat{i}} \in \skew4\hat{N}} \right)} .
\label{r_eq_phi_in}
\end{eqnarray}

Multiplying equations (\ref{r_eq_phi_in}) by 
$\phi _{\skew5\hat{i}}^{\left( r\right)}$, integrating over all values of
${\boldsymbol{c}}_{\skew5\hat{i}}$, summing over ${\skew5\hat{i}}$ and
transforming integrals, as it has been made in deriving 
(\ref{H-equality_in}), we obtain  
\begin{eqnarray}
\frac{1}{4} \sum\limits_{{\skew5\hat{i}}, {\skew5\hat{j}} \in \skew4\hat{N}}
\iiint
f_{\skew5\hat{i}}^{\left( 0\right)}f_{\skew5\hat{j}}^{\left( 0\right)}
\left( 
\phi _{\skew5\hat{i}}^{\left( r\right)} + \phi _{\skew5\hat{j}}^{\left( r\right)} -
\phi _{\skew5\hat{i}}^{\left( r\right) \prime } - \phi _{\skew5\hat{j}}^{\left( r\right)\prime } 
\right) ^2
k_{{\skew5\hat{i}}{\skew5\hat{j}}} \,d{\mathbf{k}}
\,d{\boldsymbol{c}}_{\skew5\hat{i}}
\,d{\boldsymbol{c}}_{\skew5\hat{j}} =0.
\label{phi-equality_in}
\end{eqnarray}
From (\ref{phi-equality_in}) we conclude, 
cf. with (\ref{H-equality_in}) and (\ref{ln_f_1_in}), 
that $\phi _{\skew5\hat{i}}^{\left( r\right)}$ are 
linear combinations of the summational invariants of the collision 
$\psi _{i}^{\left( l\right)}$~${\left( l=1,2,3\right)}$:
\begin{eqnarray}
\phi _{\skew5\hat{i}}^{\left( r\right)}=
\alpha _{\skew5\hat{i}}^{\left( 1,r\right)}+
\boldsymbol{\alpha }_{\skew5\hat{\phantom{i}}}^{\left( 2,r\right)}
\cdot m_{\skew5\hat{i}}\,\boldsymbol{c}_{\skew5\hat{i}}+ 
{\alpha }_{\skew5\hat{\phantom{i}}}^{\left( 3,r\right)}
\frac{1}{2}\,m_{\skew5\hat{i}}\,c_{\skew5\hat{i}}^{2} ,
\label{phi_expr_in}
\end{eqnarray}
where $\alpha _{\skew5\hat{i}}^{\left( 1,r\right)}$ and 
${\alpha }_{\skew5\hat{\phantom{i}}}^{\left( 3,r\right)}$ are some, 
independent of $\boldsymbol{c}_{\skew5\hat{i}}$, 
scalar functions of spatial coordinates, defined by the radius vector
$\boldsymbol{r}$, and time $t$, 
and $\boldsymbol{\alpha }_{\skew5\hat{\phantom{i}}}^{\left( 2,r\right)}$ is a
vector function of $\boldsymbol{r}$ and $t$
(as well as above, arbitrary functions 
$\boldsymbol{\alpha }_{\skew5\hat{\phantom{i}}}^{\left( 2,r\right)}$ and
${\alpha }_{\skew5\hat{\phantom{i}}}^{\left( 3,r\right)}$ are
identical for all inner components of the mixture), 
and, hence,
\begin{eqnarray}
\xi _{\skew5\hat{i}}^{\left( r\right)}=
f_{\skew5\hat{i}}^{\left( 0\right)} 
{\left( 
\alpha _{\skew5\hat{i}}^{\left( 1,r\right)}+
\boldsymbol{\alpha }_{\skew5\hat{\phantom{i}}}^{\left( 2,r\right)}
\cdot m_{\skew5\hat{i}}\,\boldsymbol{c}_{\skew5\hat{i}}+ 
{\alpha }_{\skew5\hat{\phantom{i}}}^{\left( 3,r\right)}
\frac{1}{2}\,m_{\skew5\hat{i}}\,c_{\skew5\hat{i}}^{2}
\right)} 
\qquad {\left( {{\skew5\hat{i}} \in \skew4\hat{N}} \right)}.
\label{xi_expr_in}
\end{eqnarray}
To simplify further evaluations according to the expression for
$f_{\skew5\hat{i}}^{\left( 0\right)}$, see (\ref{maxwell_f0_in}) and
(\ref{param_f_in}), let us rewrite (\ref{xi_expr_in}) as
\begin{eqnarray}
\xi _{\skew5\hat{i}}^{\left( r\right)}=
f_{\skew5\hat{i}}^{\left( 0\right)} 
{\left[ 
\beta _{\skew5\hat{i}}^{\left( 1,r\right)}+
\boldsymbol{\beta }_{\skew5\hat{\phantom{i}}}^{\left( 2,r\right)}
\cdot m_{\skew5\hat{i}}
\left(\boldsymbol{c}_{\skew5\hat{i}}-{\boldsymbol{u}}_{\skew5\hat{\phantom{i}}}^{\left( 0\right)}\right)
+ 
{\beta }_{\skew5\hat{\phantom{i}}}^{\left( 3,r\right)}
\frac{1}{2}\,m_{\skew5\hat{i}}
\left(\boldsymbol{c}_{\skew5\hat{i}}-
{\boldsymbol{u}}_{\skew5\hat{\phantom{i}}}^{\left(0\right)}\right)^{2}
\right]} 
\qquad {\left( {{\skew5\hat{i}} \in \skew4\hat{N}} \right)},
\label{xi_expr_ease_in}
\end{eqnarray}
where $\beta _{\skew5\hat{i}}^{\left( 1,r\right)}$, 
$\boldsymbol{\beta }_{\skew5\hat{\phantom{i}}}^{\left( 2,r\right)}$ and 
${\beta }_{\skew5\hat{\phantom{i}}}^{\left( 3,r\right)}$ are 
new functions of $\boldsymbol{r}$ and $t$.
Family of functions
$\{\mathit{\chi } _{\skew5\hat{i}}^{\left( r\right)}\}
_{{\skew5\hat{i}} \in \skew4\hat{N}}$ is
a solution of the system of inhomogeneous equations 
\begin{eqnarray}
F_{\skew5\hat{i}}^{\left( r \right)} 
= 
\sum\limits_{{\skew5\hat{j}} \in \skew4\hat{N}}
\iint
f_{\skew5\hat{i}}^{\left( 0\right)}f_{\skew5\hat{j}}^{\left( 0\right)}
\left( 
\mathit{\chi } _{\skew5\hat{i}}^{\left( r\right)} + 
\mathit{\chi } _{\skew5\hat{j}}^{\left( r\right)} -
\mathit{\chi } _{\skew5\hat{i}}^{\left( r\right) \prime } - 
\mathit{\chi } _{\skew5\hat{j}}^{\left( r\right)\prime } 
\right)
k_{{\skew5\hat{i}}{\skew5\hat{j}}} \,d{\mathbf{k}}
\,d{\boldsymbol{c}}_{\skew5\hat{j}} 
\qquad {\left( {{\skew5\hat{i}} \in \skew4\hat{N}} \right)} ,
\label{r_eq_in_F}
\end{eqnarray}
where $F_{\skew5\hat{i}}^{\left( r \right)}$
denote left-hand sides of the equations (\ref{r_eq_in}), taken with
opposite sign.

Multiplying equations (\ref{r_eq_in_F}) by 
$\psi _{\skew5\hat{i}}^{\left( l\right)}$~${\left( l=1,2,3\right)}$, 
integrating over all values of
${\boldsymbol{c}}_{\skew5\hat{i}}$ and
transforming integrals as above, 
we obtain, taking (\ref{add_inv_in}) into account, 
as necessary condition for the existence of solutions of
the system of integral equations (\ref{r_eq_in_F}), 
the necessity of the fulfillment of equalities:
\begin{subequations}
\label{r_eq_in_condiition}
 \begin{eqnarray}
\int \psi _{\skew5\hat{i}}^{\left( 1\right)} F_{\skew5\hat{i}}^{\left( r \right)} 
d{\boldsymbol{c}}_{\skew5\hat{i}} 
&=& 0
\qquad {\left( {{\skew5\hat{i}} \in \skew4\hat{N}} \right)} ,
 \label{r_eq_in_condiition_1}
 \\
\sum\limits_{{\skew5\hat{i}} \in \skew4\hat{N}}
\int \psi _{\skew5\hat{i}}^{\left( l\right)} F_{\skew5\hat{i}}^{\left( r \right)} 
d{\boldsymbol{c}}_{\skew5\hat{i}} 
&=& 0
\qquad {\left( l=2,3\right)} .
\label{r_eq_in_condiition_2_3}
 \end{eqnarray}
\end{subequations}

Among (infinitesimal) set of particular solutions of the system of
equations (\ref{r_eq_in_F}), different from each other on some
solution of the system of homogeneous equations (\ref{r_eq_phi_in}),
unique solution $\{\mathit{\Phi } _{\skew5\hat{i}}^{\left( r\right)}\}
_{{\skew5\hat{i}} \in \skew4\hat{N}}$ may be chosen such that
\begin{subequations}
\label{r_eq_in_unique}
 \begin{eqnarray}
\int \psi _{\skew5\hat{i}}^{\left( 1\right)} 
f_{\skew5\hat{i}}^{\left( 0\right)} \mathit{\Phi } _{\skew5\hat{i}}^{\left( r \right)} 
d{\boldsymbol{c}}_{\skew5\hat{i}} &=& 0
\qquad {\left( {{\skew5\hat{i}} \in \skew4\hat{N}}\right)} ,
 \label{r_eq_in_unique_1}
 \\
\sum\limits_{{\skew5\hat{i}} \in \skew4\hat{N}}
\int \psi _{\skew5\hat{i}}^{\left( l\right)} 
f_{\skew5\hat{i}}^{\left( 0\right)} \mathit{\Phi } _{\skew5\hat{i}}^{\left( r \right)} 
d{\boldsymbol{c}}_{\skew5\hat{i}} &=& 0
\qquad {\left( l=2,3\right)} .
\label{r_eq_in_unique_2_3}
 \end{eqnarray}
\end{subequations}
Having substituted expression for
$f_{\skew5\hat{i}}^{\left( r\right)} 
\left( {{\skew5\hat{i}} \in \skew4\hat{N}}\right)$
\begin{eqnarray}
f_{\skew5\hat{i}}^{\left( r\right)}
&=&
\Xi _{\skew5\hat{i}}^{\left( r\right)} + 
\xi _{\skew5\hat{i}}^{\left( r\right)}
\nonumber \\*
&=&
f_{\skew5\hat{i}}^{\left( 0\right)} 
\mathit{\Phi } _{\skew5\hat{i}}^{\left( r\right)}+
f_{\skew5\hat{i}}^{\left( 0\right)} 
{\left[ 
\beta _{\skew5\hat{i}}^{\left( 1,r\right)}+
\boldsymbol{\beta }_{\skew5\hat{\phantom{i}}}^{\left( 2,r\right)}
\cdot m_{\skew5\hat{i}}
\left(\boldsymbol{c}_{\skew5\hat{i}}-{\boldsymbol{u}}_{\skew5\hat{\phantom{i}}}^{\left( 0\right)}\right)
+ 
{\beta }_{\skew5\hat{\phantom{i}}}^{\left( 3,r\right)}
\frac{1}{2}\,m_{\skew5\hat{i}}
\left(\boldsymbol{c}_{\skew5\hat{i}}-
{\boldsymbol{u}}_{\skew5\hat{\phantom{i}}}^{\left(0\right)}\right)^{2}
\right]}
\label{f_r_expression_in}
\end{eqnarray}
in (\ref{r_condition1_in})-(\ref{r_condition3n_in}),
taking (\ref{maxwell_f0_in}), (\ref{rho_def_in})-(\ref{param_f_in}) and
(\ref{r_eq_in_unique}) into account,
we obtain a system of
${\mathrm{Card}}\left( \skew4\hat{N}\right) +4$ algebraic equations 
[constraint equations for asymptotic expansions (\ref{f_series}) and
(\ref{n_series})-(\ref{T_series_in})]:
\begin{eqnarray}
n_{\skew5\hat{i}}^{\left( 0\right)}\beta _{\skew5\hat{i}}^{\left( 1,r\right)}
+ \frac{3}{2}\,n_{\skew5\hat{i}}^{\left( 0\right)}
kT_{\skew5\hat{\phantom{i}}}^{\left( 0\right)}
{\beta }_{\skew5\hat{\phantom{i}}}^{\left( 3,r\right)}
=n_{\skew5\hat{i}}^{\left( r\right)} 
\qquad {\left( {{\skew5\hat{i}} \in \skew4\hat{N}} \right)} ,
\label{beta_r_equation1_in}
\end{eqnarray}
\begin{eqnarray}
{\boldsymbol{u}}_{\skew5\hat{\phantom{i}}}^{\left( {0} \right)}
\sum\limits_{{\skew5\hat{i}} \in \skew4\hat{N}} 
m_{\skew5\hat{i}}\,n_{\skew5\hat{i}}^{\left( 0\right)}
\beta _{\skew5\hat{i}}^{\left( 1,r\right) } 
+{\hat{\rho }^{\left( 0 \right)}}{kT_{\skew5\hat{\phantom{i}}}^{\left( 0\right)}}
\boldsymbol{\beta }_{\skew5\hat{\phantom{i}}}^{\left( 2,r\right)}
+\frac{3}{2}\,
{\hat{\rho }^{\left( 0 \right)}}{kT_{\skew5\hat{\phantom{i}}}^{\left( 0\right)}}
{\boldsymbol{u}}_{\skew5\hat{\phantom{i}}}^{\left( {0} \right)}
\beta _{\skew5\hat{\phantom{i}}}^{\left( 3,r\right) }
=\sum\limits_{s = 0}^r 
{\hat{\rho }^{\left( r - s \right)} } 
{\boldsymbol{u}}_{\skew5\hat{\phantom{i}}}^{\left( {s} \right)} ,
\label{beta_r_equation2_in}
\end{eqnarray}
\begin{eqnarray}
&&\frac{1}{2}\sum\limits_{{\skew5\hat{i}} \in \skew4\hat{N}} 
n_{\skew5\hat{i}}^{\left( 0\right)}
{\left[ 3kT_{\skew5\hat{\phantom{i}}}^{\left( 0\right)}
 +m_{\skew5\hat{i}}{\left( u_{\skew5\hat{\phantom{i}}}^{\left( 0\right) }\right)}^2\right]}
\beta _{\skew5\hat{i}}^{\left( 1,r\right) }
\nonumber \\*
&&\quad +{\hat{\rho }^{\left( 0 \right)}}{kT_{\skew5\hat{\phantom{i}}}^{\left(0\right)}}
{\boldsymbol{u}}_{\skew5\hat{\phantom{i}}}^{\left( 0\right)}\cdot 
\boldsymbol{\beta }_{\skew5\hat{\phantom{i}}}^{\left( 2,r\right)}
\nonumber \\*
&&\quad +\frac{3}{4}\,kT_{\skew5\hat{\phantom{i}}}^{\left( 0\right)}
\left[ 5{\hat{n}^{\left( 0 \right)}}kT_{\skew5\hat{\phantom{i}}}^{\left( 0\right)}
+{\hat{\rho }^{\left( 0 \right)}}{\left( u_{\skew5\hat{\phantom{i}}}^{\left( 0\right) }\right)}^2 \right]
\beta _{\skew5\hat{\phantom{i}}}^{\left( 3,r\right) }
\nonumber \\*
&&=
\frac{3}{2}\, k 
\sum\limits_{s = 0}^r 
{\hat{n}^{\left( r - s \right)} } 
{T_{\skew5\hat{\phantom{i}}}^{\left( {s} \right)}}
+ 
\frac{1}{2} 
\sum\limits_{s = 0}^{r} \sum\limits_{q = 0}^{s} 
{\hat{\rho }^{\left( r - s \right)} } 
{\boldsymbol{u}}_{\skew5\hat{\phantom{i}}}^{\left( {s-q} \right)}\cdot 
{\boldsymbol{u}}_{\skew5\hat{\phantom{i}}}^{\left( {q} \right)} ,
\label{beta_r_equation3_in}
\end{eqnarray}
from which 
we find expressions for functions
$\beta _{\skew5\hat{i}}^{\left( 1,r\right)}\left( {\boldsymbol{r}},t\right)$,
$\boldsymbol{\beta }_{\skew5\hat{\phantom{i}}}^{\left( 2,r\right)}
\left( {\boldsymbol{r}},t\right)$ and 
$\beta _{\skew5\hat{\phantom{i}}}^{\left( 3,r\right)}\left( {\boldsymbol{r}},t\right)$
through (variable) coefficients of
asymptotic expansions of the particle number
density of ${\skew5\hat{i}}$-component, of the mean mass
velocity and of the temperature of inner components of the mixture
\begin{eqnarray}
\beta _{\skew5\hat{i}}^{\left( 1,r\right) }
&=&\frac{n_{\skew5\hat{i}}^{\left( r\right)}}{n_{\skew5\hat{i}}^{\left( 0\right)}}
-\frac{3}{2}\, 
\frac{1}
{{\hat{n}}^{\left( 0\right)} T_{\skew5\hat{\phantom{i}}}^{\left( 0\right)}}
\left [
\sum\limits_{s = 0}^r 
\left ( {\hat{n}}^{\left( r - s \right)}
{T_{\skew5\hat{\phantom{i}}}^{\left( {s} \right)}}\right )
- {\hat{n}}^{\left( r \right)}
{T_{\skew5\hat{\phantom{i}}}^{\left( {0} \right)}}
\right ]
\nonumber \\*
&&-\frac{1}{2}\, 
\frac{1}
{{\hat{n}}^{\left( 0\right)} kT_{\skew5\hat{\phantom{i}}}^{\left( 0\right)}}
\left [
\sum\limits_{s = 0}^{r} \sum\limits_{q = 0}^{s} 
{\hat{\rho }}^{\left( r - s \right)}
{\boldsymbol{u}}_{\skew5\hat{\phantom{i}}}^{\left( {s-q} \right)}\cdot 
{\boldsymbol{u}}_{\skew5\hat{\phantom{i}}}^{\left( {q} \right)} 
-{\hat{\rho }}^{\left( r \right)}{\left( u_{\skew5\hat{\phantom{i}}}^{\left( 0\right) }\right)}^2
\right ]
\nonumber \\*
&&+
\frac{1}
{{\hat{n}}^{\left( 0\right)} kT_{\skew5\hat{\phantom{i}}}^{\left( 0\right)}}
\ {\boldsymbol{u}}_{\skew5\hat{\phantom{i}}}^{\left( {0} \right)} \cdot
\left [
\sum\limits_{s = 0}^r 
\left ( {\hat{\rho }}^{\left( r - s \right)}
{\boldsymbol{u}}_{\skew5\hat{\phantom{i}}}^{\left( {s} \right)} \right )
- {\hat{\rho }}^{\left( r \right)}
{\boldsymbol{u}}_{\skew5\hat{\phantom{i}}}^{\left( {0} \right)}
\right ]
,
\label{beta_1_r_in}
\end{eqnarray}
\begin{eqnarray}
\boldsymbol{\beta }_{\skew5\hat{\phantom{i}}}^{\left( 2,r\right)}
&=&
\frac{1}{{\hat{\rho }}^{\left( 0\right)}
kT_{\skew5\hat{\phantom{i}}}^{\left( 0\right)}}
\left [
\sum\limits_{s = 0}^r 
\left ( {\hat{\rho }^{\left( r - s \right)} }
{\boldsymbol{u}}_{\skew5\hat{\phantom{i}}}^{\left( {s} \right)} \right )
- {\hat{\rho }}^{\left( r \right)}
{\boldsymbol{u}}_{\skew5\hat{\phantom{i}}}^{\left( {0} \right)}
\right ] ,
\label{beta_2_r_in}
\end{eqnarray}
\begin{eqnarray}
\beta _{\skew5\hat{\phantom{i}}}^{\left( 3,r\right) }
&=&
\frac{k}{{\hat{n}}^{\left( 0\right)}
\left ( kT_{\skew5\hat{\phantom{i}}}^{\left( 0\right)}\right )^2}
\left [
\sum\limits_{s = 0}^r 
\left ( {\hat{n}^{\left( r - s \right)} } 
{T_{\skew5\hat{\phantom{i}}}^{\left( {s} \right)}}\right )
- {\hat{n}}^{\left( r \right)} 
{T_{\skew5\hat{\phantom{i}}}^{\left( {0} \right)}}
\right ]
\nonumber \\*
&&+
\,\frac{1}{3}\,
\frac{1}{{\hat{n}^{\left( 0\right)}}
\left ( kT_{\skew5\hat{\phantom{i}}}^{\left( 0\right)}\right )^2}
\left [
\sum\limits_{s = 0}^{r} \sum\limits_{q = 0}^{s} 
{\hat{\rho }}^{\left( r - s \right)} 
{\boldsymbol{u}}_{\skew5\hat{\phantom{i}}}^{\left( {s-q} \right)}\cdot 
{\boldsymbol{u}}_{\skew5\hat{\phantom{i}}}^{\left( {q} \right)} 
-{\hat{\rho }}^{\left( r \right)}{\left( u_{\skew5\hat{\phantom{i}}}^{\left( 0\right) }\right)}^2
\right ]
\nonumber \\*
&&-
\,\frac{2}{3}\,
\frac{1}{{\hat{n}^{\left( 0\right)}}
\left ( kT_{\skew5\hat{\phantom{i}}}^{\left( 0\right)}\right )^2}
\ {\boldsymbol{u}}_{\skew5\hat{\phantom{i}}}^{\left( {0} \right)} \cdot
\left [
\sum\limits_{s = 0}^r 
\left ( {\hat{\rho }^{\left( r - s \right)} }
{\boldsymbol{u}}_{\skew5\hat{\phantom{i}}}^{\left( {s} \right)} \right )
- {\hat{\rho }^{\left( r \right)} }
{\boldsymbol{u}}_{\skew5\hat{\phantom{i}}}^{\left( {0} \right)}
\right ]
.
\label{beta_3_r_in}
\end{eqnarray}
Then the fulfillment of equalities (\ref{r_eq_in_condiition}) can be
considered as the differential 
equations, the $r$-order equations of gas dynamics, for finding
$n_{\skew5\hat{i}}^{\left( r-1\right)}$, 
${\boldsymbol{u}}_{\skew5\hat{i}}^{\left( r-1\right)}$, 
$T_{\skew5\hat{i}}^{\left( r-1\right)}$ $\left( r=1,2\ldots \right)$.

For $r=1$ from (\ref{beta_1_r_in})-(\ref{beta_3_r_in}) we have 
\begin{eqnarray}
\beta _{\skew5\hat{i}}^{\left( 1,1\right) }
&=&\frac{n_{\skew5\hat{i}}^{\left( 1\right)}}{n_{\skew5\hat{i}}^{\left( 0\right)}}
-\frac{3}{2} \frac{T_{\skew5\hat{\phantom{i}}}^{\left( 1\right)}}{T_{\skew5\hat{\phantom{i}}}^{\left( 0\right)}} ,
\label{beta_1_1_in}
\\
\boldsymbol{\beta }_{\skew5\hat{\phantom{i}}}^{\left( 2,1\right)}
&=&\frac{{\boldsymbol{u}}_{\skew5\hat{\phantom{i}}}^{\left( {1} \right)}}{kT_{\skew5\hat{\phantom{i}}}^{\left( 0\right)}} ,
\label{beta_2_1_in}
\\
\beta _{\skew5\hat{\phantom{i}}}^{\left( 3,1\right) }
&=&\frac{1}{kT_{\skew5\hat{\phantom{i}}}^{\left( 0\right)}} 
\frac{T_{\skew5\hat{\phantom{i}}}^{\left( 1\right)}}{T_{\skew5\hat{\phantom{i}}}^{\left( 0\right)}} .
\label{beta_3_1_in}
\end{eqnarray}

The partial solution of the system of inhomogeneous equations
(\ref{r_eq_in_F})
$\{{\mathit{\Phi } _{\skew5\hat{i}}^{\left( r\right)}}\}_{{\skew5\hat{i}} \in
\skew4\hat{N}}$, satisfying (\ref{r_eq_in_unique}), 
may be constructed, for example, using expansion of
$\mathit{\Phi } _{\skew5\hat{i}}^{\left( r\right)}\left( \boldsymbol{c}_{\skew5\hat{i}}\right)$ 
in series in terms of Sonine polynomials with expansion coefficients, depending on 
$\boldsymbol{r}$ and $t$ (see \cite{chap52} or \cite{hirsch54});
such construction proves existence of solutions of the system of
integral equations (\ref{r_eq_in}) 
(one could simply use here and above Fredholm's theorems
\cite{fredholm03}, \cite{courant89}).

For $r \geq 1$ the velocity distribution functions of outer components of
gas mixture $f_{\skew5\check{i}}^{\left( r\right)}$
are found from the integral equations system (\ref{sys_eq_series_r_out}),
which, taking account of (\ref{J_i}) and analogous
(\ref{res_H-equality_in}) equality
\begin{eqnarray}
f_{\skew5\check{i}}^{\left( 0\right) \prime } f^{\left( 0\right) \prime }
\equiv 
f_{\skew5\check{i}}^{\left( 0\right)} f^{\left( 0\right)} ,
\label{res_H-equality_out}
\end{eqnarray}
can be rewritten in the form
\begin{eqnarray}
{\mathcal D}_{\skew5\check{i}}^{\left( r-1 \right)} 
&+& 
\sum\limits_{s = 1}^{r-1} 
 J_{\skew5\check{i}}\left(
  f_{\skew5\check{i}}^{\left( {r-s} \right)},f^{\left( s\right)}\right) 
+ \sum\limits_{j\neq {\skew5\check{i}}} 
\sum\limits_{s = 0}^{r-1} 
J_{{\skew5\check{i}}j}\left( f_{\skew5\check{i}}^{\left( r-1-s\right)},
f_{j}^{\left( s\right)}\right)
\nonumber\\
&=& 
-J_{\skew5\check{i}}\left(
  f_{\skew5\check{i}}^{\left( {0} \right)}\chi _{\skew5\check{i}}^{\left( r\right)},f^{\left( {0} \right)}\right) 
-J_{\skew5\check{i}}\left(
  f_{\skew5\check{i}}^{\left( {0} \right)},f^{\left( {0} \right)}\chi ^{\left( r\right)}\right)
\nonumber\\
&=&
-\iint
f_{\skew5\check{i}}^{\left( 0\right)}f^{\left( 0\right)}
\left( 
\chi _{\skew5\check{i}}^{\left( r\right)} + \chi ^{\left( r\right)} -
\chi _{\skew5\check{i}}^{\left( r\right) \prime } - \chi ^{\left( r\right)\prime } 
\right)
k_{{\skew5\check{i}}} \,d{\mathbf{k}}
\,d{\boldsymbol{c}} 
\qquad {\left( {{\skew5\check{i}} \in \skew4\check{N}}\right)},
\label{r_eq_out}
\end{eqnarray}
in (\ref{r_eq_out}) functions 
$f_{\skew5\check{i}}^{\left( r\right)}=f_{\skew5\check{i}}^{\left( 0\right)}
\chi _{\skew5\check{i}}^{\left( r\right)}$, where 
$\chi _{\skew5\check{i}}^{\left( r\right)}$ are new unknown functions.

Equations (\ref{r_eq_out}) differ from
equations (\ref{r_eq_in}) only in the left-hand sides, that are known from
the previous step of the successive approximations method, and in 
the absence of summation over components in the right-hand sides of
equations (\ref{r_eq_out}). 
Therefore similarly to, how it has been done above for inner
components, we obtain expression for the general solution of the
system of homogeneous integral equations,
corresponding to (\ref{r_eq_out}),
\begin{eqnarray}
\phi _{\skew5\check{i}}^{\left( r\right)}=
\alpha _{\skew5\check{i}}^{\left( 1,r\right)}+
\boldsymbol{\alpha }_{\skew5\check{{i}}}^{\left( 2,r\right)}
\cdot m_{\skew5\check{i}}\,\boldsymbol{c}_{\skew5\check{i}}+ 
{\alpha }_{\skew5\check{{i}}}^{\left( 3,r\right)}
\frac{1}{2}\,m_{\skew5\check{i}}\,c_{\skew5\check{i}}^{2} ,
\label{phi_expr_out}
\end{eqnarray}
Hence,
\begin{eqnarray}
\xi _{\skew5\check{i}}^{\left( r\right)}=
f_{\skew5\check{i}}^{\left( 0\right)} 
{\left( 
\alpha _{\skew5\check{i}}^{\left( 1,r\right)}+
\boldsymbol{\alpha }_{\skew5\check{{i}}}^{\left( 2,r\right)}
\cdot m_{\skew5\check{i}}\,\boldsymbol{c}_{\skew5\check{i}}+ 
{\alpha }_{\skew5\check{{i}}}^{\left( 3,r\right)}
\frac{1}{2}\,m_{\skew5\check{i}}\,c_{\skew5\check{i}}^{2}
\right)} 
\qquad {\left( {{\skew5\check{i}} \in \skew4\check{N}}\right)}
\label{xi_expr_out}
\end{eqnarray}
or
\begin{eqnarray}
\!\!\!\!\!\! \xi _{\skew5\check{i}}^{\left( r\right)}=
f_{\skew5\check{i}}^{\left( 0\right)} 
{\left[ 
\beta _{\skew5\check{i}}^{\left( 1,r\right)}+
\boldsymbol{\beta }_{\skew5\check{{i}}}^{\left( 2,r\right)}
\cdot m_{\skew5\check{i}}
\left(\boldsymbol{c}_{\skew5\check{i}}-{\boldsymbol{u}}_{\skew5\check{{i}}}^{\left( 0\right)}\right)
+ 
{\beta }_{\skew5\check{{i}}}^{\left( 3,r\right)}
\frac{1}{2}\,m_{\skew5\check{i}}
\left(\boldsymbol{c}_{\skew5\check{i}}-
{\boldsymbol{u}}_{\skew5\check{{i}}}^{\left(0\right)}\right)^{2}
\right]} 
\qquad {\left( {{\skew5\check{i}} \in \skew4\check{N}}\right)} .
\label{xi_expr_ease_out}
\end{eqnarray}
Necessary condition for the existence of solutions of
the system of integral equations (\ref{r_eq_out})
can be written as
\begin{eqnarray}
\int \psi _{\skew5\check{i}}^{\left( l\right)} F_{\skew5\check{i}}^{\left( r \right)} 
d{\boldsymbol{c}}_{\skew5\check{i}} 
=0
\qquad {\left( {{\skew5\check{i}} \in \skew4\check{N}},
\ l=1,2,3\right)} ,
\label{r_eq_out_condiition}
\end{eqnarray}
where $F_{\skew5\check{i}}^{\left( r \right)}$ 
denote left-hand sides of the equations (\ref{r_eq_out}), taken with
opposite sign.

Among (infinitesimal) set of particular solutions of the system of
equations (\ref{r_eq_out}) by the condition
\begin{eqnarray}
\int \psi _{\skew5\check{i}}^{\left( l\right)}
f_{\skew5\check{i}}^{\left( 0\right)} \mathit{\Phi } _{\skew5\check{i}}^{\left( r \right)} 
d{\boldsymbol{c}}_{\skew5\check{i}} =0
\qquad {\left( {{\skew5\check{i}} \in \skew4\check{N}},
\ l=1,2,3\right)} 
\label{r_eq_out_unique}
\end{eqnarray}
unique solution $\{\mathit{\Phi } _{\skew5\check{i}}^{\left( r\right)}\}
_{{\skew5\check{i}} \in \skew4\check{N}}$
may be chosen.
Having substituted expression for
$f_{\skew5\check{i}}^{\left( r\right)}
\left( {{\skew5\check{i}} \in \skew4\check{N}}\right)$
\begin{eqnarray}
f_{\skew5\check{i}}^{\left( r\right)}
&=&
\Xi _{\skew5\check{i}}^{\left( r\right)} + 
\xi _{\skew5\check{i}}^{\left( r\right)}
\nonumber \\*
&=&
f_{\skew5\check{i}}^{\left( 0\right)} 
\mathit{\Phi } _{\skew5\check{i}}^{\left( r\right)}+
f_{\skew5\check{i}}^{\left( 0\right)} 
{\left[ 
\beta _{\skew5\check{i}}^{\left( 1,r\right)}+
\boldsymbol{\beta }_{\skew5\check{{i}}}^{\left( 2,r\right)}
\cdot m_{\skew5\check{i}}
\left(\boldsymbol{c}_{\skew5\check{i}}-{\boldsymbol{u}}_{\skew5\check{{i}}}^{\left( 0\right)}\right)
+ 
{\beta }_{\skew5\check{{i}}}^{\left( 3,r\right)}
\frac{1}{2}\,m_{\skew5\check{i}}
\left(\boldsymbol{c}_{\skew5\check{i}}-
{\boldsymbol{u}}_{\skew5\check{{i}}}^{\left(0\right)}\right)^{2}
\right]}
\label{f_r_expression_out}
\end{eqnarray}
in (\ref{r_condition1_out})-(\ref{r_condition3n_out}),
taking (\ref{maxwell_f0_out}), (\ref{rho_def_out})-(\ref{param_f_out}) and
(\ref{r_eq_out_unique}) into account,
we obtain for each index ${\skew5\check{i}}$ a system of
$5$ algebraic equations [constraint equations for asymptotic
expansions (\ref{f_series}) and
(\ref{n_series}), (\ref{u_series_out})-(\ref{T_series_out})]:
\begin{eqnarray}
n_{\skew5\check{i}}^{\left( 0\right)}\beta _{\skew5\check{i}}^{\left( 1,r\right)}
+ \frac{3}{2}\,n_{\skew5\check{i}}^{\left( 0\right)}
kT_{\skew5\check{{i}}}^{\left( 0\right)}
{\beta }_{\skew5\check{{i}}}^{\left( 3,r\right)}
=n_{\skew5\check{i}}^{\left( r\right)} ,
\label{beta_r_equation1_out}
\end{eqnarray}
\begin{eqnarray}
{\boldsymbol{u}}_{\skew5\check{{i}}}^{\left( {0} \right)}
m_{\skew5\check{i}}\,n_{\skew5\check{i}}^{\left( 0\right)}
\beta _{\skew5\check{i}}^{\left( 1,r\right) } 
+{{\rho }_{\skew5\check{i}}^{\left( 0 \right)}}{kT_{\skew5\check{{i}}}^{\left( 0\right)}}
\boldsymbol{\beta }_{\skew5\check{{i}}}^{\left( 2,r\right)}
+\frac{3}{2}\,
{{\rho }_{\skew5\check{i}}^{\left( 0 \right)}}{kT_{\skew5\check{{i}}}^{\left( 0\right)}}
{\boldsymbol{u}}_{\skew5\check{{i}}}^{\left( {0} \right)}
\beta _{\skew5\check{{i}}}^{\left( 3,r\right) }
=\sum\limits_{s = 0}^r 
{{\rho }_{\skew5\check{i}}^{\left( r - s \right)} } 
{\boldsymbol{u}}_{\skew5\check{{i}}}^{\left( {s} \right)} ,
\label{beta_r_equation2_out}
\end{eqnarray}
\begin{eqnarray}
&&\frac{1}{2} 
n_{\skew5\check{i}}^{\left( 0\right)}
{\left[ 3kT_{\skew5\check{{i}}}^{\left( 0\right)}
 +m_{\skew5\check{i}}{\left( u_{\skew5\check{{i}}}^{\left( 0\right) }\right)}^2\right]}
\beta _{\skew5\check{i}}^{\left( 1,r\right) }
\nonumber \\*
&&\quad +{{\rho }_{\skew5\check{i}}^{\left( 0 \right)}}{kT_{\skew5\check{{i}}}^{\left(0\right)}}
{\boldsymbol{u}}_{\skew5\check{{i}}}^{\left( 0\right)}\cdot 
\boldsymbol{\beta }_{\skew5\check{{i}}}^{\left( 2,r\right)}
\nonumber \\*
&&\quad +\frac{3}{4}\,kT_{\skew5\check{{i}}}^{\left( 0\right)}
\left[ 5{{n}_{\skew5\check{i}}^{\left( 0 \right)}}kT_{\skew5\check{{i}}}^{\left( 0\right)}
+{{\rho }_{\skew5\check{i}}^{\left( 0 \right)}}{\left( u_{\skew5\check{{i}}}^{\left( 0\right) }\right)}^2 \right]
\beta _{\skew5\check{{i}}}^{\left( 3,r\right) }
\nonumber \\*
&&=
\frac{3}{2}\, k 
\sum\limits_{s = 0}^r 
{{n}_{\skew5\check{i}}^{\left( r - s \right)} } 
{T_{\skew5\check{{i}}}^{\left( {s} \right)}}
+ 
\frac{1}{2} 
\sum\limits_{s = 0}^{r} \sum\limits_{q = 0}^{s} 
{{\rho }_{\skew5\check{i}}^{\left( r - s \right)} } 
{\boldsymbol{u}}_{\skew5\check{{i}}}^{\left( {s-q} \right)}\cdot 
{\boldsymbol{u}}_{\skew5\check{{i}}}^{\left( {q} \right)} ,
\label{beta_r_equation3_out}
\end{eqnarray}
from which 
we find expressions for functions
$\beta _{\skew5\check{i}}^{\left( 1,r\right)}\left( {\boldsymbol{r}},t\right)$,
$\boldsymbol{\beta }_{\skew5\check{{i}}}^{\left( 2,r\right)}
\left( {\boldsymbol{r}},t\right)$ and 
$\beta _{\skew5\check{{i}}}^{\left( 3,r\right)}\left( {\boldsymbol{r}},t\right)$
through (variable) coefficients of
asymptotic expansions of the particle number
density of ${\skew5\check{i}}$-component, of the mean
velocity and of the temperature of ${\skew5\check{i}}$-component
\begin{eqnarray}
\beta _{\skew5\check{i}}^{\left( 1,r\right) }
&=&\frac{n_{\skew5\check{i}}^{\left( r\right)}}{n_{\skew5\check{i}}^{\left( 0\right)}}
-\frac{3}{2}\, 
\frac{1}
{{{n}_{\skew5\check{i}}^{\left( 0\right)} }T_{\skew5\check{{i}}}^{\left( 0\right)}}
\left [
\sum\limits_{s = 0}^r 
\left ( {{n}_{\skew5\check{i}}^{\left( r - s \right)} } 
{T_{\skew5\check{{i}}}^{\left( {s} \right)}}\right )
- {{n}_{\skew5\check{i}}^{\left( r \right)} } 
{T_{\skew5\check{{i}}}^{\left( {0} \right)}}
\right ]
\nonumber \\*
&&-\frac{1}{2}\, 
\frac{1}
{{{n}_{\skew5\check{i}}^{\left( 0\right)} }kT_{\skew5\check{{i}}}^{\left( 0\right)}}
\left [
\sum\limits_{s = 0}^{r} \sum\limits_{q = 0}^{s} 
{{\rho }_{\skew5\check{i}}^{\left( r - s \right)} } 
{\boldsymbol{u}}_{\skew5\check{{i}}}^{\left( {s-q} \right)}\cdot 
{\boldsymbol{u}}_{\skew5\check{{i}}}^{\left( {q} \right)} 
-{{\rho }_{\skew5\check{i}}^{\left( r \right)}}{\left( u_{\skew5\check{{i}}}^{\left( 0\right) }\right)}^2
\right ]
\nonumber \\*
&&+
\frac{1}
{{{n}_{\skew5\check{i}}^{\left( 0\right)} }kT_{\skew5\check{{i}}}^{\left( 0\right)}}
\ {\boldsymbol{u}}_{\skew5\check{{i}}}^{\left( {0} \right)} \cdot
\left [
\sum\limits_{s = 0}^r 
\left ( {{\rho }_{\skew5\check{i}}^{\left( r - s \right)} }
{\boldsymbol{u}}_{\skew5\check{{i}}}^{\left( {s} \right)} \right )
- {{\rho }_{\skew5\check{i}}^{\left( r \right)} }
{\boldsymbol{u}}_{\skew5\check{{i}}}^{\left( {0} \right)}
\right ]
,
\label{beta_1_r_out}
\end{eqnarray}
\begin{eqnarray}
\boldsymbol{\beta }_{\skew5\check{{i}}}^{\left( 2,r\right)}
&=&
\frac{1}{{{\rho }_{\skew5\check{i}}^{\left( 0\right)}}
kT_{\skew5\check{{i}}}^{\left( 0\right)}}
\left [
\sum\limits_{s = 0}^r 
\left ( {{\rho }_{\skew5\check{i}}^{\left( r - s \right)} }
{\boldsymbol{u}}_{\skew5\check{{i}}}^{\left( {s} \right)} \right )
- {{\rho }_{\skew5\check{i}}^{\left( r \right)} }
{\boldsymbol{u}}_{\skew5\check{{i}}}^{\left( {0} \right)}
\right ] ,
\label{beta_2_r_out}
\end{eqnarray}
\begin{eqnarray}
\beta _{\skew5\check{{i}}}^{\left( 3,r\right) }
&=&
\frac{k}{{{n}_{\skew5\check{i}}^{\left( 0\right)}}
\left ( kT_{\skew5\check{{i}}}^{\left( 0\right)}\right )^2}
\left [
\sum\limits_{s = 0}^r 
\left ( {{n}_{\skew5\check{i}}^{\left( r - s \right)} } 
{T_{\skew5\check{{i}}}^{\left( {s} \right)}}\right )
- {{n}_{\skew5\check{i}}^{\left( r \right)} } 
{T_{\skew5\check{{i}}}^{\left( {0} \right)}}
\right ]
\nonumber \\*
&&+
\,\frac{1}{3}\,
\frac{1}{{{n}_{\skew5\check{i}}^{\left( 0\right)}}
\left ( kT_{\skew5\check{{i}}}^{\left( 0\right)}\right )^2}
\left [
\sum\limits_{s = 0}^{r} \sum\limits_{q = 0}^{s} 
{{\rho }_{\skew5\check{i}}^{\left( r - s \right)} } 
{\boldsymbol{u}}_{\skew5\check{{i}}}^{\left( {s-q} \right)}\cdot 
{\boldsymbol{u}}_{\skew5\check{{i}}}^{\left( {q} \right)} 
-{{\rho }_{\skew5\check{i}}^{\left( r \right)}}{\left( u_{\skew5\check{{i}}}^{\left( 0\right) }\right)}^2
\right ]
\nonumber \\*
&&-
\,\frac{2}{3}\,
\frac{1}{{{n}_{\skew5\check{i}}^{\left( 0\right)}}
\left ( kT_{\skew5\check{{i}}}^{\left( 0\right)}\right )^2}
\ {\boldsymbol{u}}_{\skew5\check{{i}}}^{\left( {0} \right)} \cdot
\left [
\sum\limits_{s = 0}^r 
\left ( {{\rho }_{\skew5\check{i}}^{\left( r - s \right)} }
{\boldsymbol{u}}_{\skew5\check{{i}}}^{\left( {s} \right)} \right )
- {{\rho }_{\skew5\check{i}}^{\left( r \right)} }
{\boldsymbol{u}}_{\skew5\check{{i}}}^{\left( {0} \right)}
\right ]
.
\label{beta_3_r_out}
\end{eqnarray}
The fulfillment of equalities (\ref{r_eq_out_condiition}) can be
considered as the differential 
equations, the $r$-order equations of gas dynamics, for finding
$n_{\skew5\check{i}}^{\left( r-1\right)}$, 
${\boldsymbol{u}}_{\skew5\check{i}}^{\left( r-1\right)}$, 
$T_{\skew5\check{i}}^{\left( r-1\right)}$ $\left( r=1,2\ldots \right)$.

For $r=1$ from (\ref{beta_1_r_out})-(\ref{beta_3_r_out}) we have, in particular: 
\begin{eqnarray}
\beta _{\skew5\check{i}}^{\left( 1,1\right) }
&=&\frac{n_{\skew5\check{i}}^{\left( 1\right)}}{n_{\skew5\check{i}}^{\left( 0\right)}}
-\frac{3}{2} \frac{T_{\skew5\check{{i}}}^{\left( 1\right)}}{T_{\skew5\check{{i}}}^{\left( 0\right)}} ,
\label{beta_1_1_out}
\\
\boldsymbol{\beta }_{\skew5\check{{i}}}^{\left( 2,1\right)}
&=&\frac{{\boldsymbol{u}}_{\skew5\check{{i}}}^{\left( {1} \right)}}{kT_{\skew5\check{{i}}}^{\left( 0\right)}} ,
\label{beta_2_1_out}
\\
\beta _{\skew5\check{{i}}}^{\left( 3,1\right) }
&=&\frac{1}{kT_{\skew5\check{{i}}}^{\left( 0\right)}} 
\frac{T_{\skew5\check{{i}}}^{\left( 1\right)}}{T_{\skew5\check{{i}}}^{\left( 0\right)}} ,
\label{beta_3_1_out}
\end{eqnarray}
cf. with (\ref{beta_1_1_in})-(\ref{beta_3_1_in}).

The partial solution
$\{{\mathit{\Phi } _{\skew5\check{i}}^{\left( r\right)}}\}
_{{\skew5\check{i}} \in \skew4\check{N}}$ 
of the system of inhomogeneous integral equations
(\ref{r_eq_out}), satisfying (\ref{r_eq_out_unique}), 
may be constructed similarly to, how satisfying (\ref{r_eq_in_unique})
partial solution
$\{{\mathit{\Phi } _{\skew5\hat{i}}^{\left( r\right)}}\}
_{{\skew5\hat{i}} \in \skew4\hat{N}}$ 
of the system of inhomogeneous integral equations (\ref{r_eq_in_F})
is constructed, see above.

\section{The system of first order equations of multicomponent non-equilibrium gas dynamics}
\label{sec:order1gdes}

Let us consider in more detail the system of infinitesimal first order
equations (\ref{r_eq_in_condiition}), 
(\ref{r_eq_out_condiition}) ${\left( r=1 \right)}$, 
derived above as the necessary (and sufficient) condition of the
solution existence of the first order integral equations system 
(\ref{r_eq_in}), (\ref{r_eq_out}) ${\left( r=1 \right)}$.

To simplify transformations, according to the expressions for velocity
distribution functions of particles of infinitesimal zero order
(\ref{maxwell_f0_in}), (\ref{maxwell_f0_out}),
functions $\Psi _{\skew5\hat{i}}^{\left( l\right)}$, 
$\Psi _{\skew5\check{i}}^{\left( l\right)}$ may be used in
(\ref{r_eq_in_condiition}) and (\ref{r_eq_out_condiition}) 
${\left( r=1 \right)}$ rather than functions 
$\psi _{\skew5\hat{i}}^{\left( l\right)}$, 
$\psi _{\skew5\check{i}}^{\left( l\right)}$, 
respectively:
\begin{subequations}
\label{Psi_in}
 \begin{eqnarray}
 \Psi _{{i}}^{\left( 1\right)}
 &=& m_{{i}},
 \label{Psi_1_in}
 \\
 {\boldsymbol{\Psi}}_{{i}}^{\left( 2\right)}
 &=& m_{{i}}\,{\boldsymbol{C}}_{{i}},
 \label{Psi_2_in}
 \\
 \Psi _{{i}}^{\left( 3\right)}
 &=& 
 \frac{1}{2}\, m_{{i}}\, C_{{i}}^{2},
 \label{Psi_3_in}
 \end{eqnarray}
\end{subequations}
for inner components ${\boldsymbol{C}}_{\skew5\hat{i}}
={\boldsymbol{c}}_{\skew5\hat{i}}-
{\boldsymbol{u}}_{\skew5\hat{\phantom{i}}}^{\left( 0\right)}$,
for outer components ${\boldsymbol{C}}_{\skew5\check{i}}
={\boldsymbol{c}}_{\skew5\check{i}}-
{\boldsymbol{u}}_{\skew5\check{{i}}}^{\left( 0\right)}$.

At transformation of differential parts of the equations 
(\ref{r_eq_in_condiition}) and (\ref{r_eq_out_condiition}) we use equalities:
\begin{eqnarray}
\int \Psi _{i}^{\left( l\right)}\frac{\partial f_{i}^{\left( 0\right)}}
{\partial t}\,d{\boldsymbol{c}}_{i}
&=&
\frac{\partial }{\partial t}\int \Psi _{i}^{\left(
    l\right)}f_{i}^{\left( 0\right)}
\,d{\boldsymbol{c}}_{i}-
\int \frac{\partial \Psi _{i}^{\left( l\right)}}
{\partial t}f_{i}^{\left( 0\right)}\,d{\boldsymbol{c}}_{i}
\nonumber\\*
&=&
\frac{\partial \left( n_{i}
\,\overline{\Psi _{i}^{\left( l\right)}}^{\,\left( 0\right)}\right)}{\partial t}-
n_{i}\,\overline{\frac{\partial \Psi _{i}^{\left( l\right)}}{\partial t}}^{\,\left( 0\right)} ,
\label{trans1}
\end{eqnarray}
\begin{eqnarray}
\int \Psi _{i}^{\left( l\right)}{\boldsymbol{c}}_{i}\cdot 
\frac{\partial f_{i}^{\left( 0\right)}}
{\partial {\boldsymbol{r}}}\,d{\boldsymbol{c}}_{i}
&=&
\frac{\partial }{\partial {\boldsymbol{r}}}\cdot \int \Psi _{i}^{\left( l\right)}
{\boldsymbol{c}}_{i}f_{i}^{\left( 0\right)}\,d{\boldsymbol{c}}_{i} 
-\int {\boldsymbol{c}}_{i}\cdot \frac{\partial \Psi _{i}^{\left( l\right)}}
{\partial {\boldsymbol{r}}}f_{i}^{\left( 0\right)}\,d{\boldsymbol{c}}_{i} \nonumber\\*
&=&
\frac{\partial }{\partial {\boldsymbol{r}}}
\cdot n_{i}\,\overline{\Psi _{i}^{\left( l\right)}{\boldsymbol{c}}_{i}}^{\,\left( 0\right)} 
-n_{i}\,\overline{{\boldsymbol{c}}_{i}
\cdot \frac{\partial \Psi _{i}^{\left( l\right)}}{\partial {\boldsymbol{r}}}}^{\,\left( 0\right)} ,
\label{trans2}
\end{eqnarray}
\begin{eqnarray}
\int \Psi _{i}^{\left( l\right)}\frac{{\boldsymbol{X}}_{i}}{m_{i}}
\cdot \frac{\partial f_{i}^{\left( 0\right)}}{\partial {\boldsymbol{c}}_{i}}
\,d{\boldsymbol{c}}_{i}
&=&
-\int \left( \frac{\partial }{\partial {\boldsymbol{c}}_{i}}
\cdot \Psi _{i}^{\left(
    l\right)}\frac{{\boldsymbol{X}}_{i}}{m_{i}}\right) 
f_{i}^{\left( 0\right)}\,d{\boldsymbol{c}}_{i} \nonumber\\*
&=&
-n_{i}\,\overline{\frac{\partial }{\partial {\boldsymbol{c}}_{i}}
\cdot \Psi _{i}^{\left( l\right)}\frac{{\boldsymbol{X}}_{i}}{m_{i}}}^{\,\left( 0\right)} .
\label{trans3}
\end{eqnarray}
In (\ref{trans1})-(\ref{trans3}) the bar above symbol with index  
$^{\left( 0\right)}$ denotes the average of 
the value:
\begin{eqnarray}
\overline{V}^{\,\left( 0\right)}=\frac{1}{n_{i}}
\int Vf_{i}^{\left( 0\right)}\,d{\boldsymbol{c}}_{i} ;
\label{average}
\end{eqnarray}
${\boldsymbol{r}}$ and ${\boldsymbol{c}}_{i}$ are considered as
independent variables;
At averaging in (\ref{trans3}) it is assumed, that 
external force ${\boldsymbol{X}}_{i}$, acting on the particle of
species $i$, is independent of the particle velocity, 
it is assumed also, that integrals, depending on 
external forces ${\boldsymbol{X}}_{i}$, are convergent, 
and product 
$\Psi _{i}^{\left( l\right)}{\boldsymbol{X}}_{i}f_{i}^{\left( {0} \right)}$ 
tends to zero, when ${\boldsymbol{c}}_{i}$ tends to
infinity. 

After simple transformations from 
(\ref{r_eq_in_condiition}) and (\ref{r_eq_out_condiition}) 
${\left( r=1 \right)}$ we obtain following
\textit{system of infinitesimal first order equations of multicomponent
non-equilibrium gas dynamics}:
\begin{eqnarray}
\frac{\partial n_{\skew5\hat{i}}^{\left( {0} \right)}}{\partial t}
=-\frac{\partial }{\partial {\boldsymbol{r}}}
\cdot n_{\skew5\hat{i}}^{\left( {0} \right)}{\boldsymbol{u}}_{\skew5\hat{\phantom{i}}}^{\left( {0} \right)} 
\qquad {\left( {{\skew5\hat{i}} \in \skew4\hat{N}} \right)},
\label{trans_m1_in}
\end{eqnarray}
\begin{eqnarray}
{\hat{\rho}}^{\left( {0} \right)}\,\frac{\partial 
{\boldsymbol{u}}_{\skew5\hat{\phantom{i}}}^{\left( {0} \right)}}{\partial t}
+\frac{\partial }{\partial {\boldsymbol{r}}}
\cdot {\hat{\mathrm{p}}}^{\left( 0\right)}
+\sum\limits_{{{\skew5\hat{i}} \in \skew4\hat{N}},\, {{\skew5\check{j}} \in \skew4\check{N}}}
{\boldsymbol{J}}_{p,\,{\skew5\hat{i}}{\skew5\check{j}}}^{\,\left( 0\right)}
=
\sum\limits_{{{\skew5\hat{i}} \in \skew4\hat{N}}}{n_{\skew5\hat{i}}^{\left( {0} \right)}
{\boldsymbol{X}}_{\skew5\hat{i}}}-
{\hat{\rho}}^{\left( {0} \right)}{\boldsymbol{u}}_{\skew5\hat{\phantom{i}}}^{\left( {0} \right)}
\cdot \frac{\partial }{\partial {\boldsymbol{r}}}\
{\boldsymbol{u}}_{\skew5\hat{\phantom{i}}}^{\left( {0} \right)} ,
\label{trans_p1_in}
\end{eqnarray}
\begin{eqnarray}
\frac{\partial {\hat{E}}^{\left( 0\right)}}{\partial t} 
+\frac{\partial }{\partial {\boldsymbol{r}}}\cdot {\hat{\boldsymbol{q}}}^{\left( 0\right)} 
+{\hat{\mathrm{p}}}^{\left( 0\right)}:
\frac{\partial {\boldsymbol{u}}_{\skew5\hat{\phantom{i}}}^{\left( 0\right)}}
{\partial {\boldsymbol{r}}}
+\sum\limits_{{{{\skew5\hat{i}} \in \skew4\hat{N}},\,
    {{\skew5\check{j}} \in \skew4\check{N}}}}
{J}_{E,\,{\skew5\hat{i}}{\skew5\check{j}}}^{\,\left( 0\right)}
= -\frac{\partial }{\partial {\boldsymbol{r}}}
\cdot {\hat{E}}^{\left( 0\right)}{\boldsymbol{u}}_{\skew5\hat{\phantom{i}}}^{\left( 0\right)} ,
\label{trans_e1_in}
\end{eqnarray}
\begin{eqnarray}
\frac{\partial n_{\skew5\check{i}}^{\left( 0\right)}}{\partial t}
=-\frac{\partial }{\partial {\boldsymbol{r}}}
\cdot n_{\skew5\check{i}}^{\left( 0\right)}{\boldsymbol{u}}_{\skew5\check{i}}^{\left( 0\right)} 
\qquad {\left( {{\skew5\check{i}} \in \skew4\check{N}} \right)},
\label{trans_m1_out}
\end{eqnarray}
\begin{eqnarray}
\!\!\!\!\!\!\!\!\!
n_{\skew5\check{i}}^{\left( 0\right)}m_{\skew5\check{i}}\,
\frac{\partial {\boldsymbol{u}}_{\skew5\check{i}}^{\left( 0\right)}}{\partial t}
+\frac{\partial }{\partial {\boldsymbol{r}}}
\cdot {\mathrm{p}}_{\skew5\check{i}}^{\left( 0\right)}
+\sum\limits_{j\neq {\skew5\check{i}}}{\boldsymbol{J}}_{p,\,{\skew5\check{i}}j}^{\,\left( 0\right)}
=
n_{\skew5\check{i}}^{\left( 0\right)}{\boldsymbol{X}}_{\skew5\check{i}}-
n_{\skew5\check{i}}^{\left( 0\right)}m_{\skew5\check{i}}
\,{\boldsymbol{u}}_{\skew5\check{i}}^{\left( 0\right)}
\cdot \frac{\partial }{\partial {\boldsymbol{r}}}\
{\boldsymbol{u}}_{\skew5\check{i}}^{\left( 0\right)} 
\qquad {\left( {{\skew5\check{i}} \in \skew4\check{N}} \right)},
\label{trans_p1_out}
\end{eqnarray}
\begin{eqnarray}
\!\!\!\!\!\!\!\!\!
\frac{\partial {E}_{\skew5\check{i}}^{\left( 0\right)}}{\partial t} 
+\frac{\partial }{\partial {\boldsymbol{r}}}\cdot 
{\boldsymbol{q}}_{\skew5\check{i}}^{\left( 0\right)} 
+{\mathrm{p}}_{\skew5\check{i}}^{\left( 0\right)}:
\frac{\partial {\boldsymbol{u}}_{\skew5\check{i}}^{\left( 0\right)}}
{\partial {\boldsymbol{r}}}
+\sum\limits_{j\neq {\skew5\check{i}}}{J}_{E,\,{\skew5\check{i}}j}^{\,\left( 0\right)}
= -\frac{\partial }{\partial {\boldsymbol{r}}}
\cdot {E}_{\skew5\check{i}}^{\left( 0\right)}{\boldsymbol{u}}_{\skew5\check{i}}^{\left( 0\right)} 
\qquad {\left( {{\skew5\check{i}} \in \skew4\check{N}} \right)}.
\label{trans_e1_out}
\end{eqnarray}
In accordance with the general definition of pressure tensor of
$i$-component of gas mixture
\begin{eqnarray}
{\mathrm{p}}_{i}
{\overset{\mathrm{def}}{=}}
\int m_{i}
{\left( {\boldsymbol{c}}_{i}-{\boldsymbol{u}}_{i}\right)
 \left( {\boldsymbol{c}}_{i}-{\boldsymbol{u}}_{i}\right)}
f_{i}
\,d{\boldsymbol{c}}_{i} 
\label{def_p_i}
\end{eqnarray}
and with the general definition of $i$-component heat flux vector
\begin{eqnarray}
{\boldsymbol{q}}_{i}
{\overset{\mathrm{def}}{=}}
\int 
\frac{1}{2}\,m_{i}
{\left( {\boldsymbol{c}}_{i}-
{\boldsymbol{u}}_{i}\right) ^{2}
\left( {\boldsymbol{c}}_{i}-
{\boldsymbol{u}}_{i}\right)}
f_{i}
\,d{\boldsymbol{c}}_{i} 
\label{def_q_i}
\end{eqnarray}
(cf. with \cite{chap52}, Chapter~2, \S \S ~3,~4) in
(\ref{trans_m1_in})-(\ref{trans_e1_out}) 
\begin{eqnarray}
{\hat{\mathrm{p}}}^{\left( 0\right)}=
\sum\limits_{{\skew5\hat{i}} \in \skew4\hat{N}}n_{\skew5\hat{i}}^{\left( 0\right)}m_{\skew5\hat{i}}
\,\overline{\left( {\boldsymbol{c}}_{\skew5\hat{i}}-
{\boldsymbol{u}}_{\skew5\hat{\phantom{i}}}^{\left( 0\right)}\right)
\left( {\boldsymbol{c}}_{\skew5\hat{i}}-
{\boldsymbol{u}}_{\skew5\hat{\phantom{i}}}^{\left( 0\right)}\right)}
^{\,\left( 0\right)}
={\hat{n}}^{\left( 0\right)} k T_{\skew5\hat{\phantom{i}}}^{\left( 0\right)} {\mathrm{U}}=
{\hat{p}}^{\left( 0\right)} {\mathrm{U}}
\label{p_in_0}
\end{eqnarray}
is inner components pressure tensor of zero order, 
${\hat{p}}^{\left( 0\right)}$ is inner components hydrostatic pressure of zero order, 
${\mathrm{U}}$ is the unit tensor,
\textit{double product} of two second rank tensors ${\mathrm{w}}$ and 
${\mathrm{w}}^{\prime }$ 
(\cite{chap52}, Chapter~1, \S~3) is the scalar
${\mathrm{w}}:{\mathrm{w}}^{\prime }=
\sum_{\alpha }\sum_{\beta }w_{\alpha \beta }w^{\prime }_{\beta \alpha }=
{\mathrm{w}}^{\prime }:{\mathrm{w}}$,
\begin{eqnarray}
{\hat{\boldsymbol{q}}}^{\left( 0\right)}=\frac{1}{2}\,
\sum\limits_{{\skew5\hat{i}} \in \skew4\hat{N}}n_{\skew5\hat{i}}^{\left( 0\right)}m_{\skew5\hat{i}}
\,\overline{\left( {\boldsymbol{c}}_{\skew5\hat{i}}-
{\boldsymbol{u}}_{\skew5\hat{\phantom{i}}}^{\left( 0\right)}\right) ^{2}
\left( {\boldsymbol{c}}_{\skew5\hat{i}}-
{\boldsymbol{u}}_{\skew5\hat{\phantom{i}}}^{\left( 0\right)}\right)}^{\,\left( 0\right)} = 0
\label{q_in_0}
\end{eqnarray}
is inner components heat flux vector of zero order,
\begin{eqnarray}
{\hat{E}}^{\left( 0\right)}=
\frac{1}{2}\,
\sum\limits_{{\skew5\hat{i}} \in \skew4\hat{N}}n_{\skew5\hat{i}}^{\left( 0\right)}m_{\skew5\hat{i}}
\,\overline{\left( {\boldsymbol{c}}_{\skew5\hat{i}}-
{\boldsymbol{u}}_{\skew5\hat{\phantom{i}}}^{\left( 0\right)}\right) ^{2}}^{\,\left( 0\right)}
=\frac{3}{2}\,{\hat{n}}^{\left( 0\right)}k{T}_{\skew5\hat{\phantom{i}}}^{\left( 0\right)}
\label{E_in_0}
\end{eqnarray}
is zero order internal energy of particles of inner components per unit volume, 
which is equal, in this case, to energy of their translational chaotic motion, 
however, the energy transfer equations, written in form
(\ref{trans_e1_in}) and (\ref{trans_e1_out}) 
can be used in more general cases as well 
(cf. with \cite{hirsch54}, Chapter~7, \S ~6), 
in (\ref{p_in_0})-(\ref{E_in_0}) averaging (\ref{average}) is
performed with Maxwell function $f_{\skew5\hat{i}}^{\left( 0\right)}$ from 
(\ref{maxwell_f0_in}); 
\begin{eqnarray}
{\mathrm{p}}_{\skew5\check{i}}^{\left( 0\right)}=n_{\skew5\check{i}}^{\left( 0\right)}m_{\skew5\check{i}}
\,\overline{\left( {\boldsymbol{c}}_{\skew5\check{i}}-
{\boldsymbol{u}}_{\skew5\check{i}}^{\left( 0\right)}\right)
\left( {\boldsymbol{c}}_{\skew5\check{i}}-
{\boldsymbol{u}}_{\skew5\check{i}}^{\left( 0\right)}\right)}^{\,\left( 0\right)}
=n_{\skew5\check{i}}^{\left( 0\right)} k T_{\skew5\check{i}}^{\left(
    0\right)} {\mathrm{U}}
=p_{\skew5\check{i}}^{\left( 0\right)} {\mathrm{U}}
\label{p_i_0}
\end{eqnarray}
is ${\skew5\check{i}}$-component pressure tensor of zero order, 
$p_{\skew5\check{i}}^{\left( 0\right)}$ is
${\skew5\check{i}}$-component hydrostatic pressure of zero order,   
\begin{eqnarray}
{\boldsymbol{q}}_{\skew5\check{i}}^{\left( 0\right)}
=\frac{1}{2}\,n_{\skew5\check{i}}^{\left( 0\right)}m_{\skew5\check{i}}
\,\overline{\left( {\boldsymbol{c}}_{\skew5\check{i}}-
{\boldsymbol{u}}_{\skew5\check{i}}^{\left( 0\right)}\right) ^{2}\left( {\boldsymbol{c}}_{\skew5\check{i}}-
{\boldsymbol{u}}_{\skew5\check{i}}^{\left( 0\right)}\right)}^{\,\left( 0\right)} = 0
\label{q_i_0}
\end{eqnarray}
is ${\skew5\check{i}}$-component heat flux vector of zero order,
\begin{eqnarray}
{E}_{\skew5\check{i}}^{\left( 0\right)}
=\frac{1}{2}\,n_{\skew5\check{i}}^{\left( 0\right)}m_{\skew5\check{i}}
\,\overline{\left( {\boldsymbol{c}}_{\skew5\check{i}}-
{\boldsymbol{u}}_{\skew5\check{i}}^{\left( 0\right)}\right) ^{2}}^{\,\left( 0\right)}
=\frac{3}{2}\,n_{\skew5\check{i}}^{\left( 0\right)}k{T}_{\skew5\check{i}}^{\left( 0\right)}
\label{E_i_0}
\end{eqnarray}
is zero order internal energy of particles of ${\skew5\check{i}}$-component per unit volume, 
in (\ref{p_i_0})-(\ref{E_i_0}) averaging (\ref{average}) is
performed with Maxwell function $f_{\skew5\check{i}}^{\left( 0\right)}$ from 
(\ref{maxwell_f0_out}). 

Analytic expressions for integrals ${\boldsymbol{J}}_{p,\,ij}^{\,\left( 0\right)}$, 
$J_{E,\,ij}^{\,\left( 0\right)}$ from (\ref{trans_p1_in}),
(\ref{trans_e1_in}) and (\ref{trans_p1_out}), (\ref{trans_e1_out})
are given below -- see (\ref{int_impulse_result}), (\ref{int_energy_result}) and
(\ref{J_p}), (\ref{J_E}). 

In particular, if mean velocities and temperatures coincide for all
component of mixture (the set of outer components is empty), then
from (\ref{trans_m1_in})-(\ref{trans_e1_in}) we obtain the system of
infinitesimal first order gas-dynamics equations of the Enskog-Chapman
theory
[it may be noted in passing, that the sums over all
indexes $i$ of collision integrals in (\ref{r_eq_out_condiition})
are equal to zero in all orders of the method of successive
approximations, it is easy to check, using equality
(\ref{equality}), from the physical point of view this statement is
reduced to, that total impulse and (kinetic) energy of particles of
gas do not change in their collisions between themselves]:
\begin{eqnarray}
\frac{\partial n_{\skew5\hat{i}}^{\left( {0} \right)}}{\partial t}
=-\frac{\partial }{\partial {\boldsymbol{r}}}
\cdot n_{\skew5\hat{i}}^{\left( {0} \right)}{\boldsymbol{u}}_{\skew5\hat{\phantom{i}}}^{\left( {0} \right)} 
\qquad {\left( {{\skew5\hat{i}} \in \skew4\hat{N}} \right)},
\label{trans_m1_E}
\end{eqnarray}
\begin{eqnarray}
{\hat{\rho}}^{\left( {0} \right)}\,\frac{\partial 
{\boldsymbol{u}}_{\skew5\hat{\phantom{i}}}^{\left( {0} \right)}}{\partial t}
+\frac{\partial }{\partial {\boldsymbol{r}}}
\cdot {\hat{\mathrm{p}}}^{\left( 0\right)}
=\sum\limits_{{{\skew5\hat{i}} \in \skew4\hat{N}}}{n_{\skew5\hat{i}}^{\left( {0} \right)}
{\boldsymbol{X}}_{\skew5\hat{i}}}-
{\hat{\rho}}^{\left( {0} \right)}{\boldsymbol{u}}_{\skew5\hat{\phantom{i}}}^{\left( {0} \right)}
\cdot \frac{\partial }{\partial {\boldsymbol{r}}}\
{\boldsymbol{u}}_{\skew5\hat{\phantom{i}}}^{\left( {0} \right)} ,
\label{trans_p1_E}
\end{eqnarray}
\begin{eqnarray}
\frac{\partial {\hat{E}}^{\left( 0\right)}}{\partial t} 
+\frac{\partial }{\partial {\boldsymbol{r}}}\cdot {\hat{\boldsymbol{q}}}^{\left( 0\right)} 
+{\hat{\mathrm{p}}}^{\left( 0\right)}:
\frac{\partial {\boldsymbol{u}}_{\skew5\hat{\phantom{i}}}^{\left( 0\right)}}
{\partial {\boldsymbol{r}}}
= -\frac{\partial }{\partial {\boldsymbol{r}}}
\cdot {\hat{E}}^{\left( 0\right)}{\boldsymbol{u}}_{\skew5\hat{\phantom{i}}}^{\left( 0\right)} .
\label{trans_e1_E}
\end{eqnarray}
 
It follows from (\ref{trans_m1_E}), (\ref{trans_e1_E}), taking 
(\ref{p_in_0})-(\ref{E_in_0}) into account, that the flow of gas, described by 
the system of first order gas-dynamics equations of the Enskog-Chapman
theory, is \textit{adiabatic}:
\begin{eqnarray}
\frac{D }{D t} \left[ {\hat{n}}^{\left( 0\right)} 
\left( {T}_{\skew5\hat{\phantom{i}}}^{\left( 0\right)} \right)
^{-3 \left/ 2 \right.} 
\right]=0 ,
\label{adiabatic}
\end{eqnarray}
in \textit{the adiabatic equation} (\ref{adiabatic})
\begin{eqnarray}
\frac{D }{D t} 
&=& \frac{\partial }{\partial t}
+ {{\boldsymbol{u}}_{\skew5\hat{\phantom{i}}}^{\left( 0\right)}} 
\cdot \frac{\partial }{\partial {\boldsymbol{r}}} .
\label{def_Dt}
\end{eqnarray}

\section{Velocity distribution functions of infinitesimal first order}
\label{sec:order1df}

Using the system of infinitesimal first order gas-dynamics equations 
it is possible to specify the form of velocity distribution functions 
of the first order. 
In particular, we shall see, that the velocity distribution functions,
obtained by the proposed method within Enskog's
approach, i.e. when outer components are absent, and by the Enskog's
method are equivalent up to the infinitesimal first order terms of the asymptotic
expansion (inclusive). 

As 
\begin{eqnarray}
\ln f_{\skew5\hat{i}}^{\left( 0\right)}=
\mathrm{const.} + {\ln {n_{\skew5\hat{i}}^{\left( 0\right)}}}
- \frac{3}{2}\,{\ln T_{\skew5\hat{\phantom{i}}}^{\left( 0\right)}}
- \frac{m_{\skew5\hat{i}}\left( {\boldsymbol{c}}_{\skew5\hat{i}}-
{\boldsymbol{u}}_{\skew5\hat{\phantom{i}}}^{\left( 0\right)}\right) ^{2}} 
{2k T_{\skew5\hat{\phantom{i}}}^{\left( 0\right)}} ,
\label{ln_f0_in}
\end{eqnarray}
see (\ref{maxwell_f0_in}) and (\ref{param_f_in}),
the system of integral equations (\ref{r_eq_in}) ${\left( r=1 \right)}$, 
from which $\{{\mathit{\chi } _{\skew5\hat{i}}^{\left( 1\right)}}\}
_{{\skew5\hat{i}} \in \skew4\hat{N}}$ are found,
excluding time derivatives by equations 
(\ref{trans_m1_in})-(\ref{trans_e1_in}), taking
(\ref{p_in_0})-(\ref{E_in_0}) into account,
can be written in the form:
\begin{eqnarray}
\frac{\partial f_{\skew5\hat{i}}^{\left( 0\right)}}{\partial t}
&+&{\boldsymbol{c}}_{\skew5\hat{i}}\cdot 
\frac{\partial f_{\skew5\hat{i}}^{\left( 0\right)}}{\partial \boldsymbol{r}} 
+\frac{{\boldsymbol{X}}_{\skew5\hat{i}}}{m_{\skew5\hat{i}}}\cdot 
\frac{\partial f_{\skew5\hat{i}}^{\left( 0\right)}}
{\partial {\boldsymbol{c}}_{\skew5\hat{i}}}
+ \sum\limits_{{\skew5\check{j}} \in  \skew4\check{N}} 
J_{{\skew5\hat{i}}{\skew5\check{j}}}
\left( f_{\skew5\hat{i}}^{\left( 0\right)},f_{\skew5\check{j}}^{\left( 0\right)}\right)  
\nonumber\\
&=&
f_{\skew5\hat{i}}^{\left( 0\right)} 
\left(
\frac{\partial \ln f_{\skew5\hat{i}}^{\left( 0\right)}}{\partial t}
+{\boldsymbol{c}}_{\skew5\hat{i}}\cdot 
\frac{\partial \ln f_{\skew5\hat{i}}^{\left( 0\right)}}{\partial \boldsymbol{r}} 
+\frac{{\boldsymbol{X}}_{\skew5\hat{i}}}{m_{\skew5\hat{i}}}\cdot 
\frac{\partial \ln f_{\skew5\hat{i}}^{\left( 0\right)}}
{\partial {\boldsymbol{c}}_{\skew5\hat{i}}}
\right)
+ \sum\limits_{{\skew5\check{j}} \in  \skew4\check{N}} 
J_{{\skew5\hat{i}}{\skew5\check{j}}}
\left( f_{\skew5\hat{i}}^{\left( 0\right)},f_{\skew5\check{j}}^{\left( 0\right)}\right)  
\nonumber\\
&=&
f_{\skew5\hat{i}}^{\left( 0\right)} 
\left[
\left( {\mathcal C}_{\skew5\hat{i}}^{2}-\frac{5}{2}\right) 
{\boldsymbol{C}}_{\skew5\hat{i}}
\cdot \frac{\partial \ln 
T_{\skew5\hat{\phantom{i}}}^{\left( 0\right)}}{\partial \boldsymbol{r}} 
+2\,\overset{\circ }
{\left( {\boldsymbol{\mathcal C}}_{\skew5\hat{i}}\,
{\boldsymbol{\mathcal C}}_{\skew5\hat{i}}\right)}:
\frac{\partial }{\partial \boldsymbol{r}}\,
{\boldsymbol{u}}_{\skew5\hat{\phantom{i}}}^{\left( 0\right)}
+\frac{{\hat{n}}^{\left( 0\right)}}
{n_{\skew5\hat{i}}^{\left( 0\right)}}\,
{\boldsymbol{C}}_{\skew5\hat{i}}\cdot {\boldsymbol{d}}_{\skew5\hat{i}}
+ {s}_{\skew5\hat{i}}
\right]
\nonumber\\
&=&
-\sum\limits_{{\skew5\hat{j}} \in \skew4\hat{N}}
\iint
f_{\skew5\hat{i}}^{\left( 0\right)}f_{\skew5\hat{j}}^{\left( 0\right)}
\left( 
\chi _{\skew5\hat{i}}^{\left( 1\right)} + \chi _{\skew5\hat{j}}^{\left( 1\right)} -
\chi _{\skew5\hat{i}}^{\left( 1\right) \prime } - \chi _{\skew5\hat{j}}^{\left( 1\right)\prime } 
\right)
k_{{\skew5\hat{i}}{\skew5\hat{j}}} \,d{\mathbf{k}}
\,d{\boldsymbol{c}}_{\skew5\hat{j}}
\qquad {\left( {{\skew5\hat{i}} \in \skew4\hat{N}} \right)}.
\label{f1_eq_in}
\end{eqnarray}
In (\ref{f1_eq_in}), as well as in the previous section,
${\boldsymbol{C}}_{\skew5\hat{i}} = {\boldsymbol{c}}_{\skew5\hat{i}} - 
{\boldsymbol{u}}_{\skew5\hat{\phantom{i}}}^{\left( 0\right)}$,
\begin{eqnarray}
{\boldsymbol{\mathcal C}}_{\skew5\hat{i}}
= \left( \frac{m_{\skew5\hat{i}}}{2kT_{\skew5\hat{\phantom{i}}}^{\left( 0\right) }}
\right) ^{1 \left/ 2 \right.}{\boldsymbol{C}}_{\skew5\hat{i}} ,
\label{mathcalC_in}
\end{eqnarray}
${\mathcal C}_{\skew5\hat{i}}$ is the modulus of vector 
${\boldsymbol{\mathcal C}}_{\skew5\hat{i}}\,$; 
for arbitrary second rank tensor ${\mathrm{w}}$ 
\begin{eqnarray}
\overset{\circ }{\mathrm{w}}={\mathrm{w}}-\frac{1}{3}\,{\mathrm{U}}
\left( {\mathrm{U}}:{\mathrm{w}}\right)
\label{wcirc}
\end{eqnarray}
is tensor with zero trace;
\begin{eqnarray}
\!\!\!\!\!\!\!\!\!\!\!\!{\boldsymbol{d}}_{\skew5\hat{i}}=
\frac{\partial }{\partial \boldsymbol{r}}
\left( \frac{n_{\skew5\hat{i}}^{\left( 0\right)}}{{\hat{n}}^{\left( 0\right)}}
\right)
+\left( \frac{n_{\skew5\hat{i}}^{\left( 0\right)}}{{\hat{n}}^{\left( 0\right)}}-
\frac{n_{\skew5\hat{i}}^{\left( 0\right)}m_{\skew5\hat{i}}}
{{\hat{\rho }}^{\left( 0\right)}}
\right)
\frac{\partial \ln {{\hat{p}}^{\left( 0\right)}}}
{\partial \boldsymbol{r}}
-\left( \frac{n_{\skew5\hat{i}}^{\left( 0\right)}m_{\skew5\hat{i}}}
{{{\hat{p}}^{\left( 0\right)}}{\hat{\rho }}^{\left( 0\right)}}
\right)
\left[
\frac{{\hat{\rho }}^{\left( 0\right)}}{m_{\skew5\hat{i}}}
{\boldsymbol{X}}_{\skew5\hat{i}}
-\sum\limits_{{\skew5\hat{j}} \in \skew4\hat{N}}
n_{\skew5\hat{j}}^{\left( 0\right)}{\boldsymbol{X}}_{\skew5\hat{j}}
\right];
\label{d_i_in}
\end{eqnarray}
\begin{eqnarray}
{s}_{\skew5\hat{i}}
&=&
- \frac{{\hat{n}}^{\left( 0\right)} m_{\skew5\hat{i}}}
{{{\hat{p}}^{\left( 0\right)}}{\hat{\rho }}^{\left( 0\right)}}\,
{\boldsymbol{C}}_{\skew5\hat{i}}\cdot 
\sum\limits_{{{\skew5\hat{l}} \in \skew4\hat{N}},\, {{\skew5\check{j}} \in \skew4\check{N}}}
{\boldsymbol{J}}_{p,\,{\skew5\hat{l}}{\skew5\check{j}}}^{\,\left( 0\right)}
- \frac{1}{{\hat{p}}^{\left( 0\right)}}
\left( \frac{2}{3}\,{\mathcal C}_{\skew5\hat{i}}^{2}-1\right) 
\sum\limits_{{{\skew5\hat{l}} \in \skew4\hat{N}},\, {{\skew5\check{j}} \in \skew4\check{N}}}
{J}_{E,\,{\skew5\hat{l}}{\skew5\check{j}}}^{\,\left( 0\right)}\nonumber\\
&&+ \frac{1}{f_{\skew5\hat{i}}^{\left( 0\right)}}
\sum\limits_{{\skew5\check{j}} \in  \skew4\check{N}} 
J_{{\skew5\hat{i}}{\skew5\check{j}}}
\left( f_{\skew5\hat{i}}^{\left( 0\right)},f_{\skew5\check{j}}^{\left( 0\right)}\right).
\label{s_i_in}
\end{eqnarray}

The general solution of the system of homogeneous integral equations,
corresponding to the system of equations (\ref{f1_eq_in}), was
constructed above in the section \ref{sec:method}.
A partial solution
$\{{\mathit{\Phi } _{\skew5\hat{i}}^{\left( 1\right)}}\}
_{{\skew5\hat{i}} \in \skew4\hat{N}}$ of the system of linear integral
equations (\ref{f1_eq_in}) 
can be sought in the form:
\begin{eqnarray}
\mathit{\Phi } _{\skew5\hat{i}}^{\left( 1\right)}
=
-\boldsymbol{A} _{\skew5\hat{i}}\cdot 
\frac{\partial \ln T_{\skew5\hat{\phantom{i}}}^{\left( 0\right) }}{\partial \boldsymbol{r}}
-\mathrm{B} _{\skew5\hat{i}}:
\frac{\partial }{\partial \boldsymbol{r}}
{\boldsymbol{u}}_{\skew5\hat{\phantom{i}}}^{\left( 0\right)} 
- {{\hat{n}}^{\left( 0\right)}}
\sum\limits_{{{\skew5\hat{j}} \in \skew4\hat{N}}}
\boldsymbol{D} _{\skew5\hat{i}}^{\skew5\hat{j}}\cdot {\boldsymbol{d}}_{\skew5\hat{j}}
- {S}_{\skew5\hat{i}} .
\label{f1_solution_in}
\end{eqnarray}
Substituting (\ref{f1_solution_in}) in 
(\ref{f1_eq_in}), we find, that family of vector functions
$\{{\boldsymbol{A} _{\skew5\hat{i}}}\left( {\boldsymbol{c}}_{\skew5\hat{i}}\right)\}
_{{\skew5\hat{i}} \in \skew4\hat{N}}$ must satisfy the system of
integral equations
\begin{eqnarray}
\!\!\!\!\!\!\!\!\!\!\!\!f_{\skew5\hat{i}}^{\left( 0\right)} 
\left( {\mathcal C}_{\skew5\hat{i}}^{2}-\frac{5}{2}\right) 
{\boldsymbol{C}}_{\skew5\hat{i}} 
=
\sum\limits_{{\skew5\hat{j}} \in \skew4\hat{N}}
\iint
f_{\skew5\hat{i}}^{\left( 0\right)}f_{\skew5\hat{j}}^{\left( 0\right)}
\left( 
\boldsymbol{A} _{\skew5\hat{i}} + 
\boldsymbol{A} _{\skew5\hat{j}} -
\boldsymbol{A} _{\skew5\hat{i}}^{\prime } - 
\boldsymbol{A} _{\skew5\hat{j}}^{\prime } 
\right)
k_{{\skew5\hat{i}}{\skew5\hat{j}}} \,d{\mathbf{k}}
\,d{\boldsymbol{c}}_{\skew5\hat{j}}
\qquad {\left( {{\skew5\hat{i}} \in \skew4\hat{N}} \right)},
\label{A_eq_in}
\end{eqnarray}
family of tensor functions $\{{\mathrm{B} _{\skew5\hat{i}}}
\left( {\boldsymbol{c}}_{\skew5\hat{i}}\right)\}
_{{\skew5\hat{i}} \in \skew4\hat{N}}$ must satisfy the system of
integral equations
\begin{eqnarray}
2f_{\skew5\hat{i}}^{\left( 0\right)} 
\overset{\circ }
{\left( {\boldsymbol{\mathcal C}}_{\skew5\hat{i}}\,
{\boldsymbol{\mathcal C}}_{\skew5\hat{i}}\right)} 
=
\sum\limits_{{\skew5\hat{j}} \in \skew4\hat{N}}
\iint
f_{\skew5\hat{i}}^{\left( 0\right)}f_{\skew5\hat{j}}^{\left( 0\right)}
\left( 
\mathrm{B} _{\skew5\hat{i}} + 
\mathrm{B} _{\skew5\hat{j}} -
\mathrm{B} _{\skew5\hat{i}}^{\prime } - 
\mathrm{B} _{\skew5\hat{j}}^{\prime } 
\right)
k_{{\skew5\hat{i}}{\skew5\hat{j}}} \,d{\mathbf{k}}
\,d{\boldsymbol{c}}_{\skew5\hat{j}}
\qquad {\left( {{\skew5\hat{i}} \in \skew4\hat{N}} \right)}.
\label{B_eq_in}
\end{eqnarray}
Double family of vector functions 
$\{{\boldsymbol{D} _{\skew5\hat{i}}^{\skew5\hat{j}}}\left( {\boldsymbol{c}}_{\skew5\hat{i}}\right)\}
_{{\skew5\hat{i}}, {\skew5\hat{j}} \in \skew4\hat{N}}$ must satisfy the system of
integral equations 
\begin{eqnarray}
\!\!\!\!\!\!\!\!\!\!\!\!\!\!\!\frac{1}{n_{\skew5\hat{i}}^{\left( 0\right)}}\,
f_{\skew5\hat{i}}^{\left( 0\right)} 
{\boldsymbol{C}}_{\skew5\hat{i}}\cdot {\boldsymbol{d}}_{\skew5\hat{i}}
=
\sum\limits_{{\skew5\hat{j}}, {\skew5\hat{h}} \in \skew4\hat{N}}
\iint
f_{\skew5\hat{i}}^{\left( 0\right)}f_{\skew5\hat{j}}^{\left( 0\right)}
{\boldsymbol{d}}_{\skew5\hat{h}} \cdot
\left( 
\boldsymbol{D} _{\skew5\hat{i}}^{\skew5\hat{h}} + 
\boldsymbol{D} _{\skew5\hat{j}}^{\skew5\hat{h}} -
\boldsymbol{D} _{\skew5\hat{i}}^{\prime \skew5\hat{h}} - 
\boldsymbol{D} _{\skew5\hat{j}}^{\prime \skew5\hat{h}} 
\right)
k_{{\skew5\hat{i}}{\skew5\hat{j}}} \,d{\mathbf{k}}
\,d{\boldsymbol{c}}_{\skew5\hat{j}}
\quad {\left( {{\skew5\hat{i}} \in \skew4\hat{N}} \right)}.
\label{D_eq_in}
\end{eqnarray}
It follows from the definition ${\boldsymbol{d}}_{\skew5\hat{i}}$,
see (\ref{d_i_in}), that 
\begin{eqnarray}
\sum\limits_{{\skew5\hat{i}} \in \skew4\hat{N}}
{\boldsymbol{d}}_{\skew5\hat{i}}
=0,
\label{sum_d_in}
\end{eqnarray}
therefore for every ${\skew5\hat{i}}$ one of coefficients 
${\boldsymbol{D} _{\skew5\hat{i}}^{\skew5\hat{j}}}$ may be considered
as equal to zero; 
for symmetry of expressions it is usually assumed, that
${\boldsymbol{D} _{\skew5\hat{i}}^{\skew5\hat{i}}} 
=0$.
Using (\ref{sum_d_in}), the system of equations (\ref{D_eq_in}) may be
rewritten in the form: 
\begin{eqnarray}
&& \frac{1}{n_{\skew5\hat{i}}^{\left( 0\right)}}\,
f_{\skew5\hat{i}}^{\left( 0\right)} 
{\boldsymbol{C}}_{\skew5\hat{i}} 
\left( {{\delta _{{\skew5\hat{i}}{\skew5\hat{h}}}} - 
{\delta _{{\skew5\hat{i}}{\skew5\hat{k}}}}} \right)
\nonumber\\
&&=
\sum\limits_{{\skew5\hat{j}} \in \skew4\hat{N}}
\iint
f_{\skew5\hat{i}}^{\left( 0\right)}f_{\skew5\hat{j}}^{\left( 0\right)}
\left( 
  \boldsymbol{D} _{\skew5\hat{i}}^{\skew5\hat{h}} 
+ \boldsymbol{D} _{\skew5\hat{j}}^{\skew5\hat{h}} 
- \boldsymbol{D} _{\skew5\hat{i}}^{\skew5\hat{k}} 
- \boldsymbol{D} _{\skew5\hat{j}}^{\skew5\hat{k}} 
- \boldsymbol{D} _{\skew5\hat{i}}^{\prime \skew5\hat{h}} 
- \boldsymbol{D} _{\skew5\hat{j}}^{\prime \skew5\hat{h}} 
+ \boldsymbol{D} _{\skew5\hat{i}}^{\prime \skew5\hat{k}} 
+ \boldsymbol{D} _{\skew5\hat{j}}^{\prime \skew5\hat{k}} 
\right)
k_{{\skew5\hat{i}}{\skew5\hat{j}}} \,d{\mathbf{k}}
\,d{\boldsymbol{c}}_{\skew5\hat{j}}
\nonumber\\
&&
\qquad 
\qquad \qquad \qquad \qquad \qquad \qquad \qquad \qquad \qquad \qquad \qquad \qquad \qquad 
{\left( {{\skew5\hat{i}}, {\skew5\hat{h}}, 
{\skew5\hat{k}} \in \skew4\hat{N}} \right)},
\label{D_eq_wide_in}
\end{eqnarray}
cf. with \cite{hirsch54}, Chapter~7, \S ~3, (3.32).
In (\ref{D_eq_wide_in}) $\delta _{{i}{k}}$ is Kronecker's delta symbol:
\begin{eqnarray}
{\delta _{ik}} = \left\{ {\begin{array}{*{20}{c}}
  {0\quad \,\left( {i \ne k} \right),} \\ 
  {1\quad \,\left( {i = k}   \right).} 
\end{array}} \right.
\label{delta}
\end{eqnarray}
Family of scalar functions
$\{{{S} _{\skew5\hat{i}}}\left( {\boldsymbol{c}}_{\skew5\hat{i}}\right)\}
_{{\skew5\hat{i}} \in \skew4\hat{N}}$ must satisfy the system of
integral equations
\begin{eqnarray}
f_{\skew5\hat{i}}^{\left( 0\right)} 
{s}_{\skew5\hat{i}} 
=
\sum\limits_{{\skew5\hat{j}} \in \skew4\hat{N}}
\iint
f_{\skew5\hat{i}}^{\left( 0\right)}f_{\skew5\hat{j}}^{\left( 0\right)}
\left( 
{S} _{\skew5\hat{i}} + 
{S} _{\skew5\hat{j}} -
{S} _{\skew5\hat{i}}^{\prime } - 
{S} _{\skew5\hat{j}}^{\prime } 
\right)
k_{{\skew5\hat{i}}{\skew5\hat{j}}} \,d{\mathbf{k}}
\,d{\boldsymbol{c}}_{\skew5\hat{j}}
\qquad {\left( {{\skew5\hat{i}} \in \skew4\hat{N}} \right)}.
\label{S_eq_in}
\end{eqnarray}
Solving of the system of integral equations (\ref{S_eq_in}) 
(this question will be considered, possibly, in our next article) 
becomes a little simpler,
if functions ${S} _{\skew5\hat{i}}$ are written in the form:
\begin{eqnarray}
{S} _{\skew5\hat{i}} = {\boldsymbol{S}} _{\skew5\hat{i}}^v \cdot 
\sum\limits_{{{\skew5\hat{l}} \in \skew4\hat{N}},\, {{\skew5\check{j}} \in \skew4\check{N}}}
{\boldsymbol{J}}_{p,\,{\skew5\hat{l}}{\skew5\check{j}}}^{\,\left( 0\right)}
+ {S} _{\skew5\hat{i}}^s.
\label{S_in}
\end{eqnarray}

Conditions of solubility of the system of equations (\ref{A_eq_in}),
(\ref{B_eq_in}), (\ref{D_eq_in}), (\ref{S_eq_in}), following from
(\ref{r_eq_in_condiition}) [in (\ref{r_eq_in_condiition})
$F_{\skew5\hat{i}}^{\left( r \right)}$ must be replaced,
respectively, by left-hand sides of equations (\ref{A_eq_in}),
(\ref{B_eq_in}), (\ref{D_eq_in}) and (\ref{S_eq_in})], 
are fulfilled, because conditions
(\ref{r_eq_in_condiition}) in the form
(\ref{trans_m1_in})-(\ref{trans_e1_in}) were used in the derivation of
the integral equations system (\ref{f1_eq_in}).

Because $\boldsymbol{r}$ and $t$ 
do not enter explicitly in equations (\ref{A_eq_in}), (\ref{B_eq_in})
and (\ref{D_eq_in}),  
${\boldsymbol{u}}_{\skew5\hat{\phantom{i}}}^{\left( 0\right)}$ 
enters only in combination 
$\left( {\boldsymbol{c}}_{\skew5\hat{i}} - 
{\boldsymbol{u}}_{\skew5\hat{\phantom{i}}}^{\left( 0\right)}\right) =
{\boldsymbol{C}}_{\skew5\hat{i}}$ (after replacement of integration
over ${\boldsymbol{c}}_{\skew5\hat{j}}$ by integration
over ${\boldsymbol{C}}_{\skew5\hat{j}}$) and 
the left-hand sides of equations (\ref{B_eq_in}) are symmetric tensors with 
zero trace, functions 
$\{{\boldsymbol{A} _{\skew5\hat{i}}}\left( {\boldsymbol{c}}_{\skew5\hat{i}}\right)\}
_{{\skew5\hat{i}} \in \skew4\hat{N}}$,  
$\{{\mathrm{B} _{\skew5\hat{i}}}
\left( {\boldsymbol{c}}_{\skew5\hat{i}}\right)\}
_{{\skew5\hat{i}} \in \skew4\hat{N}}$ and 
$\{{\boldsymbol{D} _{\skew5\hat{i}}^{\skew5\hat{j}}}\left( {\boldsymbol{c}}_{\skew5\hat{i}}\right)\}
_{{\skew5\hat{i}}, {\skew5\hat{j}} \in \skew4\hat{N}}$ can be sought in the form:
\begin{eqnarray}
{\boldsymbol{A} _{\skew5\hat{i}}}
&=&A_{\skew5\hat{i}}\left( {n_{\skew5\hat{i}}^{\left( 0\right)}}, {\mathcal C}_{\skew5\hat{i}}, 
T_{\skew5\hat{\phantom{i}}}^{\left( 0\right) }\right) 
{\boldsymbol{\mathcal C} _{\skew5\hat{i}}} ,
\label{A_expr_in}
\\
{\mathrm{B} _{\skew5\hat{i}}}
&=&B_{\skew5\hat{i}}\left( {n_{\skew5\hat{i}}^{\left( 0\right)}}, {\mathcal C}_{\skew5\hat{i}}, 
T_{\skew5\hat{\phantom{i}}}^{\left( 0\right) }\right) 
\overset{\circ }
{\left( {\boldsymbol{\mathcal C}}_{\skew5\hat{i}}\,
{\boldsymbol{\mathcal C}}_{\skew5\hat{i}}\right)} ,
\label{B_expr_in}
\\
{\boldsymbol{D} _{\skew5\hat{i}}^{\skew5\hat{j}}}
&=&D_{\skew5\hat{i}}^{\skew5\hat{j}}
\left( {n_{\skew5\hat{i}}^{\left( 0\right)}}, {\mathcal C}_{\skew5\hat{i}}, 
T_{\skew5\hat{\phantom{i}}}^{\left( 0\right) }\right) 
{\boldsymbol{\mathcal C} _{\skew5\hat{i}}} ,
\label{D_expr_in}
\end{eqnarray}
where $A_{\skew5\hat{i}}\left( {n_{\skew5\hat{i}}^{\left( 0\right)}}, {\mathcal C}_{\skew5\hat{i}}, 
T_{\skew5\hat{\phantom{i}}}^{\left( 0\right) }\right)$,
$B_{\skew5\hat{i}}\left( {n_{\skew5\hat{i}}^{\left( 0\right)}}, {\mathcal C}_{\skew5\hat{i}}, 
T_{\skew5\hat{\phantom{i}}}^{\left( 0\right) }\right)$ and 
$D_{\skew5\hat{i}}^{\skew5\hat{j}}
\left( {n_{\skew5\hat{i}}^{\left( 0\right)}}, {\mathcal C}_{\skew5\hat{i}}, 
T_{\skew5\hat{\phantom{i}}}^{\left( 0\right) }\right)$ are
scalar functions of $n_{\skew5\hat{i}}^{\left( 0\right)}$,~${\mathcal C}_{\skew5\hat{i}}$ and 
$T_{\skew5\hat{\phantom{i}}}^{\left( 0\right) }$.

Imposing on the functions
$\{\mathit{\Phi } _{\skew5\hat{i}}^{\left( 1\right)}\}
_{{\skew5\hat{i}} \in \skew4\hat{N}}$ the condition
(\ref{r_eq_in_unique}), from (\ref{f_r_expression_in}) and 
(\ref{beta_1_1_in})-(\ref{beta_3_1_in}) we obtain the following expression
for the infinitesimal first order velocity distribution functions of
inner components particles of gas mixture
$f_{\skew5\hat{i}}^{\left( 1\right)}$~${\left( {{\skew5\hat{i}} \in \skew4\hat{N}} \right)}$:
\begin{eqnarray}
f_{\skew5\hat{i}}^{\left( 1\right)}&=& 
f_{\skew5\hat{i}}^{\left( 0\right)}
\left(
-\boldsymbol{A} _{\skew5\hat{i}}\cdot 
\frac{\partial \ln T_{\skew5\hat{\phantom{i}}}^{\left( 0\right) }}{\partial \boldsymbol{r}}
-\mathrm{B} _{\skew5\hat{i}}:
\frac{\partial }{\partial \boldsymbol{r}}
{\boldsymbol{u}}_{\skew5\hat{\phantom{i}}}^{\left( 0\right)} 
- {{\hat{n}}^{\left( 0\right)}}
\sum\limits_{{{\skew5\hat{j}} \in \skew4\hat{N}}}
\boldsymbol{D} _{\skew5\hat{i}}^{\skew5\hat{j}}\cdot {\boldsymbol{d}}_{\skew5\hat{j}}
- {S}_{\skew5\hat{i}}
\right) \nonumber\\
&&+
f_{\skew5\hat{i}}^{\left( 0\right)}
\left(
\frac{n_{\skew5\hat{i}}^{\left( 1\right)}}{n_{\skew5\hat{i}}^{\left( 0\right)}}
-\frac{3}{2} \frac{T_{\skew5\hat{\phantom{i}}}^{\left( 1\right)}}
{T_{\skew5\hat{\phantom{i}}}^{\left( 0\right)}}
+\frac{{\boldsymbol{u}}_{\skew5\hat{\phantom{i}}}^{\left( {1}\right)}}
{kT_{\skew5\hat{\phantom{i}}}^{\left( 0\right)}}
\cdot m_{\skew5\hat{i}}\,\boldsymbol{C}_{\skew5\hat{i}}
+\frac{1}{kT_{\skew5\hat{\phantom{i}}}^{\left( 0\right)}} 
\frac{T_{\skew5\hat{\phantom{i}}}^{\left( 1\right)}}
{T_{\skew5\hat{\phantom{i}}}^{\left( 0\right)}}
\,\frac{1}{2}\,m_{\skew5\hat{i}}\,C_{\skew5\hat{i}}^{2}
\right) .
\label{f_1_expression_in}
\end{eqnarray}

Similarly,
the system of integral equations (\ref{r_eq_out}) ${\left( r=1 \right)}$, 
from which  
$\{{\mathit{\chi } _{\skew5\check{i}}^{\left( 1\right)}}\}
_{{\skew5\check{i}} \in \skew4\check{N}}$ are found,
excluding time derivatives by equations
(\ref{trans_m1_out})-(\ref{trans_e1_out}) taking
(\ref{p_i_0})-(\ref{E_i_0}) into account, 
can be written in the form:
\begin{eqnarray}
\frac{\partial f_{\skew5\check{i}}^{\left( 0\right)}}{\partial t}
&+&{\boldsymbol{c}}_{\skew5\check{i}}\cdot 
\frac{\partial f_{\skew5\check{i}}^{\left( 0\right)}}{\partial \boldsymbol{r}} 
+\frac{{\boldsymbol{X}}_{\skew5\check{i}}}{m_{\skew5\check{i}}}\cdot 
\frac{\partial f_{\skew5\check{i}}^{\left( 0\right)}}
{\partial {\boldsymbol{c}}_{\skew5\check{i}}}
+ \sum\limits_{j\neq {\skew5\check{i}}} 
J_{{\skew5\check{i}}{{j}}}
\left( f_{\skew5\check{i}}^{\left( 0\right)},f_{{j}}^{\left( 0\right)}\right)  
\nonumber\\
&=&
f_{\skew5\check{i}}^{\left( 0\right)} 
\left(
\frac{\partial \ln f_{\skew5\check{i}}^{\left( 0\right)}}{\partial t}
+{\boldsymbol{c}}_{\skew5\check{i}}\cdot 
\frac{\partial \ln f_{\skew5\check{i}}^{\left( 0\right)}}{\partial \boldsymbol{r}} 
+\frac{{\boldsymbol{X}}_{\skew5\check{i}}}{m_{\skew5\check{i}}}\cdot 
\frac{\partial \ln f_{\skew5\check{i}}^{\left( 0\right)}}
{\partial {\boldsymbol{c}}_{\skew5\check{i}}}
\right)
+ \sum\limits_{j\neq {\skew5\check{i}}} 
J_{{\skew5\check{i}}{{j}}}
\left( f_{\skew5\check{i}}^{\left( 0\right)},f_{{j}}^{\left( 0\right)}\right) 
\nonumber\\
&=&
f_{\skew5\check{i}}^{\left( 0\right)} 
\left[
\left( {\mathcal C}_{\skew5\check{i}}^{2}-\frac{5}{2}\right) 
{\boldsymbol{C}}_{\skew5\check{i}}
\cdot \frac{\partial \ln 
T_{\skew5\check{{i}}}^{\left( 0\right)}}{\partial \boldsymbol{r}} 
+2\,\overset{\circ }
{\left( {\boldsymbol{\mathcal C}}_{\skew5\check{i}}\,
{\boldsymbol{\mathcal C}}_{\skew5\check{i}}\right)}:
\frac{\partial }{\partial \boldsymbol{r}}\,
{\boldsymbol{u}}_{\skew5\check{{i}}}^{\left( 0\right)}
+ {s}_{\skew5\check{i}}
\right]
\nonumber\\
&=&
-\iint
f_{\skew5\check{i}}^{\left( 0\right)}f^{\left( 0\right)}
\left( 
\chi _{\skew5\check{i}}^{\left( 1\right)} + \chi ^{\left( 1\right)} -
\chi _{\skew5\check{i}}^{\left( 1\right) \prime } - \chi ^{\left( 1\right)\prime } 
\right)
k_{{\skew5\check{i}}} \,d{\mathbf{k}}
\,d{\boldsymbol{c}}
\qquad {\left( {{\skew5\check{i}} \in \skew4\check{N}} \right)} ,
\label{f1_eq_out}
\end{eqnarray}
where 
${\boldsymbol{C}}_{\skew5\check{i}} = {\boldsymbol{c}}_{\skew5\check{i}} - 
{\boldsymbol{u}}_{\skew5\check{{i}}}^{\left( 0\right)}$,
\begin{eqnarray}
{\boldsymbol{\mathcal C}}_{\skew5\check{i}}
= \left( \frac{m_{\skew5\check{i}}}{2kT_{\skew5\check{{i}}}^{\left( 0\right) }}
\right) ^{1 \left/ 2 \right.}{\boldsymbol{C}}_{\skew5\check{i}} ,
\label{mathcalC_out}
\end{eqnarray}
${\mathcal C}_{\skew5\check{i}}$ is the modulus of vector 
${\boldsymbol{\mathcal C}}_{\skew5\check{i}}\,$; 
\begin{eqnarray}
{s}_{\skew5\check{i}}
&=&
- \frac{1}{{p}_{\skew5\check{i}}^{\left( 0\right)}}\,
{\boldsymbol{C}}_{\skew5\check{i}}\cdot 
\sum\limits_{j\neq {\skew5\check{i}}}
{\boldsymbol{J}}_{p,\,{\skew5\check{i}}{{j}}}^{\,\left( 0\right)}
- \frac{1}{{p}_{\skew5\check{i}}^{\left( 0\right)}}
\left( \frac{2}{3}\,{\mathcal C}_{\skew5\check{i}}^{2}-1\right) 
\sum\limits_{j\neq {\skew5\check{i}}}
{J}_{E,\,{\skew5\check{i}}{{j}}}^{\,\left( 0\right)}\nonumber\\
&&+ \frac{1}{f_{\skew5\check{i}}^{\left( 0\right)}}
\sum\limits_{j\neq {\skew5\check{i}}} 
J_{{\skew5\check{i}}{{j}}}
\left( f_{\skew5\check{i}}^{\left( 0\right)},f_{{j}}^{\left( 0\right)}\right).
\label{s_i_out}
\end{eqnarray}

A partial solution
$\{{\mathit{\Phi } _{\skew5\check{i}}^{\left( 1\right)}}\}
_{{\skew5\check{i}} \in \skew4\check{N}}$ of the system of 
equations (\ref{f1_eq_out}) can be sought in the form:
\begin{eqnarray}
\mathit{\Phi } _{\skew5\check{i}}^{\left( 1\right)}
=
-\boldsymbol{A} _{\skew5\check{i}}\cdot 
\frac{\partial \ln T_{\skew5\check{{i}}}^{\left( 0\right) }}{\partial \boldsymbol{r}}
-\mathrm{B} _{\skew5\check{i}}:
\frac{\partial }{\partial \boldsymbol{r}}
{\boldsymbol{u}}_{\skew5\check{{i}}}^{\left( 0\right)} 
- {S}_{\skew5\check{i}} .
\label{f1_solution_out}
\end{eqnarray}
Substituting (\ref{f1_solution_out}) in the system of 
linear integral equations (\ref{f1_eq_out}), we find, that family of
vector functions 
$\{{\boldsymbol{A} _{\skew5\check{i}}}\left( {\boldsymbol{c}}_{\skew5\check{i}}\right)\}
_{{\skew5\check{i}} \in \skew4\check{N}}$ must satisfy the system of
integral equations
\begin{eqnarray}
f_{\skew5\check{i}}^{\left( 0\right)} 
\left( {\mathcal C}_{\skew5\check{i}}^{2}-\frac{5}{2}\right) 
{\boldsymbol{C}}_{\skew5\check{i}} 
=
\iint
f_{\skew5\check{i}}^{\left( 0\right)}f^{\left( 0\right)}
\left( 
\boldsymbol{A} _{\skew5\check{i}} + 
\boldsymbol{A} -
\boldsymbol{A} _{\skew5\check{i}}^{\prime } - 
\boldsymbol{A} ^{\prime } 
\right)
k_{{\skew5\check{i}}} \,d{\mathbf{k}}
\,d{\boldsymbol{c}}
\qquad {\left( {{\skew5\check{i}} \in \skew4\check{N}} \right)},
\label{A_eq_out}
\end{eqnarray}
family of tensor functions $\{{\mathrm{B} _{\skew5\check{i}}}
\left( {\boldsymbol{c}}_{\skew5\check{i}}\right)\}
_{{\skew5\check{i}} \in \skew4\check{N}}$ must satisfy the system of
integral equations
\begin{eqnarray}
2f_{\skew5\check{i}}^{\left( 0\right)} 
\overset{\circ }
{\left( {\boldsymbol{\mathcal C}}_{\skew5\check{i}}\,
{\boldsymbol{\mathcal C}}_{\skew5\check{i}}\right)} 
=
\iint
f_{\skew5\check{i}}^{\left( 0\right)}f^{\left( 0\right)}
\left( 
\mathrm{B} _{\skew5\check{i}} + 
\mathrm{B} -
\mathrm{B} _{\skew5\check{i}}^{\prime } - 
\mathrm{B} ^{\prime } 
\right)
k_{{\skew5\check{i}}} \,d{\mathbf{k}}
\,d{\boldsymbol{c}}
\qquad {\left( {{\skew5\check{i}} \in \skew4\check{N}} \right)};
\label{B_eq_out}
\end{eqnarray}
family of scalar functions
$\{{{S} _{\skew5\check{i}}}\left( {\boldsymbol{c}}_{\skew5\check{i}}\right)\}
_{{\skew5\check{i}} \in \skew4\check{N}}$ must satisfy the system of
integral equations
\begin{eqnarray}
f_{\skew5\check{i}}^{\left( 0\right)} 
{s}_{\skew5\check{i}} 
=
\iint
f_{\skew5\check{i}}^{\left( 0\right)}f^{\left( 0\right)}
\left( 
{S} _{\skew5\check{i}} + 
{S} -
{S} _{\skew5\check{i}}^{\prime } - 
{S} ^{\prime } 
\right)
k_{{\skew5\check{i}}} \,d{\mathbf{k}}
\,d{\boldsymbol{c}}
\qquad {\left( {{\skew5\check{i}} \in \skew4\check{N}} \right)},
\label{S_eq_out}
\end{eqnarray}
solving of the system of integral equations
(\ref{S_eq_out}) becomes a little simpler,
if functions ${S} _{\skew5\check{i}}$ are written in the form:
\begin{eqnarray}
{S} _{\skew5\check{i}} = {\boldsymbol{S}} _{\skew5\check{i}}^v \cdot 
\sum\limits_{j\neq {\skew5\check{i}}}
{\boldsymbol{J}}_{p,\,{\skew5\check{i}}{{j}}}^{\,\left( 0\right)}
+ {S} _{\skew5\check{i}}^s.
\label{S_out}
\end{eqnarray}

Conditions of solubility of the systems of equations (\ref{A_eq_out}),
(\ref{B_eq_out}), (\ref{S_eq_out}), following from
(\ref{r_eq_out_condiition}) [in (\ref{r_eq_out_condiition})
$F_{\skew5\check{i}}^{\left( r \right)}$ must be replaced,
by left-hand sides of equations (\ref{A_eq_out}), (\ref{B_eq_out}) and
(\ref{S_eq_out})], are fulfilled, because conditions
(\ref{r_eq_out_condiition}) in the form
(\ref{trans_m1_out})-(\ref{trans_e1_out}) were used in the derivation of
the integral equations system (\ref{f1_eq_out}).

As $\boldsymbol{r}$ and $t$ 
do not enter explicitly in equations (\ref{A_eq_out}) and (\ref{B_eq_out}), 
${\boldsymbol{u}}_{\skew5\check{{i}}}^{\left( 0\right)}$ 
enters only in combination 
$\left( {\boldsymbol{c}}_{\skew5\check{i}} - 
{\boldsymbol{u}}_{\skew5\check{{i}}}^{\left( 0\right)}\right) =
{\boldsymbol{C}}_{\skew5\check{i}}$ (after replacement of integration
over ${\boldsymbol{c}}$ by integration
over ${\boldsymbol{C}}$) and 
the left-hand sides of equations (\ref{B_eq_out})  are symmetric tensors with 
zero trace, functions 
$\{{\boldsymbol{A} _{\skew5\check{i}}}\left( {\boldsymbol{c}}_{\skew5\check{i}}\right)\}
_{{\skew5\check{i}} \in \skew4\check{N}}$ and  
$\{{\mathrm{B} _{\skew5\check{i}}}
\left( {\boldsymbol{c}}_{\skew5\check{i}}\right)\}
_{{\skew5\check{i}} \in \skew4\check{N}}$ 
 can be sought in the form:
\begin{eqnarray}
{\boldsymbol{A} _{\skew5\check{i}}}
&=&A_{\skew5\check{i}}\left( {n_{\skew5\check{i}}^{\left( 0\right)}}, {\mathcal C}_{\skew5\check{i}}, 
T_{\skew5\check{{i}}}^{\left( 0\right) }\right) 
{\boldsymbol{\mathcal C} _{\skew5\check{i}}} ,
\label{A_expr_out}
\\
{\mathrm{B} _{\skew5\check{i}}}
&=&B_{\skew5\check{i}}\left( {n_{\skew5\check{i}}^{\left( 0\right)}}, {\mathcal C}_{\skew5\check{i}}, 
T_{\skew5\check{{i}}}^{\left( 0\right) }\right) 
\overset{\circ }
{\left( {\boldsymbol{\mathcal C}}_{\skew5\check{i}}\,
{\boldsymbol{\mathcal C}}_{\skew5\check{i}}\right)} 
\label{B_expr_out}
\end{eqnarray}
where $A_{\skew5\check{i}}\left( {n_{\skew5\check{i}}^{\left( 0\right)}}, {\mathcal C}_{\skew5\check{i}}, 
T_{\skew5\check{{i}}}^{\left( 0\right) }\right)$ and
$B_{\skew5\check{i}}\left( {n_{\skew5\check{i}}^{\left( 0\right)}}, {\mathcal C}_{\skew5\check{i}}, 
T_{\skew5\check{{i}}}^{\left( 0\right) }\right)$ are
scalar functions of $n_{\skew5\check{i}}^{\left( 0\right)}$,~${\mathcal C}_{\skew5\check{i}}$ and 
$T_{\skew5\check{{i}}}^{\left( 0\right) }$.

Imposing on the functions
$\{\mathit{\Phi } _{\skew5\check{i}}^{\left( 1\right)}\}
_{{\skew5\check{i}} \in \skew4\check{N}}$ the condition
(\ref{r_eq_out_unique}), from (\ref{f_r_expression_out}) and 
(\ref{beta_1_1_out})-(\ref{beta_3_1_out}) we obtain the expression
for the infinitesimal first order velocity distribution functions of
outer components particles of gas mixture
$f_{\skew5\check{i}}^{\left( 1\right)}$~${\left( {{\skew5\check{i}} \in \skew4\check{N}} \right)}$:
\begin{eqnarray}
f_{\skew5\check{i}}^{\left( 1\right)}&=& 
f_{\skew5\check{i}}^{\left( 0\right)}
\left(
-\boldsymbol{A} _{\skew5\check{i}}\cdot 
\frac{\partial \ln T_{\skew5\check{{i}}}^{\left( 0\right) }}{\partial \boldsymbol{r}}
-\mathrm{B} _{\skew5\check{i}}:
\frac{\partial }{\partial \boldsymbol{r}}
{\boldsymbol{u}}_{\skew5\check{{i}}}^{\left( 0\right)} 
- {S}_{\skew5\check{i}}
\right) \nonumber\\
&&+
f_{\skew5\check{i}}^{\left( 0\right)}
\left(
\frac{n_{\skew5\check{i}}^{\left( 1\right)}}{n_{\skew5\check{i}}^{\left( 0\right)}}
-\frac{3}{2} \frac{T_{\skew5\check{{i}}}^{\left( 1\right)}}
{T_{\skew5\check{{i}}}^{\left( 0\right)}}
+\frac{{\boldsymbol{u}}_{\skew5\check{{i}}}^{\left( {1}\right)}}
{kT_{\skew5\check{{i}}}^{\left( 0\right)}}
\cdot m_{\skew5\check{i}}\,\boldsymbol{C}_{\skew5\check{i}}
+\frac{1}{kT_{\skew5\check{{i}}}^{\left( 0\right)}} 
\frac{T_{\skew5\check{{i}}}^{\left( 1\right)}}
{T_{\skew5\check{{i}}}^{\left( 0\right)}}
\,\frac{1}{2}\,m_{\skew5\check{i}}\,C_{\skew5\check{i}}^{2}
\right) .
\label{f_1_expression_out}
\end{eqnarray}

Let us consider a case when outer components in a mixture are absent.
Up to the infinitesimal first order terms (inclusive -- see
\cite{bourbaki2004}, Chapter~V, \S ~2, Section~2, definition~2)
the expression, see (\ref{f_1_expression_in}),
\begin{eqnarray}
f_{\skew5\hat{i}}^{\left( 0\right)}
+f_{\skew5\hat{i}}^{\left( 0\right)}
\left(
\frac{n_{\skew5\hat{i}}^{\left( 1\right)}}{n_{\skew5\hat{i}}^{\left( 0\right)}}
-\frac{3}{2} \frac{T_{\skew5\hat{\phantom{i}}}^{\left( 1\right)}}
{T_{\skew5\hat{\phantom{i}}}^{\left( 0\right)}}
+\frac{{\boldsymbol{u}}_{\skew5\hat{\phantom{i}}}^{\left( {1}\right)}}
{kT_{\skew5\hat{\phantom{i}}}^{\left( 0\right)}}
\cdot m_{\skew5\hat{i}}\,\boldsymbol{C}_{\skew5\hat{i}}
+\frac{1}{kT_{\skew5\hat{\phantom{i}}}^{\left( 0\right)}} 
\frac{T_{\skew5\hat{\phantom{i}}}^{\left( 1\right)}}
{T_{\skew5\hat{\phantom{i}}}^{\left( 0\right)}}
\,\frac{1}{2}\,m_{\skew5\hat{i}}\,C_{\skew5\hat{i}}^{2}
\right)
\label{f0tilda_expression}
\end{eqnarray}
coincides with asymptotic expansion of the solution
$\tilde{f}_{\skew5\hat{i}}^{\left( 0 \right)}$
in the Enskog-Chapman theory
\begin{eqnarray}
\tilde{f}_{\skew5\hat{i}}^{\left( 0 \right)}
=n_{\skew5\hat{i}}^{\left[ 1\right]}
\left( \frac{m_{\skew5\hat{i}}}{2\pi kT_{\skew5\hat{\phantom{i}}}^{\left[ 1\right]}}\right) 
^{3 \left/ 2 \right.}
e^
{-\frac{m_{\skew5\hat{i}}\left( {\boldsymbol{c}}_{\skew5\hat{i}}-
{\boldsymbol{u}}_{\skew5\hat{\phantom{i}}}^{\left[ 1\right]}\right) ^{2}} 
{2kT_{\skew5\hat{\phantom{i}}}^{\left[ 1\right]}}} ,
\label{f0_tilda}
\end{eqnarray}
where 
$n_{\skew5\hat{i}}^{\left[ 1\right]}=n_{\skew5\hat{i}}^{\left( 0\right)}+n_{\skew5\hat{i}}^{\left( 1\right)}$,
${\boldsymbol{u}}_{\skew5\hat{\phantom{i}}}^{\left[ 1\right]}
={\boldsymbol{u}}_{\skew5\hat{\phantom{i}}}^{\left( 0\right)}
+{\boldsymbol{u}}_{\skew5\hat{\phantom{i}}}^{\left( 1\right)}$ and
$T_{\skew5\hat{\phantom{i}}}^{\left[ 1\right]}
=T_{\skew5\hat{\phantom{i}}}^{\left( 0\right)}
+T_{\skew5\hat{\phantom{i}}}^{\left( 1\right)}$, 
cf. with the Taylor expansion of $\tilde{f}_{\skew5\hat{i}}^{\left( 0 \right)}$ at the 
point 
$\left( n_{\skew5\hat{i}}^{\left( 0\right)}, 
\boldsymbol{u}_{\skew5\hat{\phantom{i}}}^{\left( 0\right)}, 
T_{\skew5\hat{\phantom{i}}}^{\left( 0\right)}\right)$.
This statement can be written in the form:
\begin{eqnarray}
\tilde{f}_{\skew5\hat{i}}^{\left( 0 \right)}
\overset{1}{\sim}
f_{\skew5\hat{i}}^{\left( 0\right)}
+f_{\skew5\hat{i}}^{\left( 0\right)}
\left(
\frac{n_{\skew5\hat{i}}^{\left( 1\right)}}{n_{\skew5\hat{i}}^{\left( 0\right)}}
-\frac{3}{2} \frac{T_{\skew5\hat{\phantom{i}}}^{\left( 1\right)}}
{T_{\skew5\hat{\phantom{i}}}^{\left( 0\right)}}
+\frac{{\boldsymbol{u}}_{\skew5\hat{\phantom{i}}}^{\left( {1}\right)}}
{kT_{\skew5\hat{\phantom{i}}}^{\left( 0\right)}}
\cdot m_{\skew5\hat{i}}\,\boldsymbol{C}_{\skew5\hat{i}}
+\frac{1}{kT_{\skew5\hat{\phantom{i}}}^{\left( 0\right)}} 
\frac{T_{\skew5\hat{\phantom{i}}}^{\left( 1\right)}}
{T_{\skew5\hat{\phantom{i}}}^{\left( 0\right)}}
\,\frac{1}{2}\,m_{\skew5\hat{i}}\,C_{\skew5\hat{i}}^{2}
\right) .
\label{f0tilda_equivalence}
\end{eqnarray}

Equations (\ref{A_eq_in}), (\ref{B_eq_in}) and (\ref{D_eq_wide_in})
differ from analogous equations 
\cite{hirsch54}, Chapter~7, (3.34), (3.33) and (3.32) in the Enskog-Chapman theory
essentially only in use 
$n_{\skew5\hat{i}}^{\left( 0\right)}$, 
$\boldsymbol{u}_{\skew5\hat{\phantom{i}}}^{\left( 0\right)}$ and 
$T_{\skew5\hat{\phantom{i}}}^{\left( 0\right)}$ 
instead of $n_i$, $\boldsymbol{u}$ and $T$ (i.e.
$n_{\skew5\hat{i}}^{\left[ 1\right]}$, 
$\boldsymbol{u}_{\skew5\hat{\phantom{i}}}^{\left[ 1\right]}$ and 
$T_{\skew5\hat{\phantom{i}}}^{\left[ 1\right]}$).
Because functions
$\{{\boldsymbol{A} _{\skew5\hat{i}}}\left( {\boldsymbol{c}}_{\skew5\hat{i}}\right)\}
_{{\skew5\hat{i}} \in \skew4\hat{N}}$,  
$\{{\mathrm{B} _{\skew5\hat{i}}}
\left( {\boldsymbol{c}}_{\skew5\hat{i}}\right)\}
_{{\skew5\hat{i}} \in \skew4\hat{N}}$ and 
$\{{\boldsymbol{D} _{\skew5\hat{i}}^{\skew5\hat{j}}}\left( {\boldsymbol{c}}_{\skew5\hat{i}}\right)\}
_{{\skew5\hat{i}}, {\skew5\hat{j}} \in \skew4\hat{N}}$
are of  the infinitesimal first order, in the first, second and third
terms in the right-hand side of
(\ref{f1_solution_in}) and also in (\ref{A_expr_in})-(\ref{D_expr_in})
functions 
$n_{\skew5\hat{i}}^{\left( 0\right)}$, 
$\boldsymbol{u}_{\skew5\hat{\phantom{i}}}^{\left( 0\right)}$ and 
$T_{\skew5\hat{\phantom{i}}}^{\left( 0\right)}$ 
can be, respectively, replaced by functions 
$n_{\skew5\hat{i}}^{\left[ 1\right]}$, 
$\boldsymbol{u}_{\skew5\hat{\phantom{i}}}^{\left[ 1\right]}$ and 
$T_{\skew5\hat{\phantom{i}}}^{\left[ 1\right]}$.
Therefore up to first infinitesimal order terms the expression, 
cf. with (\ref{f_1_expression_in}),
\begin{eqnarray}
f_{\skew5\hat{i}}^{\left( 0\right)}
\left(
-\boldsymbol{A} _{\skew5\hat{i}}\cdot 
\frac{\partial \ln T_{\skew5\hat{\phantom{i}}}^{\left( 0\right) }}{\partial \boldsymbol{r}}
-\mathrm{B} _{\skew5\hat{i}}:
\frac{\partial }{\partial \boldsymbol{r}}
{\boldsymbol{u}}_{\skew5\hat{\phantom{i}}}^{\left( 0\right)} 
- {{\hat{n}}^{\left( 0\right)}}
\sum\limits_{{{\skew5\hat{j}} \in \skew4\hat{N}}}
\boldsymbol{D} _{\skew5\hat{i}}^{\skew5\hat{j}}\cdot {\boldsymbol{d}}_{\skew5\hat{j}}
\right) .
\label{f1n_expression}
\end{eqnarray}
coincides with the expression for the function $\tilde{f}_{\skew5\hat{i}}^{\left( 1 \right)}$
in the Enskog-Chapman theory
\begin{eqnarray}
\tilde{f}_{\skew5\hat{i}}^{\left( 1 \right)}
=
f_{\skew5\hat{i}}^{\left( 0\right)}
\left(
-\skew5\tilde{\boldsymbol{A}}_{\skew5\hat{i}}\cdot 
\frac{\partial \ln T_{\skew5\hat{\phantom{i}}}^{\left( 0\right) }}{\partial \boldsymbol{r}}
-\tilde{\mathrm{B}}_{\skew5\hat{i}}:
\frac{\partial }{\partial \boldsymbol{r}}
{\boldsymbol{u}}_{\skew5\hat{\phantom{i}}}^{\left( 0\right)} 
- {{\hat{n}}^{\left( 0\right)}}
\sum\limits_{{{\skew5\hat{j}} \in \skew4\hat{N}}}
\skew5\tilde{\boldsymbol{D}}_{\skew5\hat{i}}^{\skew5\hat{j}}\cdot 
{\boldsymbol{d}}_{\skew5\hat{j}}
\right) . 
\label{f1_tilda}
\end{eqnarray}
Consequently, up to first infinitesimal order terms the functions  
${f}_{\skew5\hat{i}}^{\left[ 1 \right]}=
{f}_{\skew5\hat{i}}^{\left( 0 \right)}+{f}_{\skew5\hat{i}}^{\left( 1 \right)}$
\textit{coincide} with the functions
$\tilde{f}_{\skew5\hat{i}}^{\left[ 1 \right]}=
\tilde{f}_{\skew5\hat{i}}^{\left( 0 \right)}+\tilde{f}_{\skew5\hat{i}}^{\left( 1 \right)}$, 
obtained in the Enskog-Chapman theory:
\begin{eqnarray}
{f}_{\skew5\hat{i}}^{\left[ 1 \right]}
\overset{1}{\sim}
\tilde{f}_{\skew5\hat{i}}^{\left[ 1 \right]} . 
\label{f1tilda_equivalence}
\end{eqnarray}
As a result, with the same precision (cf. with \cite{chap52},
Chapter~8, \S ~4 and \cite{hirsch54}, Chapter~7, \S ~4)
the expressions for the diffusion velocities of components of gas
mixture, for the ${\skew5\hat{i}}$-component heat flux vector, \ldots
coincide, in particular, the expressions for ${\skew5\hat{i}}$-component
pressure tensor coincide:
\begin{eqnarray}
{\mathrm{p}}_{\skew5\hat{i}}^{\left[ 1\right] }
&=&\int m_{\skew5\hat{i}}
\left( {\boldsymbol{c}}_{\skew5\hat{i}}-{\boldsymbol{u}}_{\skew5\hat{\phantom{i}}}^{\left[ 1\right] }\right)
\left( {\boldsymbol{c}}_{\skew5\hat{i}}-{\boldsymbol{u}}_{\skew5\hat{\phantom{i}}}^{\left[ 1\right] }\right)
f_{\skew5\hat{i}}^{\left[ 1\right] }d\boldsymbol{c}_{\skew5\hat{i}}
\nonumber\\*
&\overset{1}{\sim}&
\left( 
 n_{\skew5\hat{i}}^{\left( 0\right) }kT_{\skew5\hat{\phantom{i}}}^{\left( 0\right) }
+n_{\skew5\hat{i}}^{\left( 1\right) }kT_{\skew5\hat{\phantom{i}}}^{\left( 0\right) }
+n_{\skew5\hat{i}}^{\left( 0\right) }kT_{\skew5\hat{\phantom{i}}}^{\left( 1\right) }
\right) {\mathrm{U}}
\nonumber\\*
&&-
\int m_{\skew5\hat{i}}
\left( {\boldsymbol{c}}_{\skew5\hat{i}}-{\boldsymbol{u}}_{\skew5\hat{\phantom{i}}}^{\left( 0\right) }\right)
\left( {\boldsymbol{c}}_{\skew5\hat{i}}-{\boldsymbol{u}}_{\skew5\hat{\phantom{i}}}^{\left( 0\right) }\right)
f_{\skew5\hat{i}}^{\left( 0\right) }
B_{\skew5\hat{i}}
\left[ 
\overset{\circ }
{\left( {\boldsymbol{\mathcal C}}_{\skew5\hat{i}}\,
{\boldsymbol{\mathcal C}}_{\skew5\hat{i}}\right)}:
\frac{\partial }{\partial \boldsymbol{r}}
{\boldsymbol{u}}_{\skew5\hat{\phantom{i}}}^{\left( 0\right)}
\right] 
d\boldsymbol{c}_{\skew5\hat{i}}
\nonumber\\*
&\overset{1}{\sim}&
n_{\skew5\hat{i}}^{\left[ 1\right] }kT_{\skew5\hat{\phantom{i}}}^{\left[ 1\right] }{\mathrm{U}}
-2
\left[ 
\frac{1}{15}
\,
\frac{m_{\skew5\hat{i}}^2}{2kT_{\skew5\hat{\phantom{i}}}^{\left[ 1\right] }}
\int 
\left( {\boldsymbol{c}}_{\skew5\hat{i}}-
{\boldsymbol{u}}_{\skew5\hat{\phantom{i}}}^{\left[ 1\right] }\right)^4
\tilde{f}_{\skew5\hat{i}}^{\left( 0\right) }
\tilde{B}_{\skew5\hat{i}} \,d\boldsymbol{c}_{\skew5\hat{i}}
\right] 
\overset{\circ }
{\overline{\overline{\frac{\partial }{\partial \boldsymbol{r}}\,
\boldsymbol{u}_{\skew5\hat{\phantom{i}}}^{\left[ 1\right] }}}}
\nonumber\\*
&=&
n_{\skew5\hat{i}}^{\left[ 1\right] }kT^{\left[ 1\right] }{\mathrm{U}}
-2{\tilde{\mu}}_{\skew5\hat{i}}\,
\overset{\circ }
{\overline{\overline{\frac{\partial }{\partial \boldsymbol{r}}
\,\boldsymbol{u}_{\skew5\hat{\phantom{i}}}^{\left[ 1\right] }}}}
=\tilde{\mathrm{p}}_{\skew5\hat{i}}^{\left[ 1\right] } ,
\label{p1}
\end{eqnarray}
-- cf. with incorrect opposite assertions 
(about Hilbert's method for the kinetic Boltzmann equation
asymptotic solution, which, in principle, should lead to the same
results, as the proposed in this article method, see section \ref{sec:remarks}),
for example, in \cite{strum64}, \cite{cercignani75}, \cite{resibois77}; 
in (\ref{p1}) the notations are used: ${\mu}_{\skew5\hat{i}}$ is
viscosity coefficient for ${\skew5\hat{i}}$-component, 
for arbitrary second rank tensor
${\mathrm{w}}$ 
\begin{eqnarray}
{\left( \overline{\overline{\mathrm{w}}}\right)}_{\alpha \beta } 
=\frac{1}{2}\left( w_{\alpha \beta }+w_{\beta \alpha } \right) 
\label{w2overline}
\end{eqnarray}
is symmetric tensor, corresponding to it.

\section{Turbulence as multicomponent gas dynamics}
\label{sec:turbulence}

As is well known, laminar flow becomes turbulent flow, when 
some parameter characterizing the flow, namely, Reynolds number 
\begin{eqnarray}
Re=\frac{\rho u L}{\mu }> 1 .
\label{Reynolds}
\end{eqnarray}
In (\ref{Reynolds}), $\rho$ is the density of gas, $u$ and $L$ are some 
characteristic macroscopic velocity and linear size of the flow, 
$\mu $ is the coefficient of viscosity. 
Having rewritten (\ref{Reynolds}) as 
\begin{eqnarray}
Re=\frac{\rho u^{2}}{\mu \frac{u}{L}} ,
\label{Reynolds_frac}
\end{eqnarray}
cf. with the expression for viscosity tensor in (\ref{p1}), 
the Reynolds number can be treated as the ratio of 
the macroscopic momentum flux, proportional to $f^{\left( 0 \right)}$, 
to the viscosity-induced microscopic momentum 
flux, proportional to $\tilde{f}^{\left( 1 \right)}$
(from the point of view of the kinetic theory of gases instead of
Reynolds number it would be more natural to consider the ratio 
$Re = {f^{\left( 0 \right)}} \left/
{f^{\left( 1 \right)}} \right.$). 
Roughly speaking, viscosity, "aligning" the gas molecules according to a 
Maxwellian distribution with the same mean 
velocities and temperatures for different components of mixture, can
"process" only the microscopic momentum flux. 
But if the macroscopic flux exceeds the microscopic, the gas 
flow, necessarily, has begun to stratify on components
[and other mechanism, leading to an equilibrium state of gas, "turns on" 
-- see integral terms in (\ref{trans_m1_in})-(\ref{trans_e1_out}),
however, from the point of view of the kinetic 
theory of gases it is the same mechanism, collisions of gas particles
with each other]. 
The flow stratification on components can be also caused by external factors. 

The gas-dynamic equations system of the second approximation order
(i.e. with diffusion, viscosity and heat conduction) 
of the Enskog-Chapman theory, in principle, can not describe
gas flow with developed turbulence, as 
\begin{eqnarray}
Re \sim \frac{f^{\left( 0 \right)} }{\tilde{f}^{\left( 1 \right)} } \to \infty , 
\label{Reynolds_paradox}
\end{eqnarray}
when ${\tilde{f}^{\left( 1 \right)} } \to 0$;
with increasing Reynolds number, according to this system of equations, 
the entropy production decreases 
until the complete cessation of growth of the entropy in the gas, 
see the adiabatic equation (\ref{adiabatic}) 
(terms of the gas-dynamic equations system of the second approximation order
of the Enskog-Chapman theory,
additional to the gas-dynamic equations system of the
first approximation order of the Enskog-Chapman theory, 
corresponding to the transition of gas to an equilibrium state and,
hence, to the increasing of its entropy, are proportional to 
${\tilde{f}^{\left( 1 \right)} }$),
whereas the entropy of gas grows with growth of Reynolds number in
experiments, i.e. theoretical (Enskog-Chapman) and experimental
dependencies of entropy of gas on a Reynolds number are 
\textit{different}. 

Mixing is absent in the gas-dynamic equations system of the
first approximation order of the Enskog-Chapman theory
(\ref{trans_m1_E})-(\ref{trans_e1_E}), 
because all components of gas mixture, according to the mass transfer
equations (\ref{trans_m1_E}), move with the same mean mass velocity of 
mixture.
The substance from some physical region (even with moving boundary)
cannot cross region boundary in any way. 
Mixing in the Enskog-Chapman system of gas-dynamic equations, i.e. a
possibility to gas particles to cross boundary of physical region, as
well as other mechanisms (viscosity and heat conduction), leading gas to
an equilibrium, and, hence, increasing entropy of gas, 
appears only in the following infinitesimal order of 
asymptotic expansion of velocity distribution functions of gas
particles and is related to diffusion velocities  
($\sim \tilde{f}_{i}^{\left( 1 \right)}$) in mass transfer equations.  
The ratio of diffusion velocity of some component of gas
mixture to mean velocity of this component 
$\sim {1} \left/
{Re} \right. \sim {\tilde{f}_{i}^{\left( 1 \right)}} \left/
{f_{i}^{\left( 0 \right)}} \right.$ tends to
zero, when ${\tilde{f}_{i}^{\left( 1 \right)} } \to 0$.  
Therefore it is also not possible within Enskog-Chapman theory to
describe observed intensive \textit{turbulent mixing} in gas-dynamic
flows with great Reynolds numbers.

If gas-dynamic equations do not describe  
turbulent gas flows, 
then either something has been missed during the transition from 
the exact solution of the system of kinetic Boltzmann equations to its
approximate solution (by the Enskog method) and next to the
gas-dynamic equations, or the system of kinetic Boltzmann equations
does not describe turbulent gas flows and requires replacement. 
However the last, i.e. necessity of replacement of the system of
kinetic Boltzmann equations on another system of kinetic equations in
transition from gas laminar flow to turbulent gas flow, seems
ill-founded. 

The gas dynamics of the components with the velocity distribution functions,
close to the Maxwell functions, with different mean 
velocities and temperatures, should be described by equations 
(\ref{trans_m1_in})-(\ref{trans_e1_out}). 
From this point of view, 
the observed \textit{chaotic character} of the turbulent flow is similar 
to the chaotic character of the Brownian motion. 
They differ in scale: in the Brownian motion that particle moves 
stochastically, whose mass is comparable to the mass of other gas 
molecules, whereas in the turbulent flow that body moves stochastically, 
whose mass is comparable to the mass of separate gas components. 
In (\ref{trans_p1_in})-(\ref{trans_e1_in}),
(\ref{trans_p1_out})-(\ref{trans_e1_out}) the integral terms 
(proportional to $n_i$, $n_j$) can be  
huge, it explains \textit{unexpected} 
(for those, who tries to describe turbulent flow by the gas-dynamic
equations system of the second approximation order 
of the Enskog-Chapman theory)
power of turbulent effects.

\appendix

\section{Calculation of collision integrals}
\label{sec:integration}

In this section we are dealing with calculation of definite multidimensional 
integrals 
\begin{eqnarray}
\iiiint
\Psi _{i}^{\left( l\right)}\left( f_{i}^{\left( 0\right) \prime }
f_{j}^{\left( 0\right) \prime }-f_{i}^{\left( 0\right)}
f_{j}^{\left( 0\right)} \right)
g_{ij}\, b\,db \,d\epsilon \,d{\boldsymbol{c}}_{i} \,d{\boldsymbol{c}}_{j} .
\label{int_common}
\end{eqnarray}
In (\ref{int_common}) 
$\Psi _{i}^{\left( 1\right)}= m_i$,
$\boldsymbol{\Psi }_{i}^{\left( 2\right)}=m_{i}{\boldsymbol{C}}_{i}$,
$\Psi _{i}^{\left( 3\right)}= \frac{1}{2}\, m_{i}C_{i}^{2}$,
${\boldsymbol{C}}_{i}={\boldsymbol{c}}_{i}-{\boldsymbol{u}}_{i}$;
\begin{eqnarray}
f_{i}^{\left( 0\right)}=
n_{i}\left( \frac{m_{i}}{2\pi kT_{i}}\right) 
^{3 \left/ 2 \right.}
e^{-\frac{m_{i}\left( {\boldsymbol{c}}_{i}-{\boldsymbol{u}}_{i}\right) ^{2}} 
{2kT_{i}}} 
\label{maxwelldist}
\end{eqnarray}
is the Maxwell velocity distribution function of $i$-component
particles, the prime in the distribution function implies, that the  
distribution of particles velocities ${\boldsymbol{c}}^{\,\prime}_{i}$ after the 
collision is considered;
to diminish a little inconvenience of notation, the upper index 
"$^{\left( 0\right)}$" at $n_{i}$, 
${\boldsymbol{u}}_{i}$, $T_{i}$ is omitted. 
The other notation is specified above. 

According to (\ref{equality}), the integral (\ref{int_common}) can be transformed 
as follows:
\begin{eqnarray}
&&\iiiint 
\Psi _{i}^{\left( l\right)}\left( f_{i}^{\left( 0\right) \prime }
f_{j}^{\left( 0\right) \prime }-f_{i}^{\left( 0\right)}
f_{j}^{\left( 0\right)} \right)
g_{ij}\, b\,db \,d\epsilon \,d{\boldsymbol{c}}_{i} \,d{\boldsymbol{c}}_{j} \nonumber\\*
&&\quad =
\iiiint
\Psi _{i}^{\left( l\right)}f_{i}^{\left( 0\right) \prime }
f_{j}^{\left( 0\right) \prime }
g_{ij}^{\prime }\, b^{\prime }\,db^{\prime } \,d\epsilon ^{\prime }\,
d{\boldsymbol{c}}^{\,\prime }_{i} \,d{\boldsymbol{c}}^{\,\prime }_{j} \nonumber\\*
&&\qquad -
\iiiint
\Psi _{i}^{\left( l\right)}f_{i}^{\left( 0\right)}
f_{j}^{\left( 0\right)}
g_{ij}\, b\,db \,d\epsilon \,d{\boldsymbol{c}}_{i} \,d{\boldsymbol{c}}_{j} \nonumber\\*
&&\quad =
\iiiint
\left(\Psi _{i}^{\left( l\right) \prime }-\Psi _{i}^{\left( l\right)} \right)
f_{i}^{\left( 0\right)}f_{j}^{\left( 0\right)}
g_{ij}\, b\,db \,d\epsilon \,d{\boldsymbol{c}}_{i} \,d{\boldsymbol{c}}_{j} .
\label{int_trans}
\end{eqnarray}

As the particle mass is conserved in the collision, 
integral (\ref{int_trans}) vanishes for 
$\Psi _{i}^{\left( 1\right)} = m_i$. 
In two other cases, 
generally speaking, this is not true.

Statements of the two following simple propositions are used 
below several times. 
The proposition \ref{prop2} is taken from 
\cite {bourbaki2004}, Chapter~II, \S ~1, Section~5. 
Regulated functions  (see \cite {bourbaki2004}, Chapter~II,
\S ~1, Section~3) in formulations of propositions can be replaced
with more known continuous functions (any continuous function on
${\boldsymbol{R}}$ is regulated). 
As complete normed vector spaces over the field of real numbers
${\boldsymbol{R}}$
further in the article vector spaces ${\boldsymbol{R}}$ or
${\boldsymbol{R}}^{3}$ over ${\boldsymbol{R}}$ are
considered, with the usual
modulus of a real number or a three-dimensional vector as norm.  

\begin{prop}
$f$ is assumed to be a regulated function on ${\boldsymbol{R}}$ with
values in 
${\boldsymbol{R}}$, 
${\boldsymbol{w}} \in {\boldsymbol{R}}^{3}$ be a fixed nonzero vector, 
${\boldsymbol {n}} \in {\boldsymbol{R}}^{3}$ be a unit vector. 
In this case
\begin{eqnarray}
\int\limits_{\Omega _{\boldsymbol{n}}}f\left( {\boldsymbol{w}}\cdot 
{\boldsymbol{n}}\right) {\boldsymbol{n}}
\,d\Omega _{\boldsymbol{n}}=
\frac{2\pi \boldsymbol{w}}{w}\int\limits_{0}^{\pi }f\left( w\cos \left( \theta \right)
\right) \cos \left( \theta \right) \sin \left( \theta \right) d\theta .
\label{prop1_1}
\end{eqnarray}
In the left-hand side of (\ref{prop1_1}) the integral is taken over all 
directions of vector ${\boldsymbol{n}}$, 
${\boldsymbol{w}}\cdot {\boldsymbol{n}}$ is the scalar product of vectors 
${\boldsymbol{w}}$ and ${\boldsymbol{n}}$. 
\begin{rem}
If ${\boldsymbol{w}}$ is the zero vector, then the right-hand side of
(\ref {prop1_1}) may be set equal to 0.
\end{rem}
\begin{proof}
Select the system of spherical coordinates, such that the 
polar axis direction be the same as the direction of the vector ${\boldsymbol{w}}$. 
Resolve the vector ${\boldsymbol{n}}$ into two components: 
parallel ${\boldsymbol{n}}_{\parallel}$ and perpendicular ${\boldsymbol{n}}_{\perp}$ 
to the vector ${\boldsymbol{w}}$, --
\begin{eqnarray}
{\boldsymbol{n}}={\boldsymbol{n}}_{\parallel }+{\boldsymbol{n}}_{\perp }=
\frac{\left( {\boldsymbol{w}}\cdot {\boldsymbol{n}}\right) {\boldsymbol{w}}}{w^{2}}+
{\boldsymbol{n}}_{\perp } .
\label{prop1_2}
\end{eqnarray}
Having substituted expression (\ref{prop1_2}) for the vector ${\boldsymbol{n}} $ 
into the left-hand side of (\ref{prop1_1}) and integrating over the azimuth 
angle, we obtain the required equality, as in the integration 
over the azimuth angle the term, 
containing ${\boldsymbol {n}}_{\perp}$, vanishes.
\end{proof}
\label{prop1}
\end{prop}

\begin{prop}
$E$ and $F$ is assumed to be two complete normed spaces 
over field ${\boldsymbol{R}}$, 
${\boldsymbol{u}}$ be a continuous linear map of $E$ into $F$. 
In this case,
if ${\boldsymbol{f}}$ is a regulated function on interval 
$I\subset {\boldsymbol{R}}$ with 
values in $E$, then ${\boldsymbol{u}\circ {\boldsymbol{f}}}$ is the
regulated function on $I$ with values in $F$ and
\begin{eqnarray}
\int\limits_{a}^{b}{\boldsymbol{u}}\left( {\boldsymbol{f}}\left( t\right) \right) dt=
{\boldsymbol{u}}\left( \int\limits_{a}^{b}{\boldsymbol{f}}\left( t\right) dt\right) .
\label{prop2_1}
\end{eqnarray}
\begin{proof}
Equality (\ref{prop2_1}) follows immediately from the expression for 
the derivative of composite function 
${\boldsymbol{u}\circ {\boldsymbol{f}}}$; 
details of the proof 
can be found in \cite{bourbaki2004}, Chapter~II, \S ~1, Section~5.
\end{proof}
\label{prop2}
\end{prop}

Major difficulties of evaluation of integral (\ref{int_trans}) are 
caused by that parameters of Maxwell functions for the 
$i$-th and the $j$-th components are not equal: 
\begin{eqnarray}
{\boldsymbol{u}}_{i}\neq {\boldsymbol{u}}_{j},\qquad T_{i}\neq T_{j} .
\label{nequality}
\end{eqnarray}
As result, it is not easy get rid of scalar products 
of vectors in the exponent (it is desirable to have the expression for the 
exponent as simple as possible). 

As the scattering angle depends on the modulus of relative velocity of 
colliding particles \{see, for example, \cite{chap52}, Chapter~3, \S ~4, Section~2 or 
\cite{hirsch54}, Chapter~1, (5.26)\}, it is natural to proceed in 
(\ref{int_trans}) to new 
variables -- center-of-mass velocity ${\boldsymbol{G}}_{ij}$ and 
relative velocity of colliding particles ${\boldsymbol{g}}_{ij}$, which are related to  
particles velocities ${\boldsymbol{c}}_{i}$ and ${\boldsymbol{c}}_{j}$ by 
\begin{eqnarray}
{\boldsymbol{c}}_{i}&=&{\boldsymbol{G}}_{ij}+\frac{m_{j}}{m_{i}+m_{j}}\,{\boldsymbol{g}}_{ij} ,
\label{ciGg}
\\
{\boldsymbol{c}}_{j}&=&{\boldsymbol{G}}_{ij}-\frac{m_{i}}{m_{i}+m_{j}}\,{\boldsymbol{g}}_{ij} ,
\label{cjGg}
\end{eqnarray}
-- cf. with \cite{chap52}, Chapter~9, \S ~2.
For further simplification of the exponent vector 
${\boldsymbol{G}}_{ij}$ can be replaced by vector $\widetilde{\boldsymbol{G}}_{ij}$ 
resulting from ${\boldsymbol{G}}_{ij}$ in an arbitrary affine
transformation, which is a composition of shift, homothety
(multiplication by a scalar) and rotation. 
The rotation arbitrariness is reduced to the freedom 
in choosing of direction of the polar axis in the transition to the 
spherical coordinate system. 
Similarly, the vector ${\boldsymbol{g}}_{ij}$ can be 
replaced by the vector $\tilde{\boldsymbol{g}}_{ij}$, resulting from 
${\boldsymbol{g}}_{ij}$ in composition of arbitrary homothety and arbitrary rotation. 
The shift of the origin of the vector ${\boldsymbol{g}}_{ij}$ would lead to parametric 
dependence of the final integral on \textit{vectors} ${\boldsymbol{u}}_{i}$ and 
${\boldsymbol{u}}_{j}$
(cf. with \cite{oraevskiy85}, Chapter~3), that is undesirable, 
as integral (\ref{int_trans}) is supposed to be reduced to 
Chapman-Cowling integral $\Omega _{ij}^{\left( {l,s} \right)}$ 
[see \cite{chap52}, Chapter~9, \S ~3, (3.29) and 
\cite{hirsch54}, Chapter~7, (4.34)], depending on the
modulus of the relative velocity of colliding particles $g_{ij}$ only. 

In view of the aforesaid, make the following substitution of variables 
${\boldsymbol{G}}_{ij}$ and ${\boldsymbol{g}}_{ij}$: 
\begin{eqnarray}
{\boldsymbol{g}}_{ij}&=&z_{1}\,\tilde{\boldsymbol{g}}_{ij} ,
\label{trans_g}
\\
{\boldsymbol{G}}_{ij}&=&z_{2}\,\widetilde{\boldsymbol{G}}_{ij}+z_{3}\,\tilde{\boldsymbol{g}}_{ij}+
\frac{{\boldsymbol{u}}_{i}+{\boldsymbol{u}}_{j}}{2} .
\label{trans_G}
\end{eqnarray}
In (\ref{trans_g})-(\ref{trans_G}) the scalar factors 
$z_{1}$, $z_{2}$, and $z_{3}$ are 
selected from the condition that the coefficients of 
$\tilde{\boldsymbol{g}}_{ij}^{2}$ and $ \widetilde{\boldsymbol{G}}_{ij}^{2}$ 
in the exponent be equal to 1 and the 
coefficient of the scalar product 
$\tilde{\boldsymbol{g}}_{ij}\cdot \widetilde{\boldsymbol{G}}_{ij}$ 
be equal to 0 (cf. with the method of separation of variables): 
\begin{eqnarray}
z_{1}&=&\sqrt{\frac{2\left( m_{i}T_{j}+m_{j}T_{i}\right)}{m_{i}m_{j}}} ,
\label{z_1}
\\
z_{2}&=&\sqrt{\frac{2T_{i}T_{j}}{m_{i}T_{j}+m_{j}T_{i}}} ,
\label{z_2}
\\
z_{3}&=&\frac{2\left( T_{i}-T_{j}\right)}{m_{i}+m_{j}}
\sqrt{\frac{m_{i}m_{j}}{2\left( m_{i}T_{j}+m_{j}T_{i}\right)}} \,.
\label{z_3}
\end{eqnarray}
Analogous substitutions of variables can be used in more complicated situations,
for example, discussed in \cite{oraevskiy85}, Chapter~3.

With new variables the exponent can be written in the following form: 
\begin{eqnarray}
-\left[ \tilde{g}_{ij}^{2}+\widetilde{G}_{ij}^{2}+a_{0}\,w^{2}+
a_{1}\tilde{\boldsymbol{g}}_{ij}\cdot {\boldsymbol{w}} +
a_{2}\widetilde{\boldsymbol{G}}_{ij}\cdot {\boldsymbol{w}}\right] ,
\label{simple_exp}
\end{eqnarray}
where
\begin{eqnarray}
{\boldsymbol{w}}&=&\frac{{\boldsymbol{u}}_{i}-{\boldsymbol{u}}_{j}}{2} ,
\label{w}
\\
a_{0}&=&\frac{m_{i}}{2T_{i}}+\frac{m_{j}}{2T_{j}} ,
\label{a_0}
\\
a_{1}&=&-2\sqrt{\frac{2m_{i}m_{j}}{m_{i}T_{j}+m_{j}T_{i}}} ,
\label{a_1}
\\
a_{2}&=&\left( \frac{m_{j}}{T_{j}}-\frac{m_{i}}{T_{i}}\right)
\sqrt{\frac{2T_{i}T_{j}}{m_{i}T_{j}+m_{j}T_{i}}} .
\label{a_2}
\end{eqnarray}

It is easy to see, that by the above-specified transformations of
variables only, without using the shift of the origin of vector
${\boldsymbol{g}}_{ij}$, it is 
impossible to get rid of the constant term 
in exponent (\ref{simple_exp}) and, hence, of the constant exponential factor, which will 
appear hereafter in all expressions containing integrals of form 
(\ref{int_common}), (\ref{int_trans}). 
Such factors are missing in \cite{strum74}, (8). 

Determine Jacobian of transformation of variables 
$\left( {\boldsymbol{c}}_{i},{\boldsymbol{c}}_{j}\right) \longrightarrow
\left( \tilde{\boldsymbol{g}}_{ij},\widetilde{\boldsymbol{G}}_{ij}\right)$ 
[see (\ref{trans_g})-(\ref{trans_G})]:
\begin{eqnarray}
\frac{\partial \left( {\boldsymbol{c}}_{i},{\boldsymbol{c}}_{j}\right)}
{\partial \left( \tilde{\boldsymbol{g}}_{ij},\widetilde{\boldsymbol{G}}_{ij}\right)}
&=&
\frac{\partial \left( {\boldsymbol{c}}_{i},{\boldsymbol{c}}_{j}\right)}
{\partial \left( {\boldsymbol{g}}_{ij},{\boldsymbol{G}}_{ij}\right)}
\,\frac{\partial \left( {\boldsymbol{g}}_{ij},{\boldsymbol{G}}_{ij}\right)}
{\partial \left( \tilde{\boldsymbol{g}}_{ij},\widetilde{\boldsymbol{G}}_{ij}\right)}
=z_{1}^{3}\,z_{2}^{3}\,\frac{\partial \left( {\boldsymbol{c}}_{i},{\boldsymbol{c}}_{j}\right)}
{\partial \left( {\boldsymbol{g}}_{ij},{\boldsymbol{c}}_{j}+\frac{m_{i}}{m_{i}+m_{j}}
\,{\boldsymbol{g}}_{ij}\right)} \nonumber\\*
&=&
z_{1}^{3}\,z_{2}^{3}\,\frac{\partial \left( {\boldsymbol{c}}_{i},{\boldsymbol{c}}_{j}\right)}
{\partial \left( {\boldsymbol{g}}_{ij},{\boldsymbol{c}}_{j}\right)}
=z_{1}^{3}\,z_{2}^{3}\,\frac{\partial \left( {\boldsymbol{c}}_{i},{\boldsymbol{c}}_{j}\right)}
{\partial \left( {\boldsymbol{c}}_{i}-{\boldsymbol{c}}_{j},{\boldsymbol{c}}_{j}\right)}
=z_{1}^{3}\,z_{2}^{3} .
\label{jacobian}
\end{eqnarray}

Now consider the case, when 
$\Psi _{i}^{\left( l\right)}=\boldsymbol{\Psi }_{i}^{\left( 2\right)}=
m_{i}\left( {\boldsymbol{c}}_{i}-{\boldsymbol{u}}_{i}\right)$. 
In view of (\ref{simple_exp}), (\ref{jacobian}), (\ref{trans_g})-(\ref{trans_G})
and the equality, following from the definition of $\mathbf{k}$ above,
\begin{eqnarray}
m_{i}\left( {\boldsymbol{c}}_{i}^{\,\prime }-{\boldsymbol{c}}_{i}\right)
=\frac{m_{i}m_{j}}{m_{i}+m_{j}}
\left( {\boldsymbol{g}}_{ij}^{\,\prime } - {\boldsymbol{g}}_{ij}\right)
= -2\,\frac{m_{i}m_{j}}{m_{i}+m_{j}}
\left( {\boldsymbol{g}}_{ij}\cdot {\mathbf{k}}\right) {\mathbf{k}} 
\end{eqnarray}
integral 
(\ref{int_trans}) can be rewritten as: 
\begin{eqnarray}
&&\iiiint
m_{i}\left( {\boldsymbol{c}}_{i}^{\,\prime }-{\boldsymbol{c}}_{i}\right)
f_{i}^{\left( 0\right)}f_{j}^{\left( 0\right)}
g_{ij}\, b\,db \,d\epsilon \,d{\boldsymbol{c}}_{i} \,d{\boldsymbol{c}}_{j} \nonumber\\*
&&\quad =
-2\,\frac{m_{i}m_{j}}{m_{i}+m_{j}}\,z_{1}^{5}\,z_{2}^{3}\,
n_{i}\left( \frac{m_{i}}{2\pi kT_{i}}\right) 
^{3 \left/ 2 \right.}
n_{j}\left( \frac{m_{j}}{2\pi kT_{j}}\right) 
^{3 \left/ 2 \right.}
\iiiint
\left( \tilde{\boldsymbol{g}}_{ij}\cdot {\mathbf{k}}\right) {\mathbf{k}} \nonumber\\*
&&\qquad \times
\exp \left( -\left\{ \tilde{g}_{ij}^{2}+\widetilde{G}_{ij}^{2}+
a_{0}\,w^{2}
+a_{1}\tilde{\boldsymbol{g}}_{ij}\cdot {\boldsymbol{w}}
+a_{2}\widetilde{\boldsymbol{G}}_{ij}\cdot {\boldsymbol{w}} \right\} \right) \nonumber\\*
&&\qquad \times
\tilde{g}_{ij}\, b\,db
\,d\epsilon \,d\widetilde{\boldsymbol{G}}_{ij} \,d\tilde{\boldsymbol{g}}_{ij} .
\label{int_impulse}
\end{eqnarray}
Integrating over $\epsilon $ in (\ref{int_impulse}) 
(with fixed $\tilde{\boldsymbol{g}}_{ij}$ and
$\widetilde{\boldsymbol{G}}_{ij}$), we resolve 
vector ${\mathbf{k}}$ into two components: 
the ones parallel and perpendicular to vector $\tilde{\boldsymbol{g}}_{ij}$ -- 
cf. with the proof of Proposition \ref{prop1}: 
\begin{eqnarray}
\int \left( \tilde{\boldsymbol{g}}_{ij}\cdot {\mathbf{k}}\right) {\mathbf{k}}\,d\epsilon
=2\pi \cos ^{2}\left( \frac{\pi -\chi }{2}\right) \tilde{\boldsymbol{g}}_{ij}
=\pi \left( 1-\cos \chi \right) \tilde{\boldsymbol{g}}_{ij} .
\label{int_epsilon}
\end{eqnarray}
When integrating over $\widetilde{\boldsymbol{G}}_{ij}$ and 
directions of vector $\tilde{\boldsymbol{g}}_{ij}$, we use 
Proposition \ref{prop1}. As a result we obtain
\begin{eqnarray}
{\boldsymbol{J}}_{p,\,ij}^{\,\left( 0\right)}&=&
-\iiiint
m_{i}{\boldsymbol{C}}_{i} \left( f_{i}^{\left( 0\right) \prime }
f_{j}^{\left( 0\right) \prime }-f_{i}^{\left( 0\right)}
f_{j}^{\left( 0\right)} \right)
g_{ij}\, b\,db \,d\epsilon \,d{\boldsymbol{c}}_{i}
\,d{\boldsymbol{c}}_{j} \nonumber\\*
&=&
- \iiiint
m_{i}\left( {\boldsymbol{c}}_{i}^{\,\prime }-{\boldsymbol{c}}_{i}\right)
f_{i}^{\left( 0\right)}f_{j}^{\left( 0\right)}
g_{ij}\, b\,db \,d\epsilon \,d{\boldsymbol{c}}_{i} \,d{\boldsymbol{c}}_{j} \nonumber\\*
&=&
- 16n_{i}n_{j}\,\frac{m_{i}T_{j}+m_{j}T_{i}}{\left( m_{i}+m_{j}\right)}
\,\frac{\boldsymbol{w}}{w}\,\frac{\sqrt{\pi }}{\xi ^{2}}
\,\,e^{-\frac{2m_{i}m_{j}w^{2}}{m_{i}T_{j}+m_{j}T_{i}}} \nonumber\\*
&& \times
\iint e^{-\tilde{g}_{ij}^{2}}
\left[ \tilde{g}_{ij}\xi \cosh \left( \tilde{g}_{ij}\xi \right) -
\sinh \left( \tilde{g}_{ij}\xi \right) \right]
\tilde{g}_{ij}^{2}\left( 1-\cos \chi \right) b\,db\,d\tilde{g}_{ij} .
\label{int_impulse_result}
\end{eqnarray}
In (\ref{int_impulse_result})
\begin{eqnarray}
\xi =a_{1}w ,
\label{xi}
\end{eqnarray}
factor $a_{1}$ is determined by formula (\ref{a_1}). 
It is easy to check, that the 
singularity at $\xi =0$, which is possible when $w=0$, is actually absent 
in the right-hand side. 
Expression (\ref{int_impulse_result}) differs substantially from 
Struminskii's expression \cite{strum74}, (8).

The case, when 
$\Psi _{i}^{\left( l\right)}=\Psi _{i}^{\left( 3\right)}= 
\frac{1}{2}\,m_{i}\left( {\boldsymbol{c}}_{i}-{\boldsymbol{u}}_{i}\right) ^{2}$, 
differs from the just considered one in the factor 
of the exponent in the right-hand side of (\ref{int_impulse}). 
Transform difference 
$\Psi _{i}^{\left( l\right) \prime}-\Psi _{i}^{\left( l\right)}$ 
according to (\ref{trans_g}), (\ref{trans_G}) and 
\cite{chap52}, Chapter~3, (4.9) and taking  
account of that only the relative particles velocity direction changes during 
the collision ($g_{ij} = g_{ij}^{\prime}$): 
\begin{eqnarray}
\Psi _{i}^{\left( 3\right) \prime }
&-&\Psi _{i}^{\left( 3\right)}=
\frac{m_{i}}{2}\left[ \left( {\boldsymbol{c}}_{i}^{\,\prime }-
{\boldsymbol{u}}_{i}\right) ^{2}-
\left( {\boldsymbol{c}}_{i}-{\boldsymbol{u}}_{i}\right) ^{2}\right]
=\frac{m_{i}}{2}\left( {\boldsymbol{c}}_{i}^{\,\prime}-{\boldsymbol{c}}_{i}\right) \cdot
\left( {\boldsymbol{c}}_{i}^{\,\prime }+
{\boldsymbol{c}}_{i}-2{\boldsymbol{u}}_{i}\right) \nonumber\\*
&&=
\frac{m_{i}m_{j}}{m_{i}+m_{j}}\left(
{\boldsymbol{g}}_{ij}^{\,\prime }-{\boldsymbol{g}}_{ij}\right)
\cdot
\left( {\boldsymbol{G}}_{ij}-{\boldsymbol{u}}_{i} \right) \nonumber\\*
&&=
-2\,z_{1}\,\frac{m_{i}m_{j}}{m_{i}+m_{j}}\left( \tilde{\boldsymbol{g}}_{ij} \cdot
{\mathbf{k}}\right) \left( {\mathbf{k}}\cdot \left\{ z_{2}\,\widetilde{\boldsymbol{G}}_{ij}+
z_{3}\,\tilde{\boldsymbol{g}}_{ij}-\frac{{\boldsymbol{u}}_{i}-{\boldsymbol{u}}_{j}}{2}
\right\} \right) .
\label{diff_psi}
\end{eqnarray}

With respect to its arguments the scalar product is a bilinear continuous 
function, therefore Proposition \ref{prop2} can be applied. 
After integration over $\epsilon $, similarly to (\ref{int_epsilon}), we arrive at: 
\begin{eqnarray}
\!\!\!\!\!\!\!\!\!\!\!\!\!\!\!&&-2\,z_{1}\,\frac{m_{i}m_{j}}{m_{i}+m_{j}}
\int \left( \tilde{\boldsymbol{g}}_{ij} \cdot
{\mathbf{k}}\right) \left( {\mathbf{k}}\cdot \left\{ z_{2}\,\widetilde{\boldsymbol{G}}_{ij}+
z_{3}\,\tilde{\boldsymbol{g}}_{ij}-\frac{{\boldsymbol{u}}_{i}-{\boldsymbol{u}}_{j}}{2}
\right\} \right) \,d\epsilon \nonumber\\*
\!\!\!\!\!\!\!\!\!\!\!\!\!\!\!&&\quad =
-2\pi \,z_{1}\,\frac{m_{i}m_{j}}{m_{i}+m_{j}}\left( 1-\cos \chi \right)
\left( \tilde{\boldsymbol{g}}_{ij}\cdot \left\{ z_{2}\,\widetilde{\boldsymbol{G}}_{ij}+
z_{3}\,\tilde{\boldsymbol{g}}_{ij}-\frac{{\boldsymbol{u}}_{i}-{\boldsymbol{u}}_{j}}{2}\right\}
\right) .
\label{int_epsilon_energy}
\end{eqnarray}
Perform the integration over $\widetilde{\boldsymbol{G}}_{ij}$ and 
directions of vector $\tilde{\boldsymbol{g}}_{ij}$, using 
proposition \ref{prop1}:
\begin{eqnarray}
{J}_{E,\,ij}^{\,\left( 0\right)}&=&
-\iiiint
\frac{1}{2}\, m_{i}C_{i}^{2} \left( f_{i}^{\left( 0\right) \prime }
f_{j}^{\left( 0\right) \prime }-f_{i}^{\left( 0\right)}
f_{j}^{\left( 0\right)} \right)
g_{ij}\, b\,db \,d\epsilon \,d{\boldsymbol{c}}_{i} \,d{\boldsymbol{c}}_{j} \nonumber\\*
&=&
- \iiiint
\frac{m_{i}}{2}\left[ \left( {\boldsymbol{c}}_{i}^{\,\prime }-
{\boldsymbol{u}}_{i}\right) ^{2}-
\left( {\boldsymbol{c}}_{i}-{\boldsymbol{u}}_{i}\right) ^{2}\right]
f_{i}^{\left( 0\right)}f_{j}^{\left( 0\right)}
g_{ij}\, b\,db \,d\epsilon \,d{\boldsymbol{c}}_{i} \,d{\boldsymbol{c}}_{j} \nonumber\\*
&=&
{J}_{e,\,ij}^{\,\left( 0\right)}
- {\boldsymbol{u}}_{i} \cdot {\boldsymbol{J}}_{p,\,ij}^{\,\left( 0\right)}
\nonumber\\*
&=&
16n_{i}n_{j}\frac{\sqrt{\pi }}{\xi }
\,e^{-\frac{2m_{i}m_{j}w^{2}}{m_{i}T_{j}+m_{j}T_{i}}}
\nonumber\\*
&&\times \iint e^{-\tilde{g}_{ij}^{2}} 
\left\{ D_{1,\,ij}\frac{w}{\xi }\left[ \tilde{g}_{ij}\xi
\cosh \left( \tilde{g}_{ij}\xi \right) -
\sinh \left( \tilde{g}_{ij}\xi \right) \right] 
+ 2D_{2,\,ij}\tilde{g}_{ij}^{2}\sinh \left( \tilde{g}_{ij}\xi \right)
\right\} \nonumber\\*
&&\qquad \qquad \ \, \times
\tilde{g}_{ij}^{2}\left( 1-\cos \chi \right) b\,db\,d\tilde{g}_{ij} .
\label{int_energy_result}
\end{eqnarray}
In (\ref{int_energy_result}):
\begin{eqnarray}
D_{1,\,ij}&=&\frac{2\,m_{j}T_{i}}{m_{i}+m_{j}} ,
\label{C_1}
\\
D_{2,\,ij}&=&\frac{m_{i}m_{j}\left( T_{i}-T_{j}\right)}{2\,
\left( m_{i}+m_{j}\right) ^{2}}
\,\sqrt{\frac{2\,T_{i}}{m_{i}}+\frac{2\,T_{j}}{m_{j}}} .
\label{C_2}
\end{eqnarray}
The other notations are the same as in (\ref {int_impulse_result}). 

It is interesting to note, that for ${\boldsymbol{u}}_{i}={\boldsymbol{u}}_{j}$ integral 
(\ref{int_impulse_result}) and the 
first term in (\ref{int_energy_result}) vanish, the second term in 
(\ref{int_energy_result}) is proportional $\left( T_{i}-T_{j}\right)$,
that corresponds to energy transfer from the "hot" components to the 
"cold", see the gas-dynamic equations system
(\ref{trans_m1_in})-(\ref{trans_e1_out}).  
In view of the sign of 
$a_{1}$, (\ref{a_1}) and definition of $\xi $ (\ref{xi}) 
the first term leads to \textit{temperature increase} when 
${\boldsymbol{u}}_{i}\neq {\boldsymbol{u}}_{j}$.

\section{Expressions of collision integrals for interaction potential of rigid spheres}
\label{sec:integrals}

Collision integrals (\ref{int_impulse_result}) and
(\ref{int_energy_result}) are complicated functions of 
mean velocities and temperatures of 
separate components, mainly, because of the complicated dependence of
deflection angle $\chi $ on (the modulus of) relative velocity of
colliding particles $g_{ij}$ -- cf. with \cite{hirsch54}, Chapter~1, (5.26):
\begin{eqnarray}
\chi \left( b,g_{ij}\right) =
\pi -2b\int\limits_{r_{min}}^{\infty }\frac{dr/r^{2}}
{\sqrt{1-\frac{2\varphi \left( r\right)}
{m_{ij} g_{ij}^{2}}-\frac{b^{2}}{r^{2}}}} .
\label{chi2}
\end{eqnarray}
In (\ref{chi2})
$m_{ij}={{m_{i} m_{j}} \left/ {\left( {{m_i}+{m_j}} \right)} \right.} $  
is reduced mass of colliding particles, 
${\varphi \left( r\right)}$ -- central interaction potential of
particles, depending on distance $r$ between them.

In the simplest case of particles, interacting as rigid 
spheres with diameters $\sigma _i$ and $\sigma _j$, the following analytical expressions for 
collision integrals have been derived from (\ref{int_impulse_result})
and (\ref{int_energy_result}): 
\begin{eqnarray}
{\boldsymbol{J}}_{p,\,ij}^{\,\left( 0\right)\,\bullet}&=&
- n_i n_j \,\frac{{m_i T_j  + m_j T_i }}
{{m_i  + m_j }}\,\frac{\boldsymbol{w}}
{w}\,\frac{{\sqrt \pi  }}
{{2\xi ^2 }}\ \sigma _{ij}^2 \nonumber\\*
&&\times 
\left[ {
e^{-\xi ^2 \left/ 4 \right.}
2\xi \left( {\xi ^2  + 2} \right) + 
\sqrt \pi  \left( {\xi ^4  + 4 \xi ^2  - 4} \right){\mathrm{erf}}
\left( {\frac{\xi }{2}} \right)} \right] ,
\label{J_p}
\end{eqnarray}
\begin{eqnarray}
{J}_{E,\,ij}^{\,\left( 0\right)\,\bullet}&=&
n_i n_j \,\frac{{\sqrt \pi  }}
{{2\xi ^2 }} \,\sigma _{ij}^2
e^{-\xi ^2 \left/ 4 \right.}
\left[ {2 D_{1,\,ij} w\xi \left( {\xi ^2  + 2} \right) +
2 D_{2,\,ij}\,\xi ^2 \left( {\xi ^2  + 10} \right)} \right] 
\nonumber\\*
&&+\, n_i n_j \,\frac{{\pi  }} {{2\xi ^2 }}\,\sigma _{ij}^2
\nonumber\\*
&&\times  
\left[ {D_{1,\,ij} w\left( {\xi ^4  + 4\xi ^2  - 4} \right) +
D_{2,\,ij} \,\xi \left( {\xi ^4  + 12\xi ^2  + 12} \right)} \right]
{\mathrm{erf}}\left( {\frac{\xi }
{2}} \right) .
\label{J_E}
\end{eqnarray}
In (\ref{J_p})-(\ref{J_E}) 
$\sigma _{ij}={\left( {{\sigma _i} + {\sigma _j}} \right)} 
{\left/  2 \right.}$ and notations from 
(\ref{w}), (\ref{a_1}), (\ref{xi}), (\ref{C_1})-(\ref{C_2}) are used.


\end{document}